\documentclass[a4paper,oneside,11pt]{article}
\usepackage{sf}
\usepackage{amsmath,amstext,amsfonts,amsbsy,amssymb,amscd,bbm,epsfig,lscape}

\newcommand{\ba}{\begin{array}}
\newcommand{\ea}{\end{array}}

\newcommand{\req}[1]{eq.~(\ref{#1})}

\newcommand{\reapp}[1]{Appendix~\ref{#1}}
\newcommand{\rep}[1]{\cite{#1}}

\newcommand{\dif}{{\rm d}}

\newcommand{\Dslash}{\relax{\kern+.25em / \kern-.70em D}}

\newcommand{\Real}{\relax{\mathsf{\Gamma\kern-.35em R}}}
\newcommand{\Int}{\relax{\mathsf{Z\kern-.40em Z}}}

\newcommand{\half}{{\scriptstyle{{1\over 2}}}}

\newcommand{\NC}{N}

\newcommand{\MSbar}{{\overline{\rm MS}}}

\newcommand{\gbar}{\kern1pt\overline{\kern-1pt g\kern-0pt}\kern1pt}
\newcommand{\gren}{g_{\rm R}}

\newcommand{\mbar}{\kern2pt\overline{\kern-1pt m\kern-1pt}\kern1pt}
\newcommand{\mbarf}[1]{\mbar_{\rm #1}}
\newcommand{\mren}[1]{m_{{\rm R} #1}}
\newcommand{\obar}[1]{\kern3pt\overline{\kern-2pt #1\kern-0pt}\kern1pt}
\newcommand{\oren}[1]{#1_{\rm R}}
\newcommand{\corrbar}[1]{\kern3pt\overline{\kern-2pt #1\kern-0pt}\kern1pt}
\newcommand{\corrren}[1]{#1_{\rm R}}

\newcommand{\orgi}[1]{\hat #1}

\newcommand{\hopc}{\kappa_{\rm cr}}

\newcommand{\oVApAV}[1]{#1_{\rm\scriptscriptstyle VA+AV}}

\newcommand{\oVVpAA}[1]{#1_{\rm\scriptscriptstyle VV+AA}}

\newcommand{\oLL}[1]{#1_{\rm\scriptscriptstyle LL}}

\newcommand{\oVApAVren}[1]{\kern3pt\overline{\kern-2pt #1\kern-0pt}\kern1pt_{\rm\scriptscriptstyle VA+AV;s}}

\newcommand{\ZVApAV}[1]{Z_{\rm\scriptscriptstyle VA+AV #1}}

\newcommand{\ZtotVApAV}[1]{\mathcal{Z}_{\rm\scriptscriptstyle VA+AV #1}}

\newcommand{\zbar}{\kern3pt\overline{\kern-2pt Z\kern-0pt}\kern1pt}
\newcommand{\zrgi}{\hat Z}
\newcommand{\zbarVApAV}[1]{\kern3pt\overline{\kern-2pt Z\kern-0pt}\kern1pt_{\rm\scriptscriptstyle VA+AV #1}}
\newcommand{\zrgiVApAV}[1]{\hat Z_{\rm VA+AV #1}}

\newcommand{\SigmaP}{\Sigma_{\rm\scriptscriptstyle P}}

\newcommand{\sigVApAV}[1]{\sigma_{\rm\scriptscriptstyle VA+AV #1}}

\newcommand{\SigVApAV}[1]{\Sigma_{\rm\scriptscriptstyle VA+AV #1}}

\newcommand{\PISigVApAV}[1]{\tilde\Sigma_{\rm\scriptscriptstyle VA+AV #1}}

\newcommand{\UVApAV}[1]{U_{\rm\scriptscriptstyle VA+AV #1}}

\newcommand{\gamVApAV}[1]{\gamma_{\rm\scriptscriptstyle VA+AV #1}}

\newcommand{\lmax}{L_{\rm max}}

\newcommand{\mumin}{\mu_{\rm min}}

\newcommand{\Oa}{\mbox{O}(a)}

\newcommand{\icsw}{c_{\rm sw}}
\newcommand{\ict}{c_{\rm t}}
\newcommand{\icttil}{\tilde c_{\rm t}}

\newcommand{\cD}{{\cal D}}

\newcommand{\cF}{{\cal F}}

\newcommand{\cO}{{\cal O}}

\newcommand{\cZ}{{\cal Z}}

\newcommand{\vx}{\mathbf{x}}
\newcommand{\vy}{\mathbf{y}}

\begin{document}


\begin{titlepage}


\vspace*{-30truemm}
\begin{flushright}
ROM2F/2005-07\\
MS-TP-05-4\\
CERN-PH-TH/2005-044\\
FTUAM-05-5\\
IFT UAM-CSIC/05-17\\
{\large May 2005}
\end{flushright}
\vspace{15truemm}


\centerline{\Bigrm Non-perturbative renormalization of left-left}
\vskip 2 true mm
\centerline{\Bigrm four-fermion operators in quenched lattice QCD}
\vskip 9 true mm
\begin{center}
\epsfig{figure=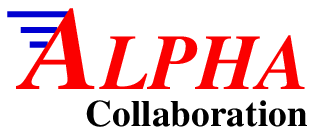, width=22 true mm}\\
\end{center}
\vskip -2 true mm
\centerline{\bigrm  M.~Guagnelli$^a$, J.~Heitger$^b$, C.~Pena$^c$, S.~Sint$^d$ and A.~Vladikas$^a$}
\vskip 4 true mm
\centerline{\it $^a$ INFN, Sezione di Roma II}
\centerline{\it c/o Dipartimento di Fisica, Universit\`a di Roma ``Tor
  Vergata''}
\centerline{\it Via della Ricerca Scientifica 1, I-00133 Rome, Italy}
\vskip 3 true mm
\centerline{\it $^b$ Westf\"alische Wilhelms-Universit\"at M\"unster,
Institut f\"ur Theoretische Physik}
\centerline{\it Wilhelm-Klemm-Strasse 9, D-48149 M\"unster, Germany}
\vskip 3 true mm
\centerline{\it $^c$ CERN, Physics Department, TH Division}
\centerline{\it CH-1211 Geneva 23, Switzerland}
\vskip 3 true mm
\centerline{\it $^d$ Departamento de F\'{\i}sica Te\'orica C-XI and}
\centerline{\it Instituto de F\'{\i}sica Te\'orica C-XVI,}
\centerline{\it Universidad Aut\'onoma de Madrid, Cantoblanco E-28049 Madrid, Spain}
\vskip 10 true mm


\thintablerule
\vskip 3 true mm
\noindent{\tenbf Abstract}
\vskip 1 true mm
\noindent
{\tenrm We define a family of Schr\"odinger Functional renormalization
schemes for the four-quark multiplicatively
renormalizable operators of the $\Delta F = 1$ and 
$\Delta F = 2$ effective weak Hamiltonians. Using the lattice
regularization with quenched Wilson quarks, we compute non-perturbatively the
renormalization group running of these operators in the continuum limit 
in a large range of renormalization scales. Continuum limit
extrapolations are well controlled thanks to the implementation of two
fermionic actions (Wilson and Clover). The ratio of the
renormalization group invariant operator to its renormalized counterpart at a low energy scale,
as well as the renormalization constant at this scale, is obtained
for all schemes.
}
\vskip 3 true mm
\thintablerule
\vspace{10truemm}

\eject
\end{titlepage}

\section{Introduction}
\label{sec:intro}

In the quest for accurate quantitative predictions
in the non-perturbative sector of the Standard Model,
non-perturbative renormalization has become
an essential element of lattice QCD
calculations. Two schemes are currently
in use in many applications, namely the RI/MOM~\cite{RIMOM} 
and the Schr\"odinger Functional (SF)~\cite{AlphaNPren}.
In the latter scheme the scale evolution of (matrix elements of)
renormalized  operators can be traced
non-perturbatively over a wide range of scales. The validity of
perturbation theory at high scales can thus be verified and
one may then convert perturbatively  to one of the commonly used continuum
schemes, such as the $\MSbar$ scheme of dimensional regularization.
Alternatively, one may use low order perturbation theory to extrapolate
from  high to infinite energies, where the so-called
renormalization-group-invariant (RGI) operators are defined.
In any case, perturbation theory is only
used in the high energy regime where it may be safely applied.

These techniques have been applied to study the scale evolution of
various physical quantities, such as the QCD gauge coupling, the
quark mass~\cite{SFcoupling,SFmassRGI,SFmassRGI2,alpha_nf2} and the
moments of pion or nucleon structure functions~\cite{GJP} (both 
in the quenched approximation and for two dynamical flavours), as
well as matrix elements of the heavy-light
axial current, with heavy quarks treated in the static
approximation~\cite{Heitger:2003xg}.
The present work is a first step towards the extension of 
this SF renormalization and renormalization group (RG) evolution programme to
four-quark operators relevant for weak matrix elements.
These arise in the OPE as the low energy QCD contribution in weak 
interaction transitions. They are key elements for the
determination of the CKM unitarity triangle (and the subsequent
understanding of CP-violation). 

We specifically investigate the renormalization of
two dimension-six operators with a ``left-left'' Dirac structure
and four fermions with distinct flavours:
\begin{gather}
  \begin{split}
    \oLL{O}^\pm(x) &= \frac{1}{2} \big[ 
      \left( \bar \psi_1(x) \gamma_\mu^L \psi_2(x) \right) \,\,
      \left(\bar \psi_3(x) \gamma_\mu^L \psi_4(x) \right) \\
      &~~~~~~~~\pm \left( \bar \psi_1(x) \gamma_\mu^L \psi_4(x) \right) \,\,
      \left(\bar \psi_3(x) \gamma_\mu^L \psi_2(x) \right) \big] \\
      &= \oVVpAA{O}^\pm - \oVApAV{O}^\pm \ ,
    \label{eq:4fermopLL}
  \end{split}
\end{gather}
where $\gamma^L_\mu = \gamma_\mu (1 - \gamma_5)$. The last
expression implicitly defines the parity-even and -odd components
of $\oLL{O}^\pm$, in fairly standard notation.
In a chirally symmetric regularization, the operators
$\oLL{O}^\pm$ are multiplicatively renormalizable.
As we opt for lattice regularizations with Wilson fermions,
the loss of chiral symmetry generates extra parity-even counterterms
with finite mixing coefficients. The parity-odd
components $\oVApAV{O}^\pm$ are protected against the generation of parity-odd
counterterms by discrete symmetries.
For a full account of these renormalization properties see 
refs.~\cite{Bernard:1987pr,Donini:1999sf}. In the present work we focus on 
these multiplicatively renormalizable, parity-odd operators $\oVApAV{O}^\pm$.

Once the four generic flavours are identified with specific
physical flavours, the corresponding weak matrix elements give rise to a 
variety of phenomenology. For example, identifying $\psi_1$ and $\psi_3$
with the strange (or bottom) quark and  $\psi_2$ and $\psi_4$ with
the down quark, we obtain the operator $\oVApAV{O}^+$ mediating
$\Delta F = 2$ transitions ($K^0 - \bar K^0$ and
$B^0 - \bar B^0$ oscillations)
in the tmQCD lattice regularization framework~\cite{tmQCD1}. If $\psi_1$
is a strange quark field and the others are suitably chosen light
and charmed quarks, we are looking into $\Delta S = 1$ operators $\oVApAV{O}^\pm$
mediating $K \rightarrow \pi \pi$ transitions. 
Our results completely determine the renormalization of
the operators mediating the $\Delta I = 3/2$ channel, whereas
for the $\Delta I= 1/2$ transitions only logarithmic divergences
are removed. This multiplicative renormalization is
sufficient in the limit of SU(4) flavour symmetry,
of which the chiral limit is a special case. Upon explicit
breaking of this symmetry by the masses of the heavier quark flavours,
the renormalization programme of the $\Delta I= 1/2$ channel
is only complete once the question of mixing with operators of
lower dimension has been addressed. This mixing is beyond the scope
of the current work.

The paper is organised as follows: In sect.~\ref{sec:RGIgen}
we present a general discussion on the RG running of
correlation functions of composite operators, leading to the
definition of the
corresponding renormalization group invariant (RGI) operators. 
In sect.~\ref{sec:RG4f} we introduce the SF renormalization
schemes used for the four-fermion operators in question,
define the operator step scaling functions (SSF), discuss their properties
and show how the operator RG running can be
obtained non-perturbatively from them. In sect.~\ref{sec:ssfNP}
we present our non-perturbative computation of the SSF.
Our results have been obtained in the quenched approximation.
We used both the standard Wilson quark action and its $O(a)$ improved
version, with the Sheikholeslami-Wohlert or clover term~\cite{impr:SW}
(henceforth referred to as Clover action).
Once extrapolated to the continuum limit, the SSF is
used in order to obtain the ratio of the RGI operator to its
renormalized counterpart at a hadronic low energy scale.
In sect.~\ref{sec:match} we compute the operator renormalization
constants at this hadronic matching scale. Finally in sect.~\ref{sec:concl}
we discuss our conclusions. Some technical points have been relegated to 
appendices. Preliminary results had already appeared in
refs.~\cite{PenVladProc}.

In a companion paper~\cite{QvapavPT} the same calculation has been
performed in perturbation theory. The lattice SF schemes have been
matched to a standard $\MSbar$ continuum scheme, at 1-loop in
perturbation theory. Combined with the known NLO
results of the operator anomalous dimension in the
continuum reference scheme, these results give the NLO estimate
of the operator step scaling function. These in turn are used in the present 
work, since part of the calculation involves the RG running of the 
operators at very high scales in
the SF schemes, where NLO perturbation theory may be safely applied.

\section{Callan-Symanzik equations and RGI operators}
\label{sec:RGIgen}

Our starting point is the Callan-Symanzik equation expressing the
RG running of correlation functions under a change of renormalization
scale $\mu$. 
Our exposition and notation follows closely that of Refs.~\cite{SFmassRGI,SFpt1}. 
We first consider an arbitrary bare $n$-point correlation function
\begin{gather}
\begin{split}
G(x_1,\ldots,x_n;g_0,m_{0,{\rm f}}) &= \langle O_1(x_1)\cdots O_n(x_n)\rangle \\
\label{corr_def}
&= \cZ^{-1}\int\cD[\psi,\bar{\psi}]\cD[U] e^{-S} O_1(x_1)\cdots O_n(x_n) \ ,
\end{split}
\end{gather}
where the $O_i$ are local gauge invariant composite operators and $\cZ$ is the QCD partition function.
A regularization such as the lattice (with ultraviolet cutoff $a^{-1}$) is implied.
For simplicity we assume that all space-time points are separated; i.e. $x_i \ne x_j$ for $i \ne j;~i,j =1, \ldots, n$. The dependence on the bare parameters $g_0,m_{0,{\rm f}}$ of the
theory has been indicated explicitly. As this is quite cumbersome,
occasionally we will omit some of the arguments, in order to simplify the notation.
The subscript ${\rm f} =1,\ldots,N_f$ borne by the mass indicates flavour (the mass matrix
will be henceforth assumed to be diagonal). 

It is adequate for the purposes of the present work to consider only multiplicatively
renormalized operators (i.e. no operator mixing occurs). Their renormalized
correlation functions at scale $\mu$ can be written as
\begin{gather}
\corrren{G}(x_1,\ldots,x_n;\mu,\gren,\mren{,\rm f})=
\left[ \prod_{i=1}^n Z_{O_i}(g_0,a\mu) \right]
G(x_1,\ldots,x_n;g_0,m_{0,{\rm f}})
\ .
\label{eq:Gren}
\end{gather}
We denote the renormalized coupling by $\gren$ and
the renormalized quark masses by $\mren{,\rm f}$. 
The operator renormalization constants $Z_{O_i}$
are determined by imposing $n$ renormalization conditions on suitably
chosen correlation functions of the operators $O_i(x)$, at scale $\mu$.
In the context of the present discussion these conditions need not
be specified; it is crucial however to keep in mind that
they are imposed in the chiral limit~\cite{Weinberg}, i.e. we are
only considering mass independent renormalization schemes.

The correlation function defined in eq.~(\ref{eq:Gren}) fulfills a
Callan-Symanzik equation which determines the RG running of the operator 
in question:
\begin{gather}
\left[
\mu\frac{\partial}{\partial\mu} +
\beta(\gren)\frac{\partial}{\partial \gren} +
\tau(\gren)\sum_{{\rm f}=1}^{N_f}\mren{,\rm f}\frac{\partial}{\partial\mren{,\rm f}}
- \sum_{i=1}^n \gamma_{O_i}(\gren)
\right]
\corrren{G} = \Oa \ ,
\label{RGE_corr1}
\end{gather}
where $\beta(\gren)$ in the Callan-Symanzik function, $\tau(\gren)$
the quark mass anomalous dimension and $\gamma_{O_i}(\gren)$
the anomalous dimension of operator $O_i$, which is related to its renormalization constant through
\begin{gather}
\label{gamma_to_Z}
\gamma_O(\gbar(\mu)) = \lim_{a \to 0} \left(\mu \frac{\partial}{\partial \mu} Z_O(g_0,a\mu)\right)
Z_O(g_0,a\mu)^{-1} \ .
\end{gather}
Since the  renormalization scheme we are working in is mass independent,
$\beta$, $\tau$ and $\gamma_O$ depend only on the renormalized
coupling. Their asymptotic expansions at small values of the coupling are given by
\begin{align}
\label{beta_exp}
\beta(g) &\stackrel{g \to 0}{\sim} -g^3\left(b_0 + b_1 g^2 + b_2 g^4 + \ldots \right) \ , \\
\label{tau_exp}
\tau(g) &\stackrel{g \to 0}{\sim} -g^2\left(d_0 + d_1 g^2 + d_2 g^4 + \ldots \right) \ , \\
\gamma_O(g) &\stackrel{g \to 0}{\sim} -g^2\left(\gamma_O^{(0)} + \gamma_O^{(1)} g^2
+ \gamma_O^{(2)} g^4 + \ldots \right) \ .
\label{ad_exp}
\end{align}
Running parameters are then defined as usual:
\begin{gather}
\label{vars_1}
q\frac{\partial\gbar}{\partial q} = \beta(\gbar(q)),~~~~~~~
q\frac{\partial\mbarf{f}}{\partial q} = \tau(\gbar(q))\,\, \mbarf{f}(q) \ ,
\end{gather}
supplemented by the boundary conditions
\begin{gather}
\gbar(\mu) = \gren,~~~~~~~\mbarf{f}(\mu) = \mren{,\rm f}  \ .
\label{vars_2}
\end{gather}

To define RGI composite operators, we start with
the formal integration of \req{RGE_corr1}, yielding
\begin{gather}
\begin{split}
\corrren{G}(x_1,\ldots,x_n;\mu',\gbar(\mu'),\mbarf{f}(\mu')) &= \\
=\Big [ \prod_{i=1}^n U_i(\mu^\prime,\mu) \Big ]
\corrren{G}&(x_1,\ldots,x_n;\mu,\gbar(\mu),\mbarf{f}(\mu)) + \Oa\ ,
\end{split}
\label{matchscales}
\end{gather}
where $U_i$ is the evolution function
\begin{gather}
\label{RG_evolution}
U_i(\mu',\mu) = \exp\left\{\int_{\gbar (\mu)}^{\gbar (\mu')}
\frac{\gamma_{O_i}(g)}{\beta(g)} \dif g
\right\} \ .
\end{gather}
This function describes the RG evolution in the continuum limit of the
renormalized operator $\oren{(O_i)}$ between the renormalization point
$\mu$ and an arbitrary scale $\mu'$, namely:
\begin{gather}
\oren{(O_i)}(x;\mu') = U_i(\mu',\mu) \,\, \oren{(O_i)}(x;\mu) \ .
\label{integrated_RGE}
\end{gather}
It can easily be seen to satisfy the RG equation
\begin{gather}
q\frac{\partial U_i(q,\mu)}{\partial q} = \gamma_O(\gbar(q)) U_i(q,\mu)
\label{eq:zbar}
\end{gather}
with initial condition
\begin{gather}
U_i(\mu,\mu) = 1 \ .
\end{gather}
From eqs.~(\ref{eq:Gren}) and~\req{matchscales} we can also express it
as a ratio of renormalization constants
\begin{gather}
U_i(\mu^\prime,\mu) = \lim_{a \to 0} \frac{Z_{O_i}(g_0,a\mu^\prime)}{Z_{O_i}(g_0,a\mu)} \ .
\label{Urat}
\end{gather}

The RGI operator could in principle be obtained by splitting the 
r.h.s. of \req{RG_evolution} into two integrals (one from $\gbar(\mu')$ to 
$\gbar = 0$ and one from $\gbar = 0$ to $\gbar(\mu)$) and subsequently bringing
the $\mu$-dependent integral on the l.h.s.~of~\req{integrated_RGE}. The
problem is that $U_i(\mu^\prime,\mu)$ diverges logarithmically in the limit $\mu^\prime\to\infty$.
This is most clearly seen upon considering the asymptotic expansions
admitted by $\beta$ (cf. \req{beta_exp}) and $\gamma_O$ (cf. \req{ad_exp})
at small values of the coupling. Hence, we proceed by casting
\req{integrated_RGE} in the form
\begin{gather}
\begin{split}
\bigg[ \frac{\gbar^2 (\mu')}{4 \pi} \bigg]^{-\gamma_O^{(0)}/(2 b_0)}
\oren{O}(x;\mu') &=
\bigg[ \frac{\gbar^2 (\mu)}{4\pi} \bigg]^{-\gamma_O^{(0)}/(2 b_0)}  \times \\
\times
\exp&\left\{-\int_{\gbar (\mu')}^{\gbar (\mu)} {\dif g} \bigg ( \frac{\gamma_O(g)}{\beta(g)}
- \frac{\gamma_O^{(0)}}{b_0 g} \bigg)
\right\} \oren{O}(x;\mu) \, . \\
\label{towards_genRGI_2}
\end{split}
\end{gather}
What has been achieved is the finiteness of the 
r.h.s.~of~\req{towards_genRGI_2} as $\mu'\to\infty$ 
(i.e. $\gbar(\mu')\to 0$)
for any value of $\mu$. Moreover, since there is no $\mu$-dependence on the 
l.h.s., also the r.h.s.~is
$\mu$-independent. Hence, taking the limit  $\mu'\to\infty$ of~\req{towards_genRGI_2}, we define
a RGI quantity as
\begin{gather}
\orgi{O}(x) =  \zrgi_O(\mu)\oren{O}(x;\mu) \ ,
\label{RGI_oper}
\end{gather}
where we have introduced
\begin{gather}
\zrgi_O(\mu) =
\bigg[ \frac{\gbar^2 (\mu)}{4\pi} \bigg]^{-\gamma_O^{(0)}/(2 b_0)}
\exp\left\{-\int_0^{\gbar (\mu)} {\dif g} \bigg ( \frac{\gamma_O(g)}{\beta(g)}
- \frac{\gamma_O^{(0)}}{b_0 g} \bigg)
\right\} \ .
\label{ZRGI}
\end{gather}

It must be stressed that the RGI operator $\orgi{O}(x)$ defined above
is (unlike $\oren{O}(x;\mu)$) independent of the renormalization scheme and scale.
The expressions (\ref{RGI_oper}) and (\ref{ZRGI}) for $\orgi{O}(x)$
are an exact result, in close analogy to the ones reported
in ref.~\cite{SFmassRGI} for the RGI quark mass
\begin{gather}
M_{\rm f} = \mbarf{f} (\mu) \left( 2 b_0 \gbar^2(\mu)\right)^{-d_0/(2b_0)} 
\exp \left\{ - \int_0^{\gbar(\mu)} {\dif g} \bigg( \frac{\tau(g)}{\beta(g)}
- \frac{d_0}{b_0 g} \bigg) \right\} \, ,
\label{RGI_mass}
\end{gather}
and the RGI scale
\begin{gather}
\begin{split}
\Lambda = \mu &\left( b_0 \gbar^2(\mu)\right)^{-b_1/(2b_0^2)} \exp\left\{ - \frac{1}{2 b_0 \gbar^2(\mu)} \right\}
\times\\
&\times \exp \left\{ - \int_0^{\gbar(\mu)} {\dif g} \bigg( \frac{1}{\beta(g)} + \frac{1}{b_0 g^3} -
\frac{b_1}{b_0^2 g} \bigg) \right\} \, .
\end{split}
\end{gather}
Notice that the only arbitrariness in the definition of these RGI quantities
is a constant overall normalization factor. For
composite operators in Eq.~(\ref{RGI_oper}) we have adopted the
normalization usually employed in the definition of the RGI kaon mixing
parameter $\hat B_K$.

\section{RG running of four-fermion operators}
\label{sec:RG4f}

Having exposed the general principles for the RG behaviour of multiplicatively renormalizable composite
operators, we now pass to the specific case of interest. This refers to the dimension-six 
composite operators of four distinct quark flavours
\begin{gather}
  \begin{split}
    \oVApAV{O}^\pm(x) =& \frac{1}{2} \big\{\big[ 
       (\bar \psi_1 \gamma_\mu \psi_2)
       (\bar \psi_3 \gamma_\mu\gamma_5 \psi_4)
      +(\bar \psi_1 \gamma_\mu\gamma_5 \psi_2)
       (\bar \psi_3 \gamma_\mu \psi_4) \big] \\
       &\pm\big[
       (\bar \psi_1 \gamma_\mu \psi_4)
       (\bar \psi_3 \gamma_\mu\gamma_5 \psi_2)
      +(\bar \psi_1 \gamma_\mu\gamma_5 \psi_4)
       (\bar \psi_3 \gamma_\mu \psi_2) \big]\big\} \, ,
  \end{split}
  \label{eq:4fermop}
\end{gather}
which are known to be multiplicatively renormalizable~\cite{Bernard:1987pr,Donini:1999sf}. 
In this section we define the correlation functions of interest,
the renormalization conditions imposed and the step scaling functions of the operators $\oVApAV{O}^\pm$.

As anticipated, we opt for the lattice Schr\"odinger functional (SF) formalism~\rep{SF,SFLNWW,SFS}.
We regularize QCD on a lattice of extension $L^3 \times T$ (here $T=L$ always) with periodic
boundary conditions in the space directions (up to a phase $\theta$ for the fermion fields) and
Dirichlet boundary conditions in the Euclidean time direction~\rep{SFLNWW,SFS}.
Otherwise the lattice gauge and fermionic field actions are of the standard Wilson
type; the Clover $\Oa$ improved version of the fermionic action is also used.
The operators $\oVApAV{O}^\pm(x)$ are defined locally on the lattice; i.e. all quark fields
live at the point $x$.

The paper follows closely the notation of ref.~\cite{SFchiral}, to
which the reader is referred for unexplained notation.

\subsection{Schr\"odinger Functional correlation functions}

Bare composite operators are defined at both time boundaries in terms of the
boundary fields $\zeta$ and $\zeta'$ of refs.~\cite{SFS,SFchiral,SFctt},
\begin{gather}
\begin{split}
\cO_{12}[\Gamma] &= a^6 \sum_{\vx, \vy} \bar \zeta_1(\vx) \Gamma \zeta_2(\vy) \, ,
\\
\cO^\prime_{12}[\Gamma] &= a^6 \sum_{\vx, \vy} \bar \zeta^\prime_1(\vx) \Gamma \zeta^\prime_2(\vy) \, .
\end{split}
\label{boundops}
\end{gather}
The indices 1,2 label distinct flavours. Unprimed fields are defined on the $x_0=0$ 
boundary, primed ones  on the  $x_0=T$ one. There are two allowed independent
choices for the Dirac matrices, namely $\Gamma = \gamma_5$ and $\Gamma = \gamma_k$ 
(with $k = 1,2,3$). This is due to the SF Dirichlet boundary conditions of the quark fields, which
involve positive and negative projection operators
$P_\pm=\half(1\pm\gamma_0)$~\cite{SFctt}. The presence of these projectors implies that
boundary sources with other Dirac matrices $\Gamma$ either vanish or
are identical  to those with $\gamma_5$ or $\gamma_k$.

\begin{figure}[!t]
\vspace{40mm}
\includegraphics{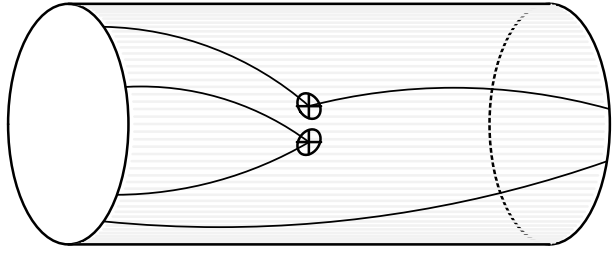}
\caption{
Four-fermion correlation functions in a finite spacetime volume with Schr\"odinger Functional boundary
conditions. The lines denote valence quark propagation of different flavours; the double dots in the bulk
denote the four-fermion operator.
}
\label{fig:traces_ren}
\end{figure}

The bare correlation functions of the four-fermion operators are now chosen as follows:
\begin{gather}
\cF^\pm_{[\Gamma_{\rm A},\Gamma_{\rm B},\Gamma_{\rm C}]} (x_0) = \frac{1}{L^3}
\langle \cO_{21}[\Gamma_{\rm A}]  \cO_{45}[\Gamma_{\rm B}] 
\,\,\,  \oVApAV{O}^\pm(x) \,\,\, 
\cO^\prime_{53}[\Gamma_{\rm C}] \rangle \, .
\label{eq:h-corr}
\end{gather}
These are depicted, in terms of valence quark lines, in Fig.~\ref{fig:traces_ren}. 
Since the boundary operators defined in eqs.~(\ref{boundops}) involve
sums over all 3-space at each
time boundary, translational invariance implies that the above
correlation functions depend only on time.
A few words are in place in order to motivate the choice of this
rather complicated quantity, involving
three composite operators at the boundary. As stated above, 
the boundary operators can either be $\cO_{12}[\gamma_5]$
or $\cO_{12}[\gamma_k]$ (and $\cO^\prime_{12}[\gamma_5]$, $\cO^\prime_{12}[\gamma_k]$). 
Moreover, the bulk operators $\oVApAV{O}^\pm(x)$ are
parity-odd. These facts, combined with the requirement that cubic
symmetry be respected by the correlation
functions, give as simplest possibilities the following five,
in principle independent correlation functions:
\begin{gather}
\begin{split}
\label{eq:4f-corr}
F^\pm_1 (x_0) &=  \cF^\pm_{[\gamma_5,\gamma_5,\gamma_5]} (x_0)  \, ,\\
F^\pm_2 (x_0) &= \frac{1}{6} \sum_{j,k,l=1}^3 \epsilon_{jkl} \cF^\pm_{[\gamma_j,\gamma_k,\gamma_l]} (x_0) \, , \\
F^\pm_3 (x_0) &= \frac{1}{3} \sum_{k=1}^3 \cF^\pm_{[\gamma_5,\gamma_k,\gamma_k]}(x_0) \, ,\\
F^\pm_4 (x_0) &= \frac{1}{3} \sum_{k=1}^3 \cF^\pm_{[\gamma_k,\gamma_5,\gamma_k]}(x_0)  \, ,\\
F^\pm_5 (x_0) &= \frac{1}{3} \sum_{k=1}^3 \cF^\pm_{[\gamma_k,\gamma_k,\gamma_5]}(x_0) \, . 
\end{split}
\end{gather}

\begin{figure}[!t]
\vspace{40mm}
\includegraphics{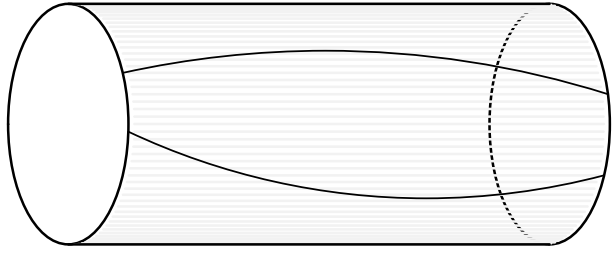}
\caption{
Boundary-to-boundary correlation function in a finite spacetime volume with Schr\"odinger Functional boundary
conditions. The lines denote valence quark propagation of different flavours.
}
\label{fig:traces_bb}
\end{figure}

We will also need the boundary-to-boundary correlation functions
\begin{gather}
\begin{split}
f_1 &= -\frac{1}{2 L^6} \langle \cO^\prime _{12}[\gamma_5] \,\,\, \cO_{21}[\gamma_5] \rangle \, ,
\\
k_1 &= -\frac{1}{6 L^6}  \sum_{k=1}^3 \langle \cO^\prime _{12}[\gamma_k] \,\,\, \cO_{21}[\gamma_k]  \rangle \, .
\end{split}
\label{eq:f-corr}
\end{gather}
In terms of valence quark propagators, these correlation functions are depicted in Fig.~\ref{fig:traces_bb}.

In practical simulations, these correlation functions are computed as traces of the boundary-to-bulk valence
quark propagators $H_{\rm f}(x)$ and $H^\prime_{\rm f}(x)$, defined in ref.~\cite{Luscher:1996jn}.

\subsection{Schr\"odinger Functional renormalization schemes}

Before turning to the renormalization of four-fermion operators, we recall that SF renormalization
schemes are mass independent; i.e.~renormalization is performed in the
chiral limit. The renormalization scale is set at $\mu = 1/L$; the renormalized coupling $\gbar(1/L)$ 
(defined in ref.~\cite{SFcoupling}) and quark mass  $\mbar(1/L)$ (defined in refs.~\cite{SFmassRGI,SFpt1})
are then only functions of this scale.

Upon removing the ultraviolet cutoff (i.e. $a \rightarrow  0$), the bare correlation functions 
defined in eqs.~(\ref{eq:h-corr}, \ref{eq:f-corr}) diverge logarithmically. The divergence due to
the boundary fields is removed by considering suitable ratios of correlation functions. Several
choices can be made, giving rise to different correlator ratios. In the present work we will be
considering the following nine specific cases
\begin{gather}
\begin{split}
h^\pm_{\rm i}(x_0) = \frac{F^\pm_{\rm i}(x_0)}{f_1^{3/2}}
&\qquad {\rm i} = 1,\dots,5 \ ,\\
\label{h-corr}
h^\pm_6(x_0) = \frac{F^\pm_2(x_0)}{k_1^{3/2}} & \ ,\\
h^\pm_{{\rm i}+4}(x_0) = \frac{F^\pm_{\rm i}(x_0)}{f_1^{1/2}
  k_1}
&\qquad {\rm i} = 3,4,5 \ ,
\end{split}
\end{gather}
which renormalize as the four-fermion operators $\oVApAV{O}^\pm$ themselves:
\begin{gather}
h^\pm_{\rm R;s}(x_0;\mu) = 
\ZVApAV{;{\rm s}}^\pm(g_0,a \mu)  h^\pm_{\rm s}(x_0;g_0) \qquad {\rm s}=1,\ldots,9 \ .
\end{gather}
The above renormalization constants are fixed by imposing the
following renormalization conditions on the correlator
$h^\pm_{\rm s}$ on time-slice $x_0 = L/2$ (for all
${\rm s} = 1, \dots , 9$)
at scale $\mu = 1/L$ and fixed
renormalized coupling $\gbar^2(1/L) = u$ in the chiral limit:
\begin{gather}
\label{eq:rencond}
\ZVApAV{;{\rm s}}^\pm(g_0,a \mu)  h^\pm_{\rm s}(x_0;g_0)=
h^\pm_{\rm s}(x_0;g_0) \bigg \vert_{g_0 = 0}  \,\, ;
\end{gather}
i.e. at tree level $\ZVApAV{;s}^\pm=1$ by construction.
We will always impose eq.~(\ref{eq:rencond}) at $\theta=0.5$~\rep{SFmassRGI,SFpt1}.
The nine correlator ratios chosen above give rise to nine 
in principle distinct SF renormalization
schemes for each of the two operators.\footnote{Some considerations
concerning the independence of the different schemes can be found
in~\reapp{app:schemes_differ}.}

The above construction hinges on a theory of five flavours, the fifth
one being a spectator quark. This, however, is only apparent. As
already mentioned in the introduction, we make contact with a specific
weak matrix element by judiciously attributing specific physical flavour
labels to the above five nominal flavours. For example, the
identifications
\begin{gather}
\psi_1 = \psi_3 = s \, ,
\qquad \psi_2 = \psi_4 = d \, , \qquad \psi_5 = u
\label{eq:phflav}
\end{gather}
lead (up to an irrelevant factor of 2 arising from a doubling of
Wick contractions) to the renormalization of the ``left-left''
operator $\oVApAV{O}^+$, which mediates $\Delta S = 2$ transitions.

\subsection{Step scaling functions}
\label{sec:ssf}

The step scaling functions (SSF) of the four-fermion operators of 
eq.~(\ref{eq:4fermop}) are defined as
\begin{gather}
\label{eq:ssfLat}
\SigVApAV{;s}^\pm (u,a/L) = \frac{\ZVApAV{;{\rm s}}^\pm(g_0,a/2L)}{\ZVApAV{;{\rm s}}^\pm(g_0,a/L)}
\Bigg \vert_{m=0,~\gbar^2(1/L) = u } \,\, .
\end{gather}
This is in close analogy to the quark mass case~\cite{SFmassRGI}; i.e. $\SigVApAV{;s}^\pm$ is defined
in the chiral limit $m(g_0)=0$, for a lattice of a given resolution $L/a$ and at fixed renormalized coupling 
$\gbar^2(1/L) = u$.
The precise definition of the current quark mass $m(g_0)$ can be found in ref.~\cite{SFmassRGI}.
The lattice SSF $\SigVApAV{;s}^\pm$ is not unique: it depends on the details of
the lattice regularization (e.g. the type of lattice action chosen,
the level of $\Oa$ improvement etc.). It has, however, a well defined
continuum limit, which should be unique (i.e. universality should hold).
We denote the continuum SSF by
\begin{gather}
\label{eq:ssfCont}
\sigVApAV{;s}^\pm(u) = \lim_{a \to 0} \SigVApAV{;s}^\pm(u,a/L)
\,\, .
\end{gather}
In terms of the operators' evolution function $\UVApAV{;s}^\pm$
and anomalous dimension $\gamVApAV{;s}^\pm$ the SSF can be written
as (cf. eqs.~(\ref{RG_evolution},\ref{Urat}))
\begin{gather}
\label{eq:rgInt}
\sigVApAV{;s}^\pm(u) = \UVApAV{;s}^\pm \Big (\frac{1}{2L},\frac{1}{L} \Big )
  = \exp \left\{\int_{\gbar(1/L)}^{\gbar(1/2L)}
  \frac{\gamVApAV{;s}^\pm(g)}{\beta(g)} \,\dif g \right\} \, .
\end{gather}
Thus the physical meaning of $\sigVApAV{;s}^\pm$ emerges readily from the above
as the operator evolution function between two scales differing by a
factor of 2. It is a quantity closely related to the anomalous
dimension of the corresponding operator.

We stress that the operator anomalous dimension is
scheme dependent. Its perturbative expansion is known to two-loop order; the
universal one-loop coefficient is
\begin{gather}
\gamVApAV{}^{\pm(0)} = \pm \frac{1}{(4\pi)^2}\frac{6(\NC \mp 1)}{\NC}
\, ,
\end{gather}
($N$ being the number of colours) while
the two-loop coefficients $\gamVApAV{;s}^{\pm(1)}$ have been calculated in \rep{QvapavPT}
for the schemes defined by eqs.~(\ref{eq:rencond}).

Finally, from eq.~(\ref{ZRGI}) we immediately obtain the following
expression for the factor relating the renormalized operator at a
scale $\mu$ with its RGI counterpart:
\begin{gather}
\begin{split}
\zrgiVApAV{;s}^\pm(\mu) =
\bigg[ \frac{\gbar^2 (\mu)}{4\pi} \bigg]^{-\gamVApAV{}^{\pm (0)}/(2b_0)} &\times \\
\times
\exp\bigg\{-\int_0^{\gbar (\mu)} &{\dif g} \bigg ( \frac{\gamVApAV{;s}^\pm(g)}{\beta(g)}
- \frac{\gamVApAV{;s}^{\pm (0)}}{b_0 g} \bigg)
\bigg\} \, .
\end{split}
\label{RGI_VApAV}
\end{gather}
It is the aim of the present work to provide accurate
estimates of the above quantity in all nine schemes and for a
large range of scales $\mu = 1/L$ (albeit in the quenched approximation).

It is useful to keep in mind that the only flavour dependence of $\zrgiVApAV{;s}^\pm(\mu)$
is through $N_f$; there is no dependence on the values of the physical
quark masses, as this quantity in defined (and computed) in the chiral
limit. Thus it can be readily used in the renormalization of various
physical matrix elements.

\subsection{RG running of four-fermion operators}
\label{subsec:RGrun}

Once the step scaling functions $\sigVApAV{;s}^\pm$ have been computed through numerical
simulation, the ratio of renormalized correlation functions involving
the four-fermion operators of interest between 
the minimum and maximum renormalization scales 
covered by these simulations can be worked out.
In order to
be consistent with the notation of ref.~\cite{SFmassRGI},
we denote the former by $\mumin = (2\lmax)^{-1}$.
The ratio in question is then obtained in two steps:

First the SSF of the gauge coupling
\begin{gather}
\sigma(u)=\left.\gbar^2(1/2L)\right|_{\gbar^2(1/L)=u} \ ,
\end{gather}
computed in~\rep{SFcoupling, SFmassRGI},
is used in order to determine the correspondence between renormalized
couplings and renormalization scales. This is done through the recursion
\begin{gather}
  \label{eq:ssf-u}
  u_{l}=\sigma(u_{l+1}) \, ,
\end{gather}
with $u_0 = \gbar^2(1/\lmax) = 3.48$ the initial value.\footnote{This initial
value $u_0=3.48$ corresponds to $\lmax/r_0=0.738(16)$; the initial
calculation was performed in ref.~\cite{gsw} while the above result is obtained in
the more recent ref.~\cite{ns}.}

Second the SSF $\sigVApAV{;s}^\pm$, known non-perturbatively,
is used for this sequence of couplings
in order to compute the quantity
\begin{gather}
  \label{eq:ratio}
  \UVApAV{;s}^\pm(\mumin,2^k\mumin)=
  \prod_{l=0}^{k-1}\left[\sigVApAV{;{\rm s}}^\pm (u_l)\right] \, .
\end{gather}
The number of recursion steps $k$ has to be chosen so that the
(large) scale $2^k\mumin$ lies in the range covered by the
computation of the SSF. In practice (cf. section~\ref{sec:ssfNP}), it
is safe to take $k=7$, which means that $2^k\mumin$ is deep in the
region where perturbation theory can be expected to apply.

The final step in our calculation is the computation of the RG running
factor of eq.~(\ref{RGI_VApAV}) at the renormalization
point $\mumin$, written as a product:
\begin{gather}
\zrgiVApAV{;s}^\pm(\mumin) =
\UVApAV{;s}^\pm(2^k \mumin,\mumin)\zrgiVApAV{;s}^\pm(2^k\mumin)
\ .
\label{eq:hrat}
\end{gather} 
The first factor on the r.h.s.~is known from eq.~(\ref{eq:ratio}). The
second factor, which involves a (presumably) perturbative scale $2^k \mumin$,
is calculated from eq.~(\ref{RGI_VApAV}) with the 
NNLO and NLO perturbative expressions of $\beta(g)$ and $\gamVApAV{;s}^\pm(g)$, respectively. Clearly 
the underlying assumption is that the truncation of the perturbative
series at NLO is safe at this scale. This is a scheme dependent
statement. The perturbative results of ref.~\cite{QvapavPT} indicate that, for
schemes $s=1,3,7$, the NLO coefficient of $\gamVApAV{;s}^+$ is of the
same sign and much smaller than the LO one; $s=1$ is the scheme with the
smallest NLO coefficient. In all other schemes the relative sign is negative.
For  $\gamVApAV{;s}^-$ the ratio is negative
for all nine schemes, with $s=8$ the smallest in absolute value.
Conservatively, we indicate $s=1$ and $s=8$ as the most suitable schemes
for operators $\oVApAV{O}^+$ and $\oVApAV{O}^-$ respectively.
We stress that these choices are only dictated by the behaviour of the
NLO perturbative results; the non-perturbative computation is equally
reliable for all nine schemes considered.
In any case, we have carried out our computations for all schemes.
\section{Non-perturbative computation of the step scaling function}
\label{sec:ssfNP}

In this section we present the computation of $\SigVApAV{;s}^\pm$ and
its extrapolation to the continuum limit.
We also obtain estimates of the corresponding RGI quantity (or, more
precisely, of the expression of eq.~(\ref{eq:hrat}) at a hadronic scale).
The method of computation parallels closely that of
refs.~\cite{SFmassRGI,SFmassRGI2} for the SSF of the quark mass $\SigmaP$.

\subsection{Wilson and Clover actions}

We have used both the standard Wilson action and its $\Oa$ improved
version (Clover) in our simulations. Our notation is fairly standard;
$\beta \equiv 6/g_0^2$ is the inverse coupling and 
$\kappa \equiv 1/[2am_0 + 8]$ is the hopping parameter. At fixed bare
coupling we define $\hopc$ as the value where the PCAC  quark mass $m(g_0)$ of ref.~\cite{SFmassRGI} vanishes.
Following~\cite{SFmassRGI}, the computation of $\hopc$ is done at $\theta = 0$.

The $\Oa$ Symanzik improvement of the Schr\"odinger Functional has been
worked out in refs.~\cite{SFLNWW,SFchiral,impr:SW}.  For the pure gauge
action, it amounts to modifying it by introducing time-boundary
counterterms proportional to $[\ict(g_0^2)-1]$. For the fermionic
action we must introduce the well-known clover counterterm in the
lattice bulk, proportional to  $\icsw(g_0^2)$, and time-boundary
counterterms proportional to $[\icttil(g_0^2)-1]$. Correlation
functions of composite operators may then also be $\Oa$ improved
by including in their lattice definition the appropriate higher
dimension counterterms. Since for dimension-six operators, such as the
ones of eq.~(\ref{eq:4fermop}), there are several dimension-seven
counterterms, we will not pursue operator improvement in this
work.\footnote{$\Oa$ improvement along the lines of
ref.~\cite{Frezzotti:2003ni} is not readily applicable in the SF
framework.}

The improvement coefficient $\icsw$ has been computed
non-per\-tur\-ba\-tively for a range of values of the bare coupling
$g_0$; see ref.~\cite{SFcSW}.
The coefficients $\ict$ and $\icttil$ are known
only in perturbation theory,  to NLO~\cite{SFct} and
LO~\cite{SFctt} respectively:
\begin{align}
\label{eq:ctPT}
\ict(g_0^2) &= 1 - 0.089 g_0^2 - 0.030 g_0^4 \,\, ,
\\
\label{eq:cttPT}
\icttil(g_0^2) &= 1 - 0.018 g_0^2 \, .
\end{align}

In the present work we will distinguish two approaches to the
continuum limit:
\begin{enumerate}
\item What we call ``Wilson action results'' (or ``Wilson
case'' for short) consists in setting $\icsw = 0\,\, $. Moreover, we
set $\icttil = 1$, while the one-loop value\footnote{This is a choice of
convenience: it is important to know for renormalization purposes (see
eq.~(\ref{eq:ssfLat}) below) the dependence of the Schr\"odinger
functional renormalized coupling $\gbar(1/L)$ on the bare coupling
$g_0$. This dependence  is known
non-perturbatively~\cite{SFcoupling,SFmassRGI} for the pure Yang-Mills
action with this $\ict$ value. In any case, the choice for $\ict$ has no
bearing on the order of leading lattice artifacts.}
(eq.~(\ref{eq:ctPT}) truncated to O($g_0^2$)) is used for
$\ict$. 
\item  What we call ``Clover action results'' (or ``Clover
case'' for short) consists in using the Clover action with a
non-perturbative $\icsw$. The one-loop value from eqs.~(\ref{eq:ctPT}) and 
(\ref{eq:cttPT}) is used for $\ict$ and
$\icttil$ respectively. 
\end{enumerate}
In both cases the four-fermion operator is left unimproved, so the dominant discretisation
effects are expected to be $\Oa$.
We note, however, that the correlation functions (\ref{eq:4f-corr}) are $\Oa$ improved at tree-level, implying that all $\Oa$ counterterms to the local four-quark operators vanish at this order.
Thus, for the
Clover case we are left with discretisation errors which are
${\cal O}(g_0^2 a)$.

\subsection{Continuum limit of the step scaling function}

For both the Wilson and Clover action the lattice SSF 
$\SigVApAV{;s}^\pm$ have been evaluated at
14 values of the renormalized coupling $\gbar(1/L)$,
each for four lattice resolutions $L/a = 6,8,12$ and $16$. 
The tuning of $\beta$ at the four $L/a$ values, corresponding to a
fixed renormalized coupling $\gbar^2(1/L) = u$, has been taken over from ref.~\cite{SFmassRGI}.
The values of $\hopc$ are taken from refs.~\cite{SFmassRGI,GJP,SFmassRGI2}.
The typical statistics accumulated for small lattices is of several hundred configurations.
For the largest lattices the number of configurations ranges from around 60 at the
weaker couplings to around 200 at the stronger ones.
It has to be stressed that Wilson and Clover data have been obtained
from independent ensembles of gauge configurations.

A full collection of our raw data for $\SigVApAV{;s}^\pm$ is
available from the authors upon request.
In Tables~\ref{tab:Z1}-\ref{tab:Z4} we present our results for
$\SigVApAV{;1}^+$ and $\SigVApAV{;8}^-$; see the discussion after
eq.~(\ref{eq:hrat}) for a motivation behind this choice.
The quality of the data for the other schemes is comparable.
The SSF $\SigVApAV{;s}^\pm$
must be extrapolated to zero lattice spacing $a/L$ (at fixed gauge coupling)
in order to obtain its continuum limit counterpart $\sigVApAV{;s}^\pm$.
Since the four-fermion operators have not been improved, we
expect the dominant discretisation effects to be $\Oa$ both for the
Wilson and Clover action data and thus a linear behaviour
in $a/L$. Nevertheless we have performed fits on both datasets with
two ans\"atze
\begin{align}
\label{eq:linfit}
\SigVApAV{;s}^\pm (u,a/L) &= \sigVApAV{;s}^\pm (u) + \rho(u)  (a/L) \ ,\\
\label{eq:quafit}
\SigVApAV{;s}^\pm (u,a/L) &= \sigVApAV{;s}^\pm (u) + \rho(u)  (a/L)^2 \ .
\end{align}
An issue raised in refs.~\cite{SFmassRGI,SFmassRGI2} is the number of data points which
should be included in each fit. In those works the $L/a=6$ results were dropped from
the fits, being too far from the continuum limit. We have performed fits with
all data (4-point fits) and also without the $L/a=6$ data (3-point fits). This
means that we have applied a total of four fitting procedures (the two ans\"atze
of eqs.~(\ref{eq:linfit},\ref{eq:quafit}), each for a 3- and a 4-point fit).

The details related to the continuum limit extrapolation are presented
in Appendix~\ref{app:CLextrap}. From that discussion we conclude that
a conservative choice consists in performing
3-point fits (i.e. drop the data computed at the largest lattice
spacing) which are linear in $(a/L)$. The 1-loop perturbative
discretisation errors have been divided out of $\SigVApAV{;s}^\pm$
in the Wilson case. Moreover, following ref.~\cite{GJP},
we constrain the fits to the Clover and Wilson action data (at a given
renormalized coupling) to have a unique continuum limit.\footnote{This
universality assumption has been thoroughly tested on our data for the
SSF of the quark mass in \cite{SFmassRGI2}.}
The outcome of this procedure is illustrated in
Figs.~\ref{fig:extrap_1},\ref{fig:extrap_2} for the two schemes of
reference, and reported in Tables~\ref{tab:CLe1},\ref{tab:CLe2} for all schemes.
We consider results obtained from these combined fits to be our best,
and use them in the next
step of the analysis.\footnote{We have also, in the spirit of
ref.~\cite{SFuniv}, studied the impact of one-loop cutoff
effects on the extrapolations.
Some details are provided in Appendix~\ref{app:CLextrap}.}

\subsection{Continuum step scaling function and RG running}

The previous analysis has yielded accurate results for the
continuum SSF $\sigVApAV{;s}^\pm$ for a wide range of renormalized
couplings. The data in this range of couplings
can be represented by a polynomial of the form
\begin{gather}
\sigVApAV{;s}^\pm(u) = 1 + \sum_{n=1}^N s^\pm_n u^n \,\,\, .
\label{eq:polyn}
\end{gather}
This ansatz is motivated by the form of the perturbative series.
In perturbation theory the first two coefficients are known:
\begin{gather}
s_1^\pm = \gamVApAV{}^{\pm (0)} \ln 2 \, , \\
s_2^\pm = \gamVApAV{;s}^{\pm (1)} \ln 2 + \Big [ \frac{1}{2} (\gamVApAV{}^{\pm (0)})^2 +
b_0 \gamVApAV{}^{\pm (0)} \Big ]  (\ln 2)^2 \, .
\end{gather}
The LO coefficient is universal, while the NLO one is scheme dependent
and has been calculated in ref.~\cite{QvapavPT}.
As a result of the rather strong scheme dependence of the NLO
anomalous dimension, also $s_2^\pm$ varies significantly between different
schemes. In Figs.~\ref{fig:sigmaVApAVp},\ref{fig:sigmaVApAVm} (see left
columns only) we compare the LO and NLO
perturbative predictions for the SSF to the non-perturbative results
of the present work. We observe that, while the LO results are close
to the non-perturbative ones at least for weak couplings, the NLO
corrections show marked disagreement for certain schemes. This
simply indicates poor convergence of the NLO perturbative series
for some schemes.

The values of the coefficients $s_n^\pm$ of~\req{eq:polyn}
have been obtained through a suitable fitting procedure 
(see below for details). We can then compute the running of the
composite operator between the scales $\mumin$ and $2^k\mumin$
as explained in
sect.~\ref{subsec:RGrun} (cf.~\req{eq:ratio}).
As input for the recursion in~\req{eq:ssf-u} we use the
SSF of the renormalized coupling and its fit to a polynomial 
\begin{gather}
\sigma(u) = u \Big[ 1 + \sum_{n=1}^4 \sigma_n u^n \Big] \,\,\, ,
\label{eq:polyng}
\end{gather}
as obtained in refs.~\rep{SFcoupling, SFmassRGI}, with
$\sigma_1$ and $\sigma_2$ fixed from PT and $\sigma_3$,
$\sigma_4$ kept as fit parameters. Then we apply~\req{eq:ratio} with
$k=7$ iteration steps (corresponding to the range of scales covered by
our simulation), and finally~\req{eq:hrat} to obtain the RGI
renormalization factor
$\zrgiVApAV{;s}^\pm(\mumin)$.
The reliability of the computation of both factors on the 
r.h.s.~of~\req{eq:hrat} may in
principle be compromised by a poor convergence of the perturbative
series at NLO, which is indeed the case in some schemes.
In particular the first factor could be affected if the coefficient
$s_2$ is kept fixed to its NLO value in the fit. 
The second factor could also clearly be affected, since it
is calculated to NLO.\footnote{In practice the
calculation of this second factor is performed by numerically
integrating the first of
eqs.~(\ref{vars_1}) (with the $\beta$-function given at 3 loops)
followed by numerical integration of the exponent in
\req{RGI_VApAV} (with the operator anomalous dimension given at
2 loops and the $\beta$-function at 3 loops).}

Before addressing these issues, we discuss how we obtain a faithful fit to
our data for $\sigVApAV{;s}^\pm(u)$, based on~\req{eq:polyn}.
We keep the first order coefficient $s_1$ fixed to its perturbative value
and perform a series of fits:
\vspace{2truemm}\\
(A) one-parameter fits with with $s_2$ a free parameter; \\
(B) two-parameter fits with $s_2$ and $s_3$ as free parameters;\\
(C) one-parameter fits with $s_2$ fixed from PT and $s_3$ a free parameter;\\
(D) three-parameter fits with $s_2$, $s_3$ and $s_4$ as free parameters;\\
(E) two-parameter fits with $s_2$ fixed from PT and $s_3$, $s_4$ as free parameters.
\vspace{2truemm}\\
The results of these fits are summarised in Tables~\ref{tab:CLfit1},\ref{tab:CLfit2}.
We see that the SSF $\sigVApAV{;s}^+$ is well fit in all cases
($\chi^2/{\rm d.o.f.} \sim 1$). Also the SSF $\sigVApAV{;s}^-$ is always modelled
well by the fitting curves, with the only exception of Fit C in
schemes~1 and~7, where the $\chi^2/{\rm d.o.f.}$ is slightly higher.
In any case, it appears that even in
those schemes where the NLO RG running does not match the NP one,
the fits are satisfactory, as the effect of the fixed NLO value of
$s^\pm_2$ is compensated by the higher order free parameters.

The results for the the RG running factor $\zrgiVApAV{;s}^\pm(\mumin)$
are also shown in Tables~\ref{tab:CLfit1},\ref{tab:CLfit2}.
The errors borne by these numbers have been computed as outlined in 
Appendix B of ref.~\rep{SFmassRGI}. They do not include the effect of the
uncertainty in the determination of $\lmax/r_0$, reported in
ref.~\cite{ns}, which is numerically well below the error already present.
The most important overall feature of these results is that
all possible fits provide numbers compatible within
$1\sigma$. This shows that the fit systematics are well under control.
We conservatively take our best
result to be that of fit D, which has the largest error, and report
our final best estimates for $\zrgiVApAV{;s}^\pm(\mumin)$
in Table~\ref{tab:final_ratios}. The result of fit D for the SSF
(errors included) is represented in the form of shaded areas in
Figs.~\ref{fig:sigmaVApAVp},\ref{fig:sigmaVApAVm} (left columns).

We now return to our earlier discussion concerning the systematic
effect on the two factors on the r.h.s.~of~\req{eq:hrat}, induced
by the poor NLO behaviour of the perturbative series in some schemes.
By comparing final results obtained with $s_2$ fixed to
the NLO value to those where $s_2$ is a free fitting parameter,
we confirm that the use of perturbative input for the fit
introduces no significant effect to the first factor (i.e.
the running between the scales $\mumin$ and $2^k\mumin$).
A proper assessment of the systematics on the
second factor could only be obtained by calculating it to NNLO, which
in turn would require knowledge of the perturbative coefficient
$\gamVApAV{;s}^{\pm (2)}$. As the former is not available,
we can estimate the size of the effect by redoing the computation with
an educated guess for $\gamVApAV{;s}^{\pm (2)}$. We have used two ans\"atze:
First, we postulate that $\gamVApAV{;s}^{\pm (2)}/\gamVApAV{;s}^{\pm (1)}=
\gamVApAV{;s}^{\pm (1)}/\gamVApAV{}^{\pm (0)}$. 
Second, $\gamVApAV{;s}^{\pm (2)}$ is obtained from the perturbative expression
\begin{gather}
\begin{split}
s_3^\pm = \gamVApAV{;s}^{\pm (2)} (\ln 2) & + \Big [ \gamVApAV{}^{\pm (0)}  \gamVApAV{;s}^{\pm (1)} 
+ 2 b_0  \gamVApAV{;s}^{\pm (1)} + b_1  \gamVApAV{}^{\pm (0)} \Big ]  (\ln 2)^2 \\
& +  \Big [ \frac{1}{6} (\gamVApAV{}^{\pm (0)})^3 + b_0 (\gamVApAV{}^{\pm (0)})^2 +
 \frac{4}{3} b_0^2 \gamVApAV{}^{\pm (0)} \Big ]  (\ln 2)^3 \, .
\end{split}
\end{gather}
with $s_3^\pm$ estimated from Fit C. The outcome of both checks is
that the running factors in Table~\ref{tab:final_ratios} for
$\oVApAV{O}^+$ remain compatible within errors for all schemes. In
the case of $\oVApAV{O}^-$ they change by more than one standard
deviation only for ${\rm s}=1,3,7$. These are indeed the schemes with
largest NLO anomalous dimensions. 
We then conclude that the systematic uncertainty
induced by the NLO matching in these three cases is not safely covered by
the quoted error, and therefore these schemes should be discarded.
As an even more conservative approach, we suggest that all schemes for
which $\vert \gamVApAV{;s}^{\pm (1)} / \gamVApAV{}^{\pm (0)} \vert > 0.2$ be discarded. This
stricter requirement would leave us with schemes 1,3,7 for
$\oVApAV{O}^+$ and schemes 2,4,5,6,8,9 for $\oVApAV{O}^-$, which is our
final choice of schemes deemed fully reliable.

The RG running of the two operators is shown in
Figs.~\ref{fig:sigmaVApAVp},\ref{fig:sigmaVApAVm} (right columns).
A few comments are in place:
\begin{enumerate}
\item What is plotted is the RG running of the {\it inverse} of
$\zrgiVApAV{;s}^\pm$, which has the same scale dependence as
the physical matrix elements of the corresponding operator
(cf. eq.~(\ref{RGI_oper})).
\item A glance at eq.~(\ref{eq:hrat}) reminds us that
$\zrgiVApAV{;s}^\pm(\mu)$ is the product of the evolution
function $\UVApAV{;s}^\pm(2^k \mumin,\mu)$ and the
quantity $\zrgiVApAV{;s}^\pm(2^k\mumin)$. While the latter quantity is
computed in PT (with a 3-loop $\beta$-function and a 2-loop
anomalous dimension), the former is the key
outcome of our non-perturbative calculation.
Thus, by construction, the non-perturbative points 
coincide at scale $2^k\mumin$ with the perturbative curve,
evaluated at the same order in PT as the quantity
$\zrgiVApAV{;s}^\pm(2^k\mumin)$.
\item The two perturbative curves in each plot are independent of
any parameters, once the scale and the coupling are fixed.
The degree of convergence of the two curves at large scales
$\mu/\Lambda$ reflects the reliability of the perturbative
estimates of the operator anomalous dimension. Clearly,
some schemes show a better perturbative behaviour than others.
\item The non-perturbative points are obtained as in eq.~(\ref{eq:hrat}),
with the factor $\UVApAV{;s}^\pm(2^k \mumin,\mu)$ calculated
from fit D of the SSF. In some cases the non-perturbative
result follows closely the NLO perturbative one up to
surprisingly small scales. This is explicitly seen to be
a scheme dependent situation.
\item These plots justify our strict criterion of scheme
selection, as detailed above.
\end{enumerate}

\begin{table}[t!]
\centering
\begin{tabular}{cll}
\Hline\\[-1.0ex]
$s$ & $\zrgiVApAV{;s}^+(\mumin)$ & $\zrgiVApAV{;s}^-(\mumin)$\\[1.0ex]
\hline\\[-1.0ex]
1 &~~ 1.111(19) &~~ 0.486(7)$^*$ \\ 
2 &~~ 1.074(24)$^*$ &~~ 0.451(10)\\
3 &~~ 1.008(19) &~~ 0.398(7)$^*$ \\
4 &~~ 1.190(24)$^*$ &~~ 0.541(9)\\
5 &~~ 1.171(23)$^*$ &~~ 0.522(10) \\
6 &~~ 1.315(24)$^*$ &~~ 0.549(9)\\
7 &~~ 1.151(19) &~~ 0.453(7)$^*$\\
8 &~~ 1.358(25)$^*$ &~~ 0.618(10)\\
9 &~~ 1.338(23)$^*$ &~~ 0.598(9) \\
[1.0ex]\Hline\\
\end{tabular}
\caption{
Final results for the RG running factors
$\zrgiVApAV{;s}^\pm(\mumin)$. The schemes which suffer from systematic
uncertainties related to perturbation theory have been indicated with
an asterisk (as argued in the text, a strict criterion adopted for
discarding a scheme is $\vert \gamVApAV{;s}^{\pm (1)} / \gamVApAV{}^{\pm (0)} \vert > 0.2$).
}
\label{tab:final_ratios}
\end{table}

\section{Connection to hadronic observables}
\label{sec:match}

The RGI operator, as defined in~\req{RGI_oper}, can be connected
to its bare counterpart via a total renormalization factor,
given by
\begin{gather}
{\oVApAV{\hat O}}^\pm(x) =  \ZtotVApAV{;s}^\pm(g_0) \oVApAV{O}^\pm(x;g_0) \ .
\label{RGI_bare}
\end{gather}
Once the RG running of the four-fermion operator from the
reference scale $\mumin=(2\lmax)^{-1}$ has been determined via the SSF,
this factor decomposes into:
\begin{gather}
  \label{eq:tot_renorm}
  \ZtotVApAV{;s}^\pm(g_0) =\zrgiVApAV{;s}^\pm(\mumin)\,
  \ZVApAV{;s}^\pm(g_0,a\mumin) \ .
\end{gather}
We stress that $\ZtotVApAV{;s}^\pm$ is a scale-independent quantity,
which furthermore depends on the renormalization scheme only via
cutoff effects. On the other hand, it depends on the particular
lattice regularization chosen, though only through the factor
$\ZVApAV{;s}^\pm(g_0,a\mumin)$, the computation of which is much less
expensive than the one of the running
$\zrgiVApAV{;s}^\pm(\mumin)$.

The non-perturbative computation of $\ZVApAV{;s}^\pm(g_0,a\mumin)$ has
been performed at four values of $\beta$ for each scheme and
four-fermion operator, both with Clover and Wilson actions. The
results are given in Tables~\ref{tab:Zm1}-\ref{tab:Zm4}. Upon multiplying by the 
corresponding ratios in
Table~\ref{tab:final_ratios}, the total renormalization factors are obtained. These
can be further fitted to polynomials of the form
\begin{gather}
  \ZtotVApAV{;s}^\pm(g_0) = a^\pm_{s} + b^\pm_{s}(\beta-6) +
  c^\pm_{s}(\beta-6)^2 \ ,
\label{eq:Ztotfit}
\end{gather}
which can be subsequently used to obtain the total renormalization factor at
any value of $\beta$ within the covered range [6.0219,6.4956], 
which comprises the
typical $\beta$-values used in the computation of bare
observables in physically large volumes.\footnote{For~$\beta=6.0$ a short extrapolation is necessary.}
We supply in Table~\ref{tab:fit_ztot1} the
resulting fit coefficients for both the Clover and the Wilson
case. These parameterisations represent our data with an accuracy of at
least~$1\%$ (this comprises the point~$\beta=6.0$).
The contribution from the error in the RGI
renormalization factors of Table~\ref{tab:final_ratios} has been neglected:
since these factors have been computed in the continuum limit, they
should be added in quadrature {\em after} the quantity renormalized
with the factor in~\req{eq:tot_renorm} has been extrapolated itself to
the continuum limit.

\clearpage
\begin{table}[!h]
\centering

\begin{tabular}{l@{\hspace{10mm}}rrr@{\hspace{10mm}}rrr}
\Hline\\[-1.0ex]
 &
\multicolumn{3}{c}{Clover action~~~~~~~~} &
\multicolumn{3}{c}{Wilson action~~~} \\
$s$ & $a_s^+~~$ & $b_s^+~$ & $c_s^+~$ & $a_s^+~~$ & $b_s^+~$ & $c_s^+~$ \\[1.0ex]
\hline\\[-1.0ex]
1     & 0.884 & 0.17 & 0.00 & 0.710 & 0.24 & $-0.05$ \\
2$^*$ & 0.929 & 0.13 & 0.06 & 0.725 & 0.26 & $-0.05$ \\
3     & 0.890 & 0.15 & 0.02 & 0.701 & 0.26 & $-0.07$ \\
4$^*$ & 0.939 & 0.14 & 0.05 & 0.736 & 0.25 & $-0.03$ \\
5$^*$ & 0.930 & 0.14 & 0.04 & 0.724 & 0.25 & $-0.01$ \\
6$^*$ & 0.925 & 0.15 & 0.04 & 0.741 & 0.21 & $ 0.02$ \\
7     & 0.886 & 0.17 & 0.00 & 0.710 & 0.23 & $-0.02$ \\
8$^*$ & 0.934 & 0.15 & 0.04 & 0.745 & 0.22 & $ 0.01$ \\
9$^*$ & 0.926 & 0.15 & 0.03 & 0.734 & 0.21 & $ 0.03$ \\
[1.0ex]\Hline\\
\end{tabular}

\vspace{10mm}

\begin{tabular}{l@{\hspace{10mm}}rrr@{\hspace{10mm}}rrr}
\Hline\\[-1.0ex]
 &
\multicolumn{3}{c}{Clover action~~~~~~~~} &
\multicolumn{3}{c}{Wilson action~~~} \\
$s$ & $a_s^-~~$ & $b_s^-~$ & $c_s^-~$ & $a_s^-~~$ & $b_s^-~$ & $c_s^-~$ \\[1.0ex]
\hline\\[-1.0ex]
1$^*$ & 0.267 & $ 0.02$ & $-0.03$ & 0.311 & $-0.06$ & 0.05 \\
2     & 0.293 & $-0.01$ & $ 0.03$ & 0.324 & $-0.06$ & 0.06 \\
3$^*$ & 0.269 & $ 0.01$ & $-0.01$ & 0.308 & $-0.04$ & 0.01 \\
4     & 0.301 & $-0.01$ & $ 0.02$ & 0.329 & $-0.06$ & 0.08 \\
5     & 0.288 & $-0.01$ & $ 0.02$ & 0.318 & $-0.06$ & 0.08 \\
6     & 0.290 & $-0.01$ & $ 0.02$ & 0.329 & $-0.08$ & 0.09 \\
7$^*$ & 0.266 & $ 0.02$ & $-0.02$ & 0.311 & $-0.05$ & 0.03 \\
8     & 0.299 & $-0.01$ & $ 0.01$ & 0.334 & $-0.08$ & 0.10 \\
9     & 0.288 & $ 0.00$ & $ 0.01$ & 0.323 & $-0.08$ & 0.09 \\
[1.0ex]\Hline\\
\end{tabular}

\caption{
Fits to the total renormalization factor of~\req{eq:tot_renorm}.
The schemes which suffer from systematic
uncertainties related to perturbation theory have been indicated with
an asterisk (cf. Section~\ref{sec:ssfNP}).
}
\label{tab:fit_ztot1}
\end{table}
\clearpage

\section{Conclusions}
\label{sec:concl}

The present work is the first non-perturbative calculation
of the RG evolution function of four-fermion operators for scales 
ranging from the hadronic to the perturbative regime.
We limit ourselves to operators with a ``left-left'' Dirac
structure, which are multiplicatively renormalizable. This is
the simplest possible case, as operators
with other Dirac structures mix under renormalization.

The method employed is the finite size scaling approach
based on the Schr\"o\-din\-ger Functional. The Wilson lattice
regularization has been used for both gluon and fermion fields.
Combining lattice results from Wilson and
Clover fermion actions enhances our control of continuum limit
extrapolations, when obtaining  the continuum step scaling function
for a large range of scales. From the step scaling function and
the perturbative estimate
of the operator anomalous dimension at NLO, we obtain the ratio
of the RGI operator to its renormalized counterpart at a hadronic
scale. Nine different renormalization schemes have
been used for each operator. Some of these schemes have turned out to be unstable,
but this is only due to the bad convergence of the perturbative
result for the anomalous dimension at NLO. 

We envisage that our results will be used as follows:
\begin{enumerate}
\item In simulations using Wilson type fermions and the
bare operators of eq.~\req{eq:4fermop} the matrix elements
of these bare operators at fixed $\beta$ should be
multiplied by the renormalization factors given
in~\req{eq:Ztotfit}, including a 1 percent error in quadrature.
After continuum extrapolation, an additional error
should be included in quadrature,
corresponding to the errors quoted in table~\ref{tab:final_ratios}.
\item In simulations using some variant of Ginsparg-Wilson quarks
(overlap quarks, domain-wall quarks, etc.) the results of
table~\ref{tab:final_ratios}
can still be used, as these are obtained in the continuum limit.
What needs to be re-done is the calculation of the renormalization
factor at the low energy matching scale $L=1.436\,r_0$ (the equivalent
of tables 13 and 14). There are two ways of achieving this:
\begin{itemize}
\item Via a direct evaluation of the renormalization
conditions of~\req{eq:rencond} at the matching scale. Obviously, this
requires the formulation of the Schr\"odinger functional for
Ginsparg-Wilson type quarks, e.g. along the lines of ref.~\cite{Tanig}.
\item Via an indirect matching, as done in~\cite{ChCondGW} for the 
chiral condensate. In order to achieve this
one just needs to compute in both regularizations a matrix element
of the four-quark operator at matched physical conditions.
The ratio between the bare matrix element computed with Ginsparg-Wilson
fermions and the renormalized matrix element with Wilson quarks
(in a given SF scheme) then yields the desired matching factor.
\end{itemize}
\end{enumerate}

A first application using Wilson type fermions
consists in the computation of $B_K$ in a tmQCD
framework. Preliminary results have appeared in
ref.~\cite{tmQCD-BK-proc}.
\section*{Acknowledgements}
We thank M.~L\"uscher, F.~Palombi, G.C.~Rossi
and R.~Sommer for useful discussions.
In various stages of this project, we have
enjoyed the hospitality of several Institutes.
In this respect C.P. and A.V. thank DESY-Zeuthen, CERN
and the IFT-UAM/CSIC at Madrid;
J.H., C.P. and S.S. thank the INFN-Rome2 at the
University of Rome ``Tor Vergata''.
C.P. acknowledges
the financial support provided through the European
Community's Human Potential Programme under contract
HPRN-CT-2000-00145, Hadrons/Lattice QCD.
Last but not least, we wish to thank the Computing
Centre of DESY-Zeuthen, for its continuous support
throughout the project.

\appendix

\section{A note on the difference between SF schemes}
\label{app:schemes_differ}

The renormalization constants computed perturbatively at one-loop
in~\cite{QvapavPT} are equal for
some of the SF schemes considered in the present work, namely for
schemes $s=1,7$ for both $\oVApAV{O}^+$ and
$\oVApAV{O}^-$. Consequently, the same is true for the respective 
NLO anomalous dimensions. This may suggest that the two schemes are identical.

The non-perturbative results of this work show no such identity at the
level of renormalization constants: at the largest values of the
renormalized coupling, the values of $\ZVApAV{1,7}^\pm$ typically
differ by several standard deviations. The difference is however
less marked for the step scaling functions $\SigVApAV{1,7}^\pm$, which
even at the strongest couplings differ only by around $1\sigma$.

This leaves us with the possibility that the two anomalous dimensions
(which are defined in the continuum limit) are identical.
Our data do not allow to discard this
possibility, since the two SSFs exhibit good compatibility
(cf. Tables~\ref{tab:CLe1},~\ref{tab:CLe2}).
It has to be noted, however, that the continuum limit result is
remarkably similar for many of the schemes considered. Therefore, the
question whether identities between different schemes take place in
the continuum limit cannot be strictly decided based on the available data.


\section{Continuum limit extrapolation}
\label{app:CLextrap}

The results of the fitting procedures adopted can be summarised as follows:
\begin{enumerate}
\item The statistical accuracy of our result for $\sigVApAV{;s}^\pm$ is
always better than $2\%$ and typically of $O(1-2\%)$ for the largest couplings.
The results for the linear or quadratic coefficients
$\rho$ have large statistical uncertainties (up to $100\%$), reflecting
an overall weak cutoff dependence of $\SigVApAV{;s}^\pm$.
\item The results for $\sigVApAV{;s}^+$  obtained by a 3-point fit are
compatible to those obtained by a 4-point fit (at fixed coupling $u$),
when the Clover action is used. This is also true for
$\sigVApAV{;s}^-$, with only a few exceptions (schemes
$s=3,7$), where for one or two couplings there are discrepancies
of at most $1.3 \sigma$.

With the Wilson action the situation tends to worsen. For
$\sigVApAV{;s}^+$ we have (for each scheme) up to two or three couplings
which show discrepancies of at most $1.6 \sigma$, while for
$\sigVApAV{;s}^-$ we have up to six couplings (depending on the
scheme) which show discrepancies typically ranging from $1\sigma$ to
$3.3 \sigma$. 

We do not see any systematic trend related to the fitting ansatz
(linear or quadratic).

Naturally, 3-point fit results have a larger error.

\item The results for $\sigVApAV{;s}^+$ obtained by fitting 3-points linearly are
always compatible to those obtained by a quadratic fit (at fixed
coupling $u$) when the Clover action is used. When 4 points are fitted,
there is occasional disagreement (at worst for two couplings and
$3\sigma$ for most schemes). With the Wilson action things are less stable: with 4-point
fits there are discrepancies for up to seven couplings per
scheme (worst case is $6 \sigma$ at strong
coupling). With 3-point fits we have up to three
discrepancies per scheme (worst case is $4\sigma$).

The results for $\sigVApAV{;s}^-$ with the Clover action show
marked disagreement for up to nine couplings per scheme between linear
and quadratic fitting (worst case is $6\sigma$), when 4-point fits are
used. With 3-point fits we have at worst discrepancies at three
couplings (schemes 3,7) at the $2 \sigma$ level. The Wilson results show discrepancies (up
to $6 \sigma$) for most couplings irrespective of the number of fitted points.

\item The goodness of fit is satisfactory ($\chi^2/{\rm d.o.f.} < 3$)
in most cases, while in a limited number of couplings the value
tends to rise considerably. This does not depend systematically on the number of
fitted points and choice of fitting ansatz. In any case, given the small number of fitted data points,
$\chi^2/{\rm d.o.f.}$ is a goodness-of-fit criterion of relatively limited value. Instead, the 
total $\chi^2/{\rm d.o.f.}$ varies mostly between 1 and 2, indicating
satisfactory overall quality of the fits, save for a few exceptions
for $\sigVApAV{;s}^-$ (Wilson case with 4-point fits) where the value
is as high as 5.
\end{enumerate}

One-loop discretisation effects can be divided out of the lattice SSF
by defining the quantity
\begin{gather}
  \PISigVApAV{;s}^\pm(u,a/L) = \frac{\SigVApAV{;s}^\pm(u,a/L)}{1 + u \,\, k^\pm_{1;s}(a/L)} \,\, .
  \label{eq:Sigmatil}
\end{gather}
The coefficient $k^\pm_{1;s}(a/L)$ is defined by expanding the ratio
$\SigVApAV{;s}^\pm/\sigVApAV{;s}^\pm$ in perturbation theory as
\begin{gather}
  \frac{\SigVApAV{;s}^\pm(u,a/L)}{\sigVApAV{;s}^\pm(u)} =
  1 + u \,\, k^\pm_{1;s}(a/L) + u^2 \,\, k^\pm_{2;s}(a/L) + \ldots
\end{gather}
and has been computed at various values of $a/L$
in~\cite{QvapavPT}, where it is given in terms of the quantity
$\delta^\pm_{s} = k^\pm_{1;s}/(\gamVApAV{}^{\pm (0)} \ln 2)$.
The continuum limit of $\PISigVApAV{;s}^\pm$ is trivially the same as that of $\SigVApAV{;s}^\pm$,
but the former quantity may approach it faster, as it has discretisation errors which are of 
order $u^2$.

We find that the above procedure has significant impact on the Wilson
case. The fits become more stable in several ways which are discussed
here in correspondence to the criteria listed above:
\begin{enumerate}
\addtocounter{enumi}{1}
\item The results for $\sigVApAV{;s}^-$  obtained by a 3-point fit are
incompatible to those obtained by a 4-point fit (at fixed coupling $u$)
only in three schemes for at most 5 couplings and with a 1.5$\sigma$
discrepancy.
\item The results for $\sigVApAV{;s}^+$ obtained by fitting 3-points
linearly show discrepancies to those obtained by a quadratic fit (at fixed
coupling $u$) for at most 3 couplings per scheme. These discrepancies
are typically  2$\sigma$ (in one case 3$\sigma$). For $\sigVApAV{;s}^-$
and for two schemes only, we have discrepancies of less than
2$\sigma$ for a few couplings.
\item  The total $\chi^2/{\rm d.o.f.}$ is always below 1.5.
\end{enumerate}

For the Clover case the continuum extrapolation of $\PISigVApAV{}^\pm$
is always compatible to that of $\SigVApAV{}^\pm$. Furthermore, no
significant change in the error size of the extrapolated results has
been observed. For the Wilson case, where perturbative cutoff effects
are in general large, the slope of the extrapolation decreases quite
significantly in most cases and certainly at strong couplings.
However, the extrapolated values from
$\PISigVApAV{}^\pm$ and $\SigVApAV{}^\pm$ are again
compatible and bear similar errors, but for a few exceptions
(strongest couplings in schemes 1,3,7), where the difference between
the extrapolated values from
$\PISigVApAV{}^-$ and $\SigVApAV{}^-$ is slightly larger than
$1\sigma$. We conclude on the grounds of the above considerations,
that the best result for $\sigVApAV{;s}^\pm(u)$ in the Wilson
case is that obtained by extrapolating $\PISigVApAV{}^\pm$.

Finally, in the spirit of ref.~\cite{SFmassRGI2}, we perform combined
fits of Clover and Wilson data (at fixed renormalized coupling),
constrained to a common continuum limit. This is expected to reduce
the uncertainty of the results for $\sigVApAV{}^\pm$.
To muster support for this procedure we have checked the compatibility
of the values of $\sigVApAV{}^\pm$, obtained from linear three-point
fits to the Clover data, to those obtained from linear three-point fits to 
perturbatively $\Oa$ improved Wilson data; recall that these are our
best fits for each of the two datasets. In each renormalization
scheme the two results only disagreed (typically by $1$ to $2\,\sigma$ and
at worst by $2.5\,\sigma$) in a few cases (for one, two or three
couplings).
These rare discrepancies appear both at weak and strong couplings. 
Overall, this is supportive of the universality of the continuum limit
and justifies the option of constrained fits. For these fits the typical
$\chi^2/{\rm d.o.f.}$ range is between 1 and 2 and at worst 4, while the
total $\chi^2/{\rm d.o.f.}$ is around 1.2 for $\SigVApAV{}^+$ and 1.0 for
$\SigVApAV{}^-$.


\section{Tables and figures}
\label{app:tab_fig}


\clearpage

\begin{table}
\centering
\begin{tabular}{rrll@{\hspace{5mm}}lll}
\Hline \\[-1.0ex]
$\beta~~~$ & $\frac{L}{a}$ & $~~~\gbar^2(L)$ &
$~~~~~~\hopc$ & $Z_1^+\left(g_0,\frac{L}{a}\right)$ & $Z_1^+\left(g_0,\frac{2L}{a}\right)$ & $\Sigma_1^+\left(u,\frac{a}{L}\right)$ \\[1.0ex]
\hline \\[-1.0ex]
10.7503 & 6 & 0.8873(5) & 0.130591(4) & 0.8822(13) & 0.8892(24) & 1.0079(31)\\
11.0000 & 8 & 0.8873(10) & 0.130439(3) & 0.8893(14) & 0.8998(24) & 1.0118(31)\\
11.3384 & 12 & 0.8873(30) & 0.130251(2) & 0.8964(22) & 0.9136(31) & 1.0192(43)\\
11.5736 & 16 & 0.8873(25) & 0.130125(2) & 0.9033(20) & 0.9211(38) & 1.0197(48)\\
  [1.0ex]
10.0500 & 6 & 0.9944(7) & 0.131073(5) & 0.8743(14) & 0.8817(23) & 1.0085(31)\\
10.3000 & 8 & 0.9944(13) & 0.130889(3) & 0.8799(19) & 0.8921(23) & 1.0139(34)\\
10.6086 & 12 & 0.9944(30) & 0.130692(2) & 0.8943(24) & 0.9134(31) & 1.0214(44)\\
10.8910 & 16 & 0.9944(28) & 0.130515(2) & 0.8980(20) & 0.9153(38) & 1.0193(48)\\
  [1.0ex]
9.5030 & 6 & 1.0989(8) & 0.131514(5) & 0.8654(15) & 0.8793(28) & 1.0161(37)\\
9.7500 & 8 & 1.0989(13) & 0.131312(3) & 0.8714(16) & 0.8906(25) & 1.0220(34)\\
10.0577 & 12 & 1.0989(40) & 0.131079(3) & 0.8816(24) & 0.9050(31) & 1.0265(45)\\
10.3419 & 16 & 1.0989(44) & 0.130876(2) & 0.8984(25) & 0.9102(36) & 1.0131(49)\\
  [1.0ex]
8.8997 & 6 & 1.2430(13) & 0.132072(9) & 0.8523(12) & 0.8685(21) & 1.0190(29)\\
9.1544 & 8 & 1.2430(14) & 0.131838(4) & 0.8622(15) & 0.8846(31) & 1.0260(40)\\
9.5202 & 12 & 1.2430(35) & 0.131503(3) & 0.8777(20) & 0.8928(38) & 1.0172(49)\\
9.7350 & 16 & 1.2430(34) & 0.131335(3) & 0.8868(38) & 0.9168(40) & 1.0338(63)\\
  [1.0ex]
8.6129 & 6 & 1.3293(12) & 0.132380(6) & 0.8463(17) & 0.8627(31) & 1.0194(42)\\
8.8500 & 8 & 1.3293(21) & 0.132140(5) & 0.8572(17) & 0.8806(32) & 1.0273(43)\\
9.1859 & 12 & 1.3293(60) & 0.131814(3) & 0.8735(27) & 0.8875(41) & 1.0160(56)\\
9.4381 & 16 & 1.3293(40) & 0.131589(2) & 0.8861(24) & 0.9122(52) & 1.0295(65)\\
  [1.0ex]
8.3124 & 6 & 1.4300(20) & 0.132734(10) & 0.8409(13) & 0.8570(22) & 1.0191(31)\\
8.5598 & 8 & 1.4300(21) & 0.132453(5) & 0.8508(17) & 0.8722(30) & 1.0252(41)\\
8.9003 & 12 & 1.4300(50) & 0.132095(3) & 0.8658(29) & 0.8987(41) & 1.0380(59)\\
9.1415 & 16 & 1.4300(58) & 0.131855(3) & 0.8819(21) & 0.9091(57) & 1.0308(69)\\
  [1.0ex]
7.9993 & 6 & 1.5553(15) & 0.133118(7) & 0.8324(11) & 0.8490(33) & 1.0199(42)\\
8.2500 & 8 & 1.5553(24) & 0.132821(5) & 0.8440(18) & 0.8719(36) & 1.0331(48)\\
8.5985 & 12 & 1.5533(70) & 0.132427(3) & 0.8639(31) & 0.8916(45) & 1.0321(64)\\
8.8323 & 16 & 1.5533(70) & 0.132169(3) & 0.8764(30) & 0.9060(58) & 1.0338(75)\\
  [1.0ex]
\Hline
\end{tabular}
\caption{
Results for the step scaling function $\SigVApAV{;1}^+$ with Clover action.
}
\label{tab:Z1}
\end{table}\addtocounter{table}{-1}
\clearpage
\begin{table}
\centering
\begin{tabular}{rrll@{\hspace{5mm}}lll}
\Hline \\[-1.0ex]
$\beta~~~$ & $\frac{L}{a}$ & $~~~\gbar^2(L)$ &
$~~~~~~\hopc$ & $Z_1^+\left(g_0,\frac{L}{a}\right)$ & $Z_1^+\left(g_0,\frac{2L}{a}\right)$ & $\Sigma_1^+\left(u,\frac{a}{L}\right)$ \\[1.0ex]
\hline \\[-1.0ex]
7.7170 & 6 & 1.6950(26) & 0.133517(8) & 0.8247(17) & 0.8501(12) & 1.0308(26)\\
7.9741 & 8 & 1.6950(28) & 0.133179(5) & 0.8349(15) & 0.8702(36) & 1.0423(47)\\
8.3218 & 12 & 1.6950(79) & 0.132756(4) & 0.8612(11) & 0.8923(36) & 1.0361(44)\\
8.5479 & 16 & 1.6950(90) & 0.132485(3) & 0.8713(30) & 0.9121(51) & 1.0468(69)\\
  [1.0ex]
7.4082 & 6 & 1.8811(22) & 0.133961(8) & 0.8136(18) & 0.8386(12) & 1.0307(27)\\
7.6547 & 8 & 1.8811(28) & 0.133632(6) & 0.8304(16) & 0.8673(35) & 1.0444(47)\\
7.9993 & 12 & 1.8811(38) & 0.133159(4) & 0.8553(12) & 0.8847(47) & 1.0344(57)\\
8.2415 & 16 & 1.8811(99) & 0.132847(3) & 0.8691(46) & 0.9056(44) & 1.0420(75)\\
  [1.0ex]
7.1214 & 6 & 2.1000(39) & 0.134423(9) & 0.8040(18) & 0.8316(13) & 1.0343(28)\\
7.3632 & 8 & 2.1000(45) & 0.134088(6) & 0.8223(18) & 0.8591(38) & 1.0448(52)\\
7.6985 & 12 & 2.1000(80) & 0.133599(4) & 0.8484(12) & 0.8909(37) & 1.0501(46)\\
7.9560 & 16 & 2.100(11) & 0.133229(3) & 0.8661(32) & 0.9050(42) & 1.0449(62)\\
  [1.0ex]
6.7807 & 6 & 2.4484(37) & 0.134994(11) & 0.7928(19) & 0.8259(15) & 1.0418(31)\\
7.0197 & 8 & 2.4484(45) & 0.134639(7) & 0.8121(19) & 0.8483(40) & 1.0446(55)\\
7.3551 & 12 & 2.4484(80) & 0.134141(5) & 0.8407(13) & 0.8925(46) & 1.0616(57)\\
7.6101 & 16 & 2.448(17) & 0.133729(4) & 0.8634(37) & 0.9171(52) & 1.0622(75)\\
  [1.0ex]
6.5512 & 6 & 2.770(7) & 0.135327(12) & 0.7877(20) & 0.8249(11) & 1.0472(30)\\
6.7860 & 8 & 2.770(7) & 0.135056(8) & 0.8067(20) & 0.8590(45) & 1.0648(62)\\
7.1190 & 12 & 2.770(11) & 0.134513(5) & 0.8361(14) & 0.9027(35) & 1.0797(46)\\
7.3686 & 16 & 2.770(14) & 0.134114(3) & 0.8556(40) & 0.9234(53) & 1.0792(80)\\
  [1.0ex]
6.3665 & 6 & 3.111(4) & 0.135488(6) & 0.7791(24) & 0.8203(39) & 1.0529(60)\\
6.6100 & 8 & 3.111(6) & 0.135339(3) & 0.8011(24) & 0.8540(52) & 1.0660(72)\\
6.9322 & 12 & 3.111(12) & 0.134855(3) & 0.8332(32) & 0.9155(46) & 1.0988(69)\\
7.1911 & 16 & 3.111(16) & 0.134411(3) & 0.8575(30) & 0.9316(63) & 1.0864(83)\\
  [1.0ex]
6.2204 & 6 & 3.480(8) & 0.135470(15) & 0.7759(11) & 0.8355(34) & 1.0768(46)\\
6.4527 & 8 & 3.480(14) & 0.135543(9) & 0.7955(14) & 0.8668(56) & 1.0896(73)\\
6.7750 & 12 & 3.480(39) & 0.135121(5) & 0.8358(28) & 0.9143(57) & 1.0939(77)\\
7.0203 & 16 & 3.480(21) & 0.134707(4) & 0.8620(31) & 0.9472(59) & 1.0988(79)\\
  [1.0ex]
\Hline
\end{tabular}
\caption{
(continued)
}
\end{table}

\clearpage

\begin{table}
\centering
\begin{tabular}{rrll@{\hspace{5mm}}lll}
\Hline \\[-1.0ex]
$\beta~~~$ & $\frac{L}{a}$ & $~~~\gbar^2(L)$ &
$~~~~~~\hopc$ & $Z_1^+\left(g_0,\frac{L}{a}\right)$ & $Z_1^+\left(g_0,\frac{2L}{a}\right)$ & $\Sigma_1^+\left(u,\frac{a}{L}\right)$ \\[1.0ex]
\hline \\[-1.0ex]
10.7503 & 6 & 0.8873(5) & 0.134696(7) & 0.8386(15) & 0.8299(19) & 0.9896(29) \\ 
11.0000 & 8 & 0.8873(10) & 0.134548(6) & 0.8440(14) & 0.8381(22) & 0.9930(31) \\ 
11.3384 & 12 & 0.8873(30) & 0.134277(5) & 0.8515(20) & 0.8517(28) & 1.0002(40) \\ 
11.5736 & 16 & 0.8873(25) & 0.134068(6) & 0.8565(21) & 0.8578(44) & 1.0015(57) \\ 
  [1.0ex]
10.0500 & 6 & 0.9944(7) & 0.135659(8) & 0.8238(18) & 0.8175(20) & 0.9924(33) \\ 
10.3000 & 8 & 0.9944(13) & 0.135457(5) & 0.8297(15) & 0.8218(22) & 0.9905(32) \\ 
10.6086 & 12 & 0.9944(30) & 0.135160(4) & 0.8396(23) & 0.8416(33) & 1.0024(48) \\ 
10.8910 & 16 & 0.9944(28)& 0.134849(6) & 0.8510(23) & 0.8508(52) & 0.9998(67) \\ 
  [1.0ex]
9.5030 & 6 & 1.0989(8) & 0.136520(5) & 0.8157(19) & 0.8042(22) & 0.9859(35) \\ 
9.7500 & 8 & 1.0989(13) & 0.136310(3) & 0.8159(17) & 0.8182(21) & 1.0028(33) \\ 
10.0577 & 12 & 1.0989(40) & 0.135949(4) & 0.8284(23) & 0.8256(33) & 0.9966(49) \\ 
10.3419 & 16 & 1.0989(44) & 0.135572(4) & 0.8414(32) & 0.8467(34) & 1.0063(56) \\ 
  [1.0ex]
8.8997 & 6 & 1.2430(13) & 0.137706(5) & 0.7977(19) & 0.7921(23) & 0.9930(37) \\ 
9.1544 & 8 & 1.2430(14) & 0.137400(4) & 0.8053(18) & 0.7981(26) & 0.9911(39) \\ 
9.5202 & 12 & 1.2430(35)  & 0.136855(2) & 0.8192(24) & 0.8199(25) & 1.0009(42) \\ 
9.7350 & 16 & 1.2430(34)  & 0.136523(4) & 0.8215(27) & 0.8305(45) & 1.0110(64) \\ 
  [1.0ex]
8.6129 & 6 & 1.3293(12) & 0.138346(6) & 0.7903(23) & 0.7808(24) & 0.9880(42) \\ 
8.8500 & 8 & 1.3293(21) & 0.138057(4) & 0.7964(18) & 0.7884(28) & 0.9900(42) \\ 
9.1859 & 12 & 1.3293(60)  & 0.137503(2) & 0.8090(27) & 0.8135(30) & 1.0056(50) \\ 
9.4381 & 16 & 1.3293(40)  & 0.137061(4) & 0.8183(39) & 0.8265(38) & 1.0100(67) \\ 
  [1.0ex]
8.3124 & 6 & 1.4300(20) & 0.139128(11) & 0.7777(20) & 0.7727(23) & 0.9936(39) \\ 
8.5598 & 8 & 1.4300(21) & 0.138742(7) & 0.7878(19) & 0.7834(31) & 0.9944(46) \\ 
8.9003 & 12 & 1.4300(50)  & 0.138120(8) & 0.8041(27) & 0.8049(38) & 1.0010(58) \\ 
9.1415 & 16 & 1.4300(58)  & 0.137655(5) & 0.8176(27) & 0.8165(50) & 0.9987(69) \\ 
  [1.0ex]
7.9993 & 6 & 1.5553(15) & 0.140003(11) & 0.7687(21) & 0.7570(24) & 0.9848(41) \\ 
8.2500 & 8 & 1.5553(24) & 0.139588(8) & 0.7773(19) & 0.7724(29) & 0.9937(45) \\ 
8.5985 & 12 & 1.5533(70)  & 0.138847(6) & 0.7949(29) & 0.7997(42) & 1.0060(64) \\ 
8.8323 & 16 & 1.5533(70)  & 0.138339(7) & 0.8095(34) & 0.8170(55) & 1.0093(80) \\ 
  [1.0ex]
\Hline
\end{tabular}
\caption{
Results for the step scaling function $\SigVApAV{;1}^+$ with Wilson action.
}
\label{tab:Z2}
\end{table}\addtocounter{table}{-1}
\clearpage
\begin{table}
\centering
\begin{tabular}{rrll@{\hspace{5mm}}lll}
\Hline \\[-1.0ex]
$\beta~~~$ & $\frac{L}{a}$ & $~~~\gbar^2(L)$ &
$~~~~~~\hopc$ & $Z_1^+\left(g_0,\frac{L}{a}\right)$ & $Z_1^+\left(g_0,\frac{2L}{a}\right)$ & $\Sigma_1^+\left(u,\frac{a}{L}\right)$ \\[1.0ex]
\hline \\[-1.0ex]
7.7170 & 6 & 1.6950(26) & 0.140954(12) & 0.7628(22) & 0.7451(24) & 0.9768(42) \\ 
7.9741 & 8 & 1.6950(28) & 0.140438(8) & 0.7646(20) & 0.7638(42) & 0.9990(61) \\ 
8.3218 & 12 & 1.6950(79)  & 0.139589(6) & 0.7853(30) & 0.8001(45) & 1.0188(69) \\ 
8.5479 & 16 & 1.6950(90)  & 0.139058(6) & 0.7971(35) & 0.8141(55) & 1.0213(82) \\ 
  [1.0ex]
7.4082 & 6 & 1.8811(22) & 0.142145(11) & 0.7474(23) & 0.7251(27) & 0.9702(47) \\ 
7.6547 & 8 & 1.8811(28) & 0.141572(9) & 0.7529(22) & 0.7550(29) & 1.0028(48) \\ 
7.9993 & 12 & 1.8811(38) & 0.140597(6) & 0.7758(31) & 0.7783(43) & 1.0032(68) \\ 
8.2415 & 16 & 1.8811(99) & 0.139900(6) & 0.7877(33) & 0.7990(46) & 1.0143(72) \\ 
  [1.0ex]
7.1214 & 6 & 2.1000(39) & 0.143416(11) & 0.7187(25) & 0.7101(28) & 0.9880(52) \\ 
7.3632 & 8 & 2.1000(45) & 0.142749(9) & 0.7346(21) & 0.7351(42) & 1.0007(64) \\ 
7.6985 & 12 & 2.1000(80) & 0.141657(6) & 0.7658(22) & 0.7690(35) & 1.0042(54) \\ 
7.9560 & 16 & 2.100(11) & 0.140817(7) & 0.7824(36) & 0.7958(46) & 1.0171(75) \\ 
  [1.0ex]
6.7807 & 6 & 2.4484(37) & 0.145286(11) & 0.7044(25) & 0.6894(27) & 0.9787(52) \\ 
7.0197 & 8 & 2.4484(45) & 0.144454(7) & 0.7210(24) & 0.7209(31) & 0.9999(54) \\ 
7.3551 & 12 & 2.4484(80) & 0.143113(6) & 0.7527(27) & 0.7556(48) & 1.0039(73) \\ 
7.6101 & 16 & 2.448(17) & 0.142107(6) & 0.7635(36) & 0.7853(48) & 1.0286(79) \\ 
  [1.0ex]
6.5512 & 6 & 2.770(7) & 0.146825(11) & 0.6886(28) & 0.6702(26) & 0.9733(55) \\ 
6.7860 & 8 & 2.770(7) & 0.145859(7) & 0.7080(25) & 0.6942(43) & 0.9805(70) \\ 
7.1190 & 12 & 2.770(11) & 0.144299(8) & 0.7359(33) & 0.7543(37) & 1.0250(68) \\ 
7.3686 & 16 & 2.770(14) & 0.143175(7) & 0.7638(47) & 0.7860(47) & 1.0291(88) \\ 
  [1.0ex]
6.3665 & 6 & 3.111(4) & 0.148317(10) & 0.6779(30) & 0.6478(24) & 0.9556(55) \\ 
6.6100 & 8 & 3.111(6) & 0.147112(7) & 0.6962(27) & 0.6882(34) & 0.9885(62) \\ 
6.9322 & 12 & 3.111(12 & 0.145371(7) & 0.7294(30) & 0.7444(46) & 1.0206(76) \\ 
7.1911 & 16 & 3.111(16 & 0.144060(8) & 0.7589(43) & 0.7896(55) & 1.0405(93) \\ 
  [1.0ex]
6.2204 & 6 & 3.480(8) & 0.149685(15) & 0.6583(32) & 0.6295(27) & 0.9563(62) \\ 
6.4527 & 8 & 3.480(14) & 0.148391(9) & 0.6814(27) & 0.6681(51) & 0.9805(84) \\ 
6.7750 & 12 & 3.480(39) & 0.146408(7) & 0.7254(27) & 0.7355(59) & 1.0139(90) \\ 
7.0203 & 16 & 3.480(21) & 0.145025(8) & 0.7511(35) & 0.7990(48) & 1.0638(81) \\ 
  [1.0ex]
\Hline
\end{tabular}
\caption{
(continued)
}
\end{table}

\clearpage

\begin{table}
\centering
\begin{tabular}{rrll@{\hspace{5mm}}lll}
\Hline \\[-1.0ex]
$\beta~~~$ & $\frac{L}{a}$ & $~~~\gbar^2(L)$ &
$~~~~~~\hopc$ & $Z_8^-\left(g_0,\frac{L}{a}\right)$ & $Z_8^-\left(g_0,\frac{2L}{a}\right)$ & $\Sigma_8^-\left(u,\frac{a}{L}\right)$ \\[1.0ex]
\hline \\[-1.0ex]
10.7503 & 6 & 0.8873(5) & 0.130591(4)  & 0.8147(10) & 0.7852(16) & 0.9638(23) \\
11.0000 & 8 & 0.8873(10) & 0.130439(3)) & 0.8077(10) & 0.7811(17) & 0.9671(24)\\
11.3384 & 12 & 0.8873(30) & 0.130251(22) & 0.8032(13) & 0.7776(19) & 0.9681(28)\\
11.5736 & 16 & 0.8873(25) & 0.130125(22) & 0.7960(15) & 0.7724(34) & 0.9704(46)\\
  [1.0ex]			                                              
10.0500 & 6 & 0.9944(7) & 0.131073(5)  & 0.7968(10) & 0.7653(16) & 0.9605(23) \\
10.3000 & 8 & 0.9944(13) & 0.130889(3) & 0.7919(14) & 0.7621(17) & 0.9624(27)\\
10.6086 & 12 & 0.9944(30) & 0.130692(22) & 0.7828(15) & 0.7533(24) & 0.9623(36)\\
10.8910 & 16 & 0.9944(28) & 0.130515(22) & 0.7798(13) & 0.7508(28) & 0.9628(39)\\
  [1.0ex]			                                              
9.5030 & 6 & 1.0989(8) & 0.131514(5) & 0.7827(12) & 0.7457(19) & 0.9527(28) \\
9.7500 & 8 & 1.0989(13) & 0.131312(3)  & 0.7736(11) & 0.7419(17) & 0.9590(26) \\
10.0577 & 12 & 1.0989(40) & 0.131079(33) & 0.7654(17) & 0.7359(21) & 0.9615(35)\\
10.3419 & 16 & 1.0989(44) & 0.130876(22) & 0.7645(18) & 0.7320(21) & 0.9575(36)\\
  [1.0ex]			                                              
8.8997 & 6 & 1.2430(13) & 0.132072(9)  & 0.7598(8) & 0.7256(15) & 0.9550(22) \\
9.1544 & 8 & 1.2430(14) & 0.131838(4)  & 0.7562(11) & 0.7188(21) & 0.9505(31) \\
9.5202 & 12 & 1.2430(35) & 0.131503(3) & 0.7473(13) & 0.7159(23) & 0.9580(35)\\
9.7350 & 16 & 1.2430(34) & 0.131335(3) & 0.7444(22) & 0.7145(24) & 0.9598(43)\\
  [1.0ex]			                                              
8.6129 & 6 & 1.3293(12) & 0.132380(6)  & 0.7473(13) & 0.7092(21) & 0.9490(33) \\
8.8500 & 8 & 1.3293(21) & 0.132140(5)  & 0.7416(13) & 0.7096(21) & 0.9569(33) \\
9.1859 & 12 & 1.3293(60) & 0.131814(3) & 0.7367(19) & 0.6985(24) & 0.9481(41)\\
9.4381 & 16 & 1.3293(40) & 0.131589(2) & 0.7303(17) & 0.6983(34) & 0.9562(52)\\
  [1.0ex]			                                              
8.3124 & 6 & 1.4300(20) & 0.132734(10) & 0.7371(10) & 0.6939(16) & 0.9414(25)\\
8.5598 & 8 & 1.4300(21) & 0.132453(5)  & 0.7308(12) & 0.6901(21) & 0.9443(33) \\
8.9003 & 12 & 1.4300(50) & 0.132095(3) & 0.7219(19) & 0.6869(27) & 0.9515(45)\\
9.1415 & 16 & 1.4300(58) & 0.131855(3) & 0.7196(15) & 0.6850(27) & 0.9519(42)\\
  [1.0ex]			                                              
7.9993 & 6 & 1.5553(15) & 0.133118(7)  & 0.7187(9) & 0.6723(24) & 0.9354(35) \\
8.2500 & 8 & 1.5553(24) & 0.132821(5)  & 0.7135(14) & 0.6747(24) & 0.9456(38) \\
8.5985 & 12 & 1.5533(70) & 0.132427(3) & 0.7109(20) & 0.6681(29) & 0.9398(49)\\
8.8323 & 16 & 1.5533(70) & 0.132169(3) & 0.7023(23) & 0.6667(34) & 0.9493(58) \\
  [1.0ex]
\Hline
\end{tabular}
\caption{
\label{tab:Z3}
Results for the step scaling function $\SigVApAV{;8}^-$ with Clover action.
}
\end{table}\addtocounter{table}{-1}
\clearpage
\begin{table}
\centering
\begin{tabular}{rrll@{\hspace{5mm}}lll}
\Hline \\[-1.0ex]
$\beta~~~$ & $\frac{L}{a}$ & $~~~\gbar^2(L)$ &
$~~~~~~\hopc$ & $Z_8^-\left(g_0,\frac{L}{a}\right)$ & $Z_8^-\left(g_0,\frac{2L}{a}\right)$ & $\Sigma_8^-\left(u,\frac{a}{L}\right)$ \\[1.0ex]
\hline \\[-1.0ex]
7.7170 & 6 & 1.6950(26) & 0.133517(8) & 0.7045(13) & 0.6558(8) & 0.9309(21) \\
7.9741 & 8 & 1.6950(28) & 0.133179(5) & 0.6998(11) & 0.6551(19) & 0.9361(31)\\
8.3218 & 12 & 1.6950(79) & 0.132756(4) & 0.6964(7) & 0.6533(25) & 0.9381(37)\\
8.5479 & 16 & 1.6950(90) & 0.132485(3) & 0.6887(23) & 0.6472(33) & 0.9397(57)\\
  [1.0ex]			                                            
7.4082 & 6 & 1.8811(22) & 0.133961(8) & 0.6828(14) & 0.6307(9) & 0.9237(23) \\
7.6547 & 8 & 1.8811(28) & 0.133632(6) & 0.6801(13) & 0.6336(23) & 0.9316(38)\\
7.9993 & 12 & 1.8811(38) & 0.133159(4) & 0.6775(8) & 0.6273(30) & 0.9259(46)\\
8.2415 & 16 & 1.8811(99) & 0.132847(3) & 0.6762(32) & 0.6268(27) & 0.9269(59)\\
  [1.0ex]			                                            
7.1214 & 6 & 2.1000(39) & 0.134423(9) & 0.6622(14) & 0.6036(10) & 0.9115(24)\\
7.3632 & 8 & 2.1000(45) & 0.134088(6) & 0.6583(13) & 0.6053(24) & 0.9195(41)\\
7.6985 & 12 & 2.1000(80) & 0.133599(4) & 0.6568(8) & 0.6060(27) & 0.9227(43)\\
7.9560 & 16 & 2.100(11) & 0.133229(3) & 0.6500(21) & 0.6027(24) & 0.9272(48)\\
  [1.0ex]			                                            
6.7807 & 6 & 2.4484(37) & 0.134994(11) & 0.6330(16) & 0.5639(10) & 0.8908(28)\\
7.0197 & 8 & 2.4484(45) & 0.134639(7) & 0.6295(14) & 0.5667(27) & 0.9002(47)\\
7.3551 & 12 & 2.4484(80) & 0.134141(5) & 0.6287(9) & 0.5669(29) & 0.9017(48)\\
7.6101 & 16 & 2.448(17) & 0.133729(4) & 0.6280(23) & 0.5737(26) & 0.9135(53)\\
  [1.0ex]			                                            
6.5512 & 6 & 2.770(7) & 0.135327(12) & 0.6079(17) & 0.5311(8) & 0.8737(28) \\
6.7860 & 8 & 2.770(7) & 0.135056(8) & 0.6056(15) & 0.5403(30) & 0.8922(54) \\
7.1190 & 12 & 2.770(11) & 0.134513(5) & 0.6057(10) & 0.5398(21) & 0.8912(38)\\
7.3686 & 16 & 2.770(14) & 0.134114(3) & 0.6059(27) & 0.5396(30) & 0.8906(63)\\
  [1.0ex]			                                            
6.3665 & 6 & 3.111(4) & 0.135488(6) & 0.5830(18) & 0.4976(26) & 0.8535(52) \\
6.6100 & 8 & 3.111(6) & 0.135339(3) & 0.5868(17) & 0.5144(32) & 0.8766(60) \\
6.9322 & 12 & 3.111(12) & 0.134855(3) & 0.5860(22) & 0.5090(29) & 0.8686(59)\\
7.1911 & 16 & 3.111(16) & 0.134411(3) & 0.5931(21) & 0.5226(34) & 0.8811(65)\\
  [1.0ex]			                                            
6.2204 & 6 & 3.480(8) & 0.135470(15) & 0.5615(9) & 0.4652(24) & 0.8285(45) \\
6.4527 & 8 & 3.480(14) & 0.135543(9) & 0.5629(11) & 0.4739(35) & 0.8419(64) \\
6.7750 & 12 & 3.480(39) & 0.135121(5) & 0.5703(20) & 0.4877(31) & 0.8552(62)\\
7.0203 & 16 & 3.480(21) & 0.134707(4) & 0.5671(20) & 0.4811(32) & 0.8484(64)\\
  [1.0ex]
\Hline
\end{tabular}
\caption{
(continued)
}
\end{table}

\begin{table}
\centering
\begin{tabular}{rrll@{\hspace{5mm}}lll}
\Hline \\[-1.0ex]
$\beta~~~$ & $\frac{L}{a}$ & $~~~\gbar^2(L)$ &
$~~~~~~\hopc$ & $Z_8^-\left(g_0,\frac{L}{a}\right)$ & $Z_8^-\left(g_0,\frac{2L}{a}\right)$ & $\Sigma_8^-\left(u,\frac{a}{L}\right)$ \\[1.0ex]
\hline \\[-1.0ex]
10.7503 & 6 & 0.8873(5) & 0.134696(7) & 0.8410(11) & 0.7891(14) & 0.9383(21) \\ 
11.0000 & 8 & 0.8873(10) & 0.134548(6) & 0.8248(10) & 0.7809(15) & 0.9468(22)\\ 
11.3384 & 12 & 0.8873(30) & 0.134277(5) & 0.8110(13) & 0.7709(20) & 0.9506(29) \\ 
11.5736 & 16 & 0.8873(25) & 0.134068(6) & 0.8024(13) & 0.7691(24) & 0.9585(34) \\ 
  [1.0ex]			                                             
10.0500 & 6 & 0.9944(7) & 0.135659(8) & 0.8243(12) & 0.7698(15) & 0.9339(23) \\ 
10.3000 & 8 & 0.9944(13) & 0.135457(5) & 0.8093(10) & 0.7586(17) & 0.9374(24)\\ 
10.6086 & 12 & 0.9944(30) & 0.135160(4) & 0.7956(14) & 0.7531(26) & 0.9466(37) \\ 
10.8910 & 16 & 0.9944(28) & 0.134849(6) & 0.7876(15) & 0.7510(30) & 0.9535(42) \\ 
  [1.0ex]			                                             
9.5030 & 6 & 1.0989(8) & 0.136520(5) & 0.8121(12) & 0.7521(17) & 0.9261(25) \\ 
9.7500 & 8 & 1.0989(13) & 0.136310(3) & 0.7966(11) & 0.7439(16) & 0.9338(24) \\ 
10.0577 & 12 & 1.0989(40) & 0.135949(4) & 0.7818(15) & 0.7341(25) & 0.9390(37) \\ 
10.3419 & 16 & 1.0989(44) & 0.135572(4) & 0.7739(22) & 0.7293(23) & 0.9424(40) \\ 
  [1.0ex]			                                             
8.8997 & 6 & 1.2430(13) & 0.137706(5) & 0.7933(13) & 0.7321(17) & 0.9229(26) \\ 
9.1544 & 8 & 1.2430(14) & 0.137400(4) & 0.7786(11) & 0.7188(20) & 0.9232(29) \\ 
9.5202 & 12 & 1.2430(35) & 0.136855(2) & 0.7609(15) & 0.7120(22) & 0.9357(34)\\ 
9.7350 & 16 & 1.2430(34) & 0.136523(4) & 0.7514(17) & 0.7084(28) & 0.9428(43)\\ 
  [1.0ex]			                                             
8.6129 & 6 & 1.3293(12) & 0.138346(6) & 0.7872(15) & 0.7158(15) & 0.9093(26) \\ 
8.8500 & 8 & 1.3293(21) & 0.138057(4) & 0.7694(12) & 0.7038(20) & 0.9147(30) \\ 
9.1859 & 12 & 1.3293(60) & 0.137503(2) & 0.7503(16) & 0.6980(23) & 0.9303(37)\\ 
9.4381 & 16 & 1.3293(40) & 0.137061(4) & 0.7430(25) & 0.6891(29) & 0.9275(50)\\ 
  [1.0ex]			                                             
8.3124 & 6 & 1.4300(20)  & 0.139128(11) & 0.7693(15) & 0.7017(18) & 0.9121(29) \\ 
8.5598 & 8 & 1.4300(21) & 0.138742(7) & 0.7569(13) & 0.6885(21) & 0.9096(32) \\ 
8.9003 & 12 & 1.4300(50) & 0.138120(8) & 0.7419(18) & 0.6877(32) & 0.9269(49)\\ 
9.1415 & 16 & 1.4300(58) & 0.137655(5) & 0.7316(19) & 0.6818(31) & 0.9319(49)\\ 
  [1.0ex]			                                             
7.9993 & 6 & 1.5553(15) & 0.140003(11)& 0.7600(16) & 0.6797(19) & 0.8943(31) \\ 
8.2500 & 8 & 1.5553(24) & 0.139588(8) & 0.7423(13) & 0.6739(22) & 0.9079(34) \\ 
8.5985 & 12 & 1.5533(70) & 0.138847(6) & 0.7300(19) & 0.6681(35) & 0.9152(54)\\ 
8.8323 & 16 & 1.5533(70) & 0.138339(7) & 0.7208(28) & 0.6594(31) & 0.9148(56)\\ 
  [1.0ex]
\Hline
\end{tabular}
\caption{
Results for the step scaling function $\SigVApAV{;8}^-$ with Wilson action.
}
\label{tab:Z4}
\end{table}\addtocounter{table}{-1}
\clearpage
\begin{table}
\centering
\begin{tabular}{rrll@{\hspace{5mm}}lll}
\Hline \\[-1.0ex]
$\beta~~~$ & $\frac{L}{a}$ & $~~~\gbar^2(L)$ &
$~~~~~~\hopc$ & $Z_8^-\left(g_0,\frac{L}{a}\right)$ & $Z_8^-\left(g_0,\frac{2L}{a}\right)$ & $\Sigma_8^-\left(u,\frac{a}{L}\right)$ \\[1.0ex]
\hline \\[-1.0ex]
7.7170 & 6 & 1.6950(26) & 0.140954(12) & 0.7443(16) & 0.6638(20) & 0.8918(33)  \\ 
7.9741 & 8 & 1.6950(28) & 0.140438(8) & 0.7292(15) & 0.6636(30) & 0.9100(45)  \\ 
8.3218 & 12 & 1.6950(79)& 0.139589(6) & 0.7131(20) & 0.6535(30) & 0.9164(49)  \\ 
8.5479 & 16 & 1.6950(90) & 0.139058(6)  & 0.7038(22) & 0.6445(41) & 0.9157(65) \\ 
  [1.0ex]			                                              
7.4082 & 6 & 1.8811(22) & 0.142145(11) & 0.7280(17) & 0.6404(21) & 0.8797(35)  \\ 
7.6547 & 8 & 1.8811(28) & 0.141572(9) & 0.7123(16) & 0.6339(25) & 0.8899(40)  \\ 
7.9993 & 12 & 1.8811(38)& 0.140597(6) & 0.6965(21) & 0.6269(33) & 0.9001(55)  \\ 
8.2415 & 16 & 1.8811(99) & 0.139900(6)  & 0.6855(26) & 0.6303(36) & 0.9195(63) \\ 
  [1.0ex]			                                              
7.1214 & 6 & 2.1000(39)  & 0.143416(11) & 0.7040(18) & 0.6116(23) & 0.8688(40) \\ 
7.3632 & 8 & 2.1000(45)  & 0.142749(9) & 0.6886(16) & 0.6009(32) & 0.8726(51)  \\ 
7.6985 & 12 & 2.1000(80) & 0.141657(6) & 0.6754(16) & 0.5992(22) & 0.8872(39)  \\ 
7.9560 & 16 & 2.100(11)  & 0.140817(7) & 0.6688(24) & 0.5995(29) & 0.8964(54)  \\ 
  [1.0ex]			                                              
6.7807 & 6 & 2.4484(37) & 0.145286(11) & 0.6818(18) & 0.5807(22) & 0.8517(39) \\ 
7.0197 & 8 & 2.4484(45)  & 0.144454(7) & 0.6644(17) & 0.5678(25) & 0.8546(44)  \\ 
7.3551 & 12 & 2.4484(80) & 0.143113(6) & 0.6537(19) & 0.5576(35) & 0.8530(59)  \\ 
7.6101 & 16 & 2.448(17)  & 0.142107(6) & 0.6452(25) & 0.5735(42) & 0.8889(74)  \\ 
  [1.0ex]			                                              
6.5512 & 6 & 2.770(7) & 0.146825(11) &  0.6596(21) & 0.5459(23) & 0.8276(44) \\ 
6.7860 & 8 & 2.770(7) & 0.145859(7) & 0.6430(20) & 0.5370(34) & 0.8351(59) \\ 
7.1190 & 12 & 2.770(11) & 0.144299(8) & 0.6303(24) & 0.5366(28) & 0.8513(55)  \\ 
7.3686 & 16 & 2.770(14) & 0.143175(7) & 0.6273(31) & 0.5423(38) & 0.8645(74)  \\ 
  [1.0ex]			                                              
6.3665 & 6 & 3.111(4) & 0.148317(10) &  0.6420(22) & 0.5124(21) & 0.7981(43) \\ 
6.6100 & 8 & 3.111(6) & 0.147112(7) & 0.6276(21) & 0.5100(25) & 0.8126(48) \\ 
6.9322 & 12 & 3.111(12) & 0.145371(7) & 0.6124(22) & 0.5053(38) & 0.8251(69)  \\ 
7.1911 & 16 & 3.111(16) & 0.144060(8) & 0.6059(27) & 0.5151(33) & 0.8501(66)  \\ 
  [1.0ex]			                                              
6.2204 & 6 & 3.480(8) & 0.149685(15) &  0.6236(24) & 0.4871(25) & 0.7811(50) \\ 
6.4527 & 8 & 3.480(14) & 0.148391(9) &  0.6017(22) & 0.4759(43) & 0.7909(77) \\ 
6.7750 & 12 & 3.480(39) & 0.146408(7) & 0.5955(21) & 0.4808(46) & 0.8074(82)  \\ 
7.0203 & 16 & 3.480(21) & 0.145025(8) & 0.5863(25) & 0.4917(45) & 0.8386(85)  \\ 
  [1.0ex]
\Hline
\end{tabular}
\caption{
(continued)
}
\end{table}

\begin{figure}[p]
\centering
\hspace*{-3mm}\psfig{file=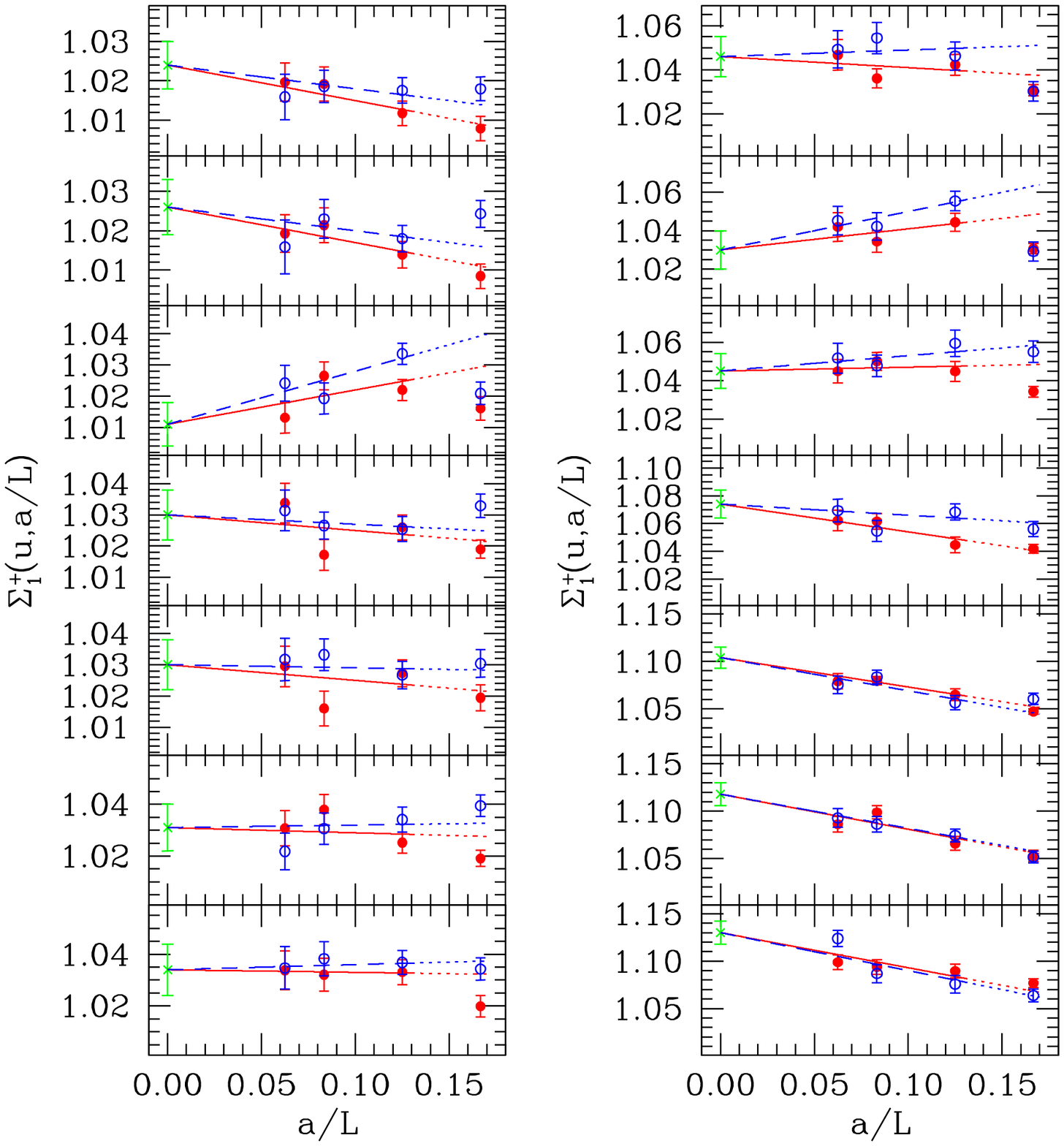,width=15.0cm}
\caption{
Continuum extrapolations of $\SigVApAV{;1}^+$ at fixed renormalized coupling $u$
for the improved action (full symbols, solid line) and the unimproved action
(open symbols, dashed line).
The $L/a=6$ data points have not been included in the fits. 
The value of $u$ increases from top to bottom and from left to right.
}
\label{fig:extrap_1}
\end{figure}

\begin{figure}[p]
\centering
\hspace*{-3mm}\psfig{file=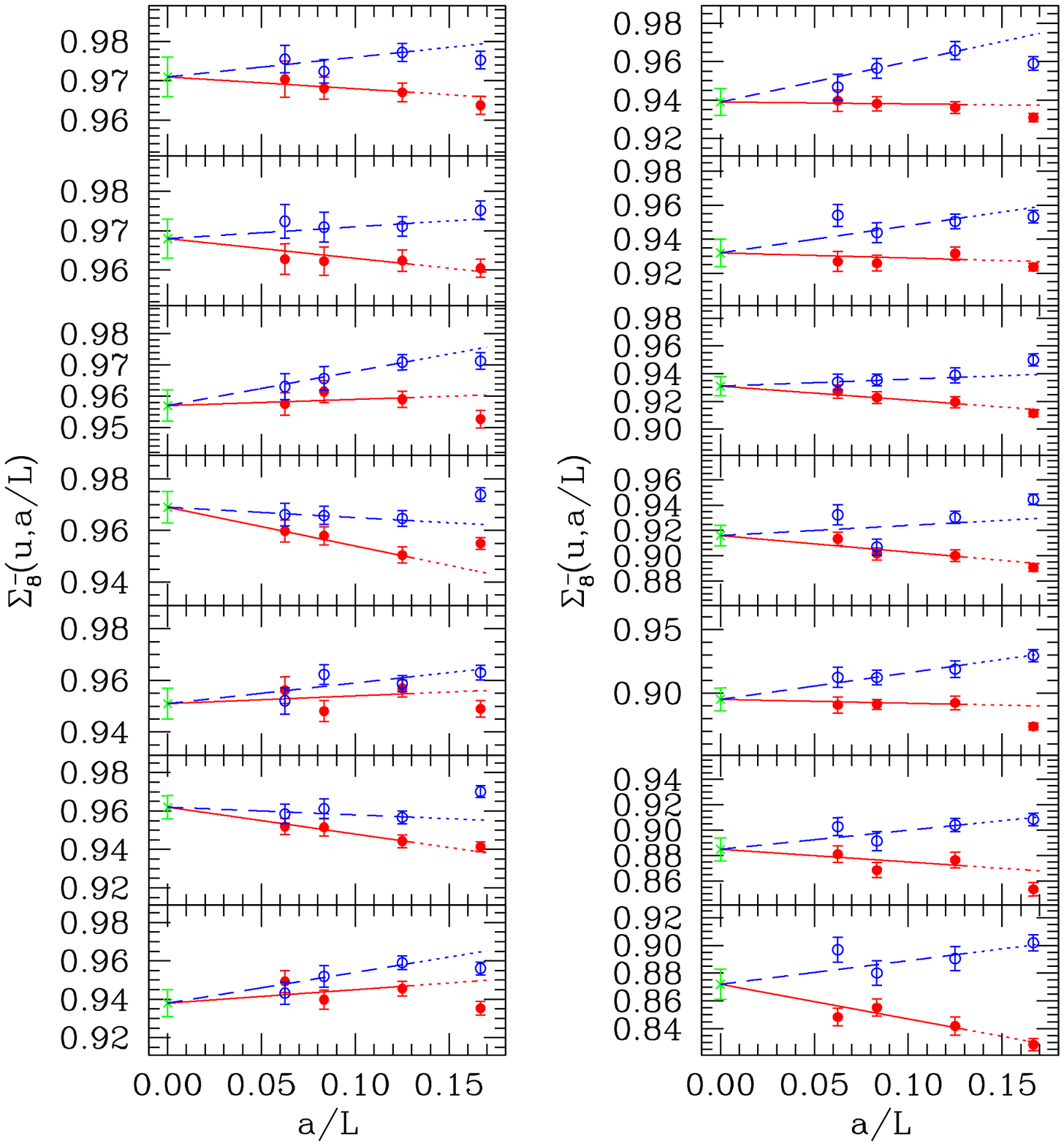,width=15.0cm}
\caption{
Continuum extrapolations of $\SigVApAV{;8}^-$ at fixed renormalized coupling $u$
for the improved action (full symbols, solid line) and the unimproved action
(open symbols, dashed line).
The $L/a=6$ data points have not been included in the fits. 
The value of $u$ increases from top to bottom and from left to right.
}
\label{fig:extrap_2}
\end{figure}

\begin{table}
\centering
\begin{tabular}{l@{\hspace{10mm}}lllll}
\Hline\\[-1.0ex]
$~u$ &
$~\sigma^+_1(u)$ &
$~\sigma^+_2(u)$ &
$~\sigma^+_3(u)$ &
$~\sigma^+_4(u)$ &
$~\sigma^+_5(u)$ \\[1.0ex]
\hline\\[-1.0ex]
0.8873 & 1.024(6)  & 1.022(7)  & 1.026(7)  & 1.020(7)  & 1.020(7)  \\
0.9944 & 1.026(7)  & 1.028(8)  & 1.029(7)  & 1.025(8)  & 1.025(8)  \\
1.0989 & 1.011(7)  & 1.008(8)  & 1.012(7)  & 1.006(7)  & 1.007(7)  \\
1.2430 & 1.030(8)  & 1.031(9)  & 1.034(8)  & 1.028(9)  & 1.028(8)  \\
1.3293 & 1.030(8)  & 1.031(10) & 1.034(9)  & 1.027(9)  & 1.027(9)  \\
1.4300 & 1.031(9)  & 1.035(11) & 1.035(9)  & 1.031(10) & 1.031(10) \\
1.5553 & 1.034(10) & 1.038(12) & 1.040(10) & 1.032(11) & 1.033(11) \\
1.6950 & 1.046(9)  & 1.052(11) & 1.055(10) & 1.043(10) & 1.044(10) \\
1.8811 & 1.030(10) & 1.035(12) & 1.039(10) & 1.027(11) & 1.028(11) \\
2.1000 & 1.045(9)  & 1.060(12) & 1.058(10) & 1.048(10) & 1.050(10) \\
2.4484 & 1.074(10) & 1.093(13) & 1.095(11) & 1.073(12) & 1.074(12) \\ 
2.770  & 1.104(11) & 1.143(15) & 1.138(12) & 1.110(13) & 1.112(12) \\ 
3.111  & 1.118(12) & 1.157(17) & 1.157(14) & 1.120(14) & 1.124(14) \\
3.480  & 1.130(12) & 1.186(18) & 1.180(15) & 1.140(15) & 1.146(14) \\
[1.0ex]\Hline
\end{tabular}

\vspace{10truemm}

\begin{tabular}{l@{\hspace{10mm}}llll}
\Hline\\[-1.0ex]
$~u$ &
$~\sigma^+_6(u)$ &
$~\sigma^+_7(u)$ &
$~\sigma^+_8(u)$ &
$~\sigma^+_9(u)$ \\[1.0ex]
\hline\\[-1.0ex]
0.8873 & 1.018(7)  & 1.023(6)  & 1.017(7)  & 1.017(7)  \\
0.9944 & 1.022(8)  & 1.025(7)  & 1.021(8)  & 1.021(8)  \\
1.0989 & 1.005(7)  & 1.010(7)  & 1.004(7)  & 1.004(7)  \\
1.2430 & 1.024(8)  & 1.029(8)  & 1.023(8)  & 1.024(8)  \\
1.3293 & 1.024(9)  & 1.029(8)  & 1.023(9)  & 1.022(8)  \\
1.4300 & 1.025(9)  & 1.029(8)  & 1.025(9)  & 1.024(9)  \\
1.5553 & 1.027(10) & 1.033(9)  & 1.025(10) & 1.026(10) \\
1.6950 & 1.035(10) & 1.044(9)  & 1.031(10) & 1.033(9)  \\
1.8811 & 1.018(10) & 1.028(9)  & 1.016(10) & 1.017(10) \\
2.1000 & 1.034(9)  & 1.041(9)  & 1.031(9)  & 1.033(9)  \\
2.4484 & 1.050(11) & 1.066(9)  & 1.045(10) & 1.047(10) \\ 
2.770  & 1.072(11) & 1.090(10) & 1.063(11) & 1.065(10) \\ 
3.111  & 1.073(12) & 1.100(11) & 1.064(11) & 1.068(11) \\
3.480  & 1.084(12) & 1.111(11) & 1.074(12) & 1.079(11) \\
[1.0ex]\Hline
\end{tabular}
\caption{
Continuum extrapolations of $\SigVApAV{;s}^+$ combining Clover and
(perturbatively $\Oa$ improved) Wilson data.
Linear dependence on $(a/L)$ is assumed for both actions.
The $L/a=6$ data have not been taken into account.
}
\label{tab:CLe1}
\end{table}

\clearpage

\begin{table}
\centering
\begin{tabular}{l@{\hspace{10mm}}lllll}
\Hline\\[-1.0ex]
$~u$ &
$~\sigma^-_1(u)$ &
$~\sigma^-_2(u)$ &
$~\sigma^-_3(u)$ &
$~\sigma^-_4(u)$ &
$~\sigma^-_5(u)$ \\[1.0ex]
\hline\\[-1.0ex]

0.8873 & 0.981(5)  & 0.978(6)  & 0.985(6)  & 0.973(5)  & 0.974(6)  \\
0.9944 & 0.971(6)  & 0.975(7)  & 0.976(7)  & 0.971(6)  & 0.970(6)  \\
1.0989 & 0.968(5)  & 0.963(7)  & 0.972(7)  & 0.959(6)  & 0.961(6)  \\
1.2430 & 0.976(6)  & 0.980(8)  & 0.984(7)  & 0.974(6)  & 0.975(7)  \\
1.3293 & 0.961(6)  & 0.960(9)  & 0.967(8)  & 0.955(7)  & 0.955(8)  \\
1.4300 & 0.963(6)  & 0.975(8)  & 0.972(8)  & 0.968(7)  & 0.969(8)  \\
1.5553 & 0.950(8)  & 0.952(11) & 0.960(10) & 0.944(8)  & 0.944(9)  \\
1.6950 & 0.954(7)  & 0.962(10) & 0.970(9)  & 0.949(8)  & 0.951(9)  \\
1.8811 & 0.953(7)  & 0.956(11) & 0.966(9)  & 0.943(9)  & 0.942(10) \\
2.1000 & 0.934(7)  & 0.967(11) & 0.959(8)  & 0.947(8)  & 0.951(9)  \\
2.4484 & 0.943(8)  & 0.971(13) & 0.982(10) & 0.941(10) & 0.942(10) \\
2.770  & 0.911(8)  & 0.986(14) & 0.974(10) & 0.934(10) & 0.938(11) \\
3.111  & 0.908(8)  & 0.987(15) & 0.982(11) & 0.930(11) & 0.939(12) \\
3.480  & 0.896(10) & 0.995(16) & 0.983(12) & 0.926(12) & 0.935(13) \\
[1.0ex]\Hline
\end{tabular}

\vspace{10truemm}

\begin{tabular}{l@{\hspace{10mm}}llll}
\Hline\\[-1.0ex]
$~u$ &
$~\sigma^-_6(u)$ &
$~\sigma^-_7(u)$ &
$~\sigma^-_8(u)$ &
$~\sigma^-_9(u)$ \\[1.0ex]
\hline\\[-1.0ex]

0.8873 & 0.974(5)  & 0.982(6)  & 0.971(5)  & 0.971(5)  \\
0.9944 & 0.971(6)  & 0.973(6)  & 0.968(5)  & 0.967(6)  \\
1.0989 & 0.960(5)  & 0.970(6)  & 0.957(5)  & 0.958(6)  \\
1.2430 & 0.973(6)  & 0.980(6)  & 0.969(6)  & 0.970(6)  \\
1.3293 & 0.954(7)  & 0.963(7)  & 0.951(6)  & 0.950(7)  \\
1.4300 & 0.967(7)  & 0.967(7)  & 0.962(6)  & 0.963(7)  \\
1.5553 & 0.943(9)  & 0.953(8)  & 0.938(7)  & 0.938(8)  \\
1.6950 & 0.947(8)  & 0.960(7)  & 0.939(7)  & 0.940(8)  \\
1.8811 & 0.940(9)  & 0.956(8)  & 0.932(8)  & 0.932(9)  \\
2.1000 & 0.944(8)  & 0.944(7)  & 0.931(7)  & 0.935(8)  \\
2.4484 & 0.933(10) & 0.956(8)  & 0.916(8)  & 0.917(9)  \\
2.770  & 0.925(10) & 0.932(8)  & 0.895(9)  & 0.898(10) \\
3.111  & 0.917(11) & 0.935(9)  & 0.885(9)  & 0.893(10) \\
3.480  & 0.910(12) & 0.927(10) & 0.872(11) & 0.880(12) \\
[1.0ex]\Hline
\end{tabular}
\caption{
Continuum extrapolations of $\SigVApAV{;s}^-$ combining Clover and
(perturbatively $\Oa$ improved) Wilson data.
Linear dependence on $(a/L)$ is assumed for both actions.
The $L/a=6$ data have not been taken into account.
}
\label{tab:CLe2}
\end{table}

\clearpage

\begin{table}
\centering
\begin{tabular}{cclr@{\hspace{0pt}}llr@{\hspace{0pt}}lll}
\Hline\\[-1.0ex]
$s$ & Fit & ~~~~~~$s_1$ & \multicolumn{2}{c}{~~$s_2$} & ~~~~~~$s_3$ &  \multicolumn{2}{c}{~~$s_4$} &
$\frac{\chi^2}{{\rm d.o.f.}}$ & $\zrgi_{\rm s}^+(\mumin)$\\[1.0ex]
\hline\\[-1.0ex]
1 & A & 0.01755762 & 0.&00529(56) & & &  & 1.27 & 1.095(16) \\ 
  & B & 0.01755762 & 0.&0006(23) & 0.00167(81) &&  & 1.02 & 1.108(18) \\ 
  & C & 0.01755762 & 0.&00137(2) & 0.00142(19) &&  & 0.95 & 1.105(15) \\ 
  & D & 0.01755762 & $-$0.&0021(70) & 0.0041(59) & $-$0.&0005(12) & 1.10 & 1.111(19) \\ 
  & E & 0.01755762 & 0.&00137(2) & 0.0012(12) & 0.&00006(38) & 1.03 & 1.106(16) \\ 
[1.0ex]\hline\\
2 & A & 0.01755762 & 0.&00873(76) &  &&  & 1.75 & 1.050(21) \\ 
  & B & 0.01755762 & $-$0.&0006(29) & 0.0035(11) &&  & 0.99 & 1.067(22) \\ 
  & C & 0.01755762 & $-$0.&00320(2) & 0.00438(27) &&  & 0.97 & 1.076(20) \\ 
  & D & 0.01755762 & $-$0.&0068(87) & 0.0091(75) & $-$0.&0011(15) & 1.02 & 1.074(24) \\ 
  & E & 0.01755762 & $-$0.&00320(2) & 0.0060(16) & $-$0.&00055(51) & 0.95 & 1.069(21) \\ 
[1.0ex]\hline\\
3 & A & 0.01755762 & 0.&00884(63) &  &&  & 1.83 & 0.990(17) \\ 
  & B & 0.01755762 & 0.&0014(25) & 0.00275(90) & & & 1.22 & 1.004(18) \\ 
  & C & 0.01755762 & 0.&00218(2) & 0.00248(23) &  && 1.13 & 1.001(15) \\ 
  & D & 0.01755762 & $-$0.&0030(76) & 0.0066(65) & $-$0.&0008(13) & 1.29 & 1.008(19) \\ 
  & E & 0.01755762 & 0.&00218(2) & 0.0024(13) & 0.&00004(43) & 1.22 & 1.001(16) \\ 
[1.0ex]\hline\\
4 & A & 0.01755762 & 0.&00537(66) &  &&  & 1.57 & 1.163(20) \\ 
  & B & 0.01755762 & $-$0.&0023(26) & 0.00281(94) &&  & 0.95 & 1.182(22) \\ 
  & C & 0.01755762 & $-$0.&00451(2) & 0.00356(23) & & & 0.93 & 1.191(19) \\ 
  & D & 0.01755762 & $-$0.&0086(80) & 0.0084(68) & $-$0.&0011(14) & 0.97 & 1.190(24) \\ 
  & E & 0.01755762 & $-$0.&00451(2) & 0.0050(14) & $-$0.&00047(45) & 0.91 & 1.183(20) \\ 
[1.0ex]\hline\\
5 & A & 0.01755762 & 0.&00584(63) &  &&  & 1.60 & 1.143(19) \\ 
  & B & 0.01755762 & $-$0.&0020(26) & 0.00285(91) &  && 0.92 & 1.164(21) \\ 
  & C & 0.01755762 & $-$0.&00404(1) & 0.00353(22) &  && 0.89 & 1.173(18) \\ 
  & D & 0.01755762 & $-$0.&0084(78) & 0.0085(66) & $-$0.&0011(13) & 0.93 & 1.171(23) \\ 
  & E & 0.01755762 & $-$0.&00404(1) & 0.0049(13) & $-$0.&00044(43) & 0.88 & 1.165(19) \\ 
[1.0ex]\Hline\\
\end{tabular}
\caption{
Fits to the continuum step scaling functions $\sigVApAV{;s}^+$ and
results for the ratio $\zrgiVApAV{;s}^+(\mumin)$.
}
\label{tab:CLfit1}
\end{table}\addtocounter{table}{-1}

\clearpage

\begin{table}
\centering
\begin{tabular}{cclr@{\hspace{0pt}}llr@{\hspace{0pt}}lll}
\Hline\\[-1.0ex]
$s$ & Fit & ~~~~~~$s_1$ & \multicolumn{2}{c}{~~$s_2$} & ~~~~~~$s_3$ &  \multicolumn{2}{c}{~~$s_4$} &
$\frac{\chi^2}{{\rm d.o.f.}}$ & $\zrgi_{\rm s}^+(\mumin)$\\[1.0ex]
\hline\\[-1.0ex]
6 & A & 0.01755762 & 0.&00136(56) &  & & & 0.96 & 1.294(20) \\ 
  & B & 0.01755762 & $-$0.&0030(24) & 0.00154(83) &&  & 0.75 & 1.309(22) \\ 
  & C & 0.01755762 & $-$0.&00442(2) & 0.00203(19) & & & 0.72 & 1.317(18) \\ 
  & D & 0.01755762 & $-$0.&0071(74) & 0.0052(62) & $-$0.&0007(12) & 0.79 & 1.315(24) \\ 
  & E & 0.01755762 & $-$0.&00442(2) & 0.0030(12) & $-$0.&00030(39) & 0.74 & 1.310(20) \\ 
[1.0ex]\hline\\
7 & A & 0.01755762 & 0.&00387(51) &  &  && 1.06 & 1.138(16) \\ 
  & B & 0.01755762 & 0.&0004(22) & 0.00122(77) &&  & 0.93 & 1.148(17) \\ 
  & C & 0.01755762 & 0.&00137(2) & 0.00091(18) & & & 0.87 & 1.144(14) \\ 
  & D & 0.01755762 & $-$0.&0019(68) & 0.0033(56) & $-$0.&0004(11) & 1.00 & 1.151(19) \\ 
  & E & 0.01755762 & 0.&00137(2) & 0.0006(11) & 0.&00010(36) & 0.94 & 1.146(16) \\ 
[1.0ex]\hline\\
8 & A & 0.01755762 & 0.&00049(55) &  & & & 0.87 & 1.338(20) \\ 
  & B & 0.01755762 & $-$0.&0037(24) & 0.00147(83) &&  & 0.67 & 1.353(22) \\ 
  & C & 0.01755762 & $-$0.&00533(1) & 0.00204(19) & & & 0.66 & 1.362(18) \\ 
  & D & 0.01755762 & $-$0.&0070(73) & 0.0044(61) & $-$0.&0006(12) & 0.71 & 1.358(25) \\ 
  & E & 0.01755762 & $-$0.&00533(1) & 0.0030(12) & $-$0.&00032(39) & 0.66 & 1.355(20) \\ 
[1.0ex]\hline\\
9 & A & 0.01755762 & 0.&00092(52) &  &&  & 0.92 & 1.316(19) \\ 
  & B & 0.01755762 & $-$0.&0033(23) & 0.00150(79) &&  & 0.70 & 1.332(21) \\ 
  & C & 0.01755762 & $-$0.&00485(2) & 0.00200(18) & & & 0.67 & 1.341(17) \\ 
  & D & 0.01755762 & $-$0.&0075(71) & 0.0050(59) & $-$0.&0007(11) & 0.72 & 1.338(23) \\ 
  & E & 0.01755762 & $-$0.&00485(2) & 0.0029(12) & $-$0.&00030(37) & 0.68 & 1.333(19) \\ 
[1.0ex]\Hline\\
\end{tabular}
\caption{
(continued)
}
\end{table}

\begin{table}
\centering
\begin{tabular}{cclr@{\hspace{0pt}}lr@{\hspace{0pt}}lr@{\hspace{0pt}}lll}
\Hline\\[-1.0ex]
$s$ & Fit & ~~~~~~$s_1$ & \multicolumn{2}{c}{~~$s_2$} & \multicolumn{2}{c}{~~$s_3$} &  \multicolumn{2}{c}{~~$s_4$} &
$\frac{\chi^2}{{\rm d.o.f.}}$ & $\zrgi_{\rm s}^-(\mumin)$\\[1.0ex]
\hline\\[-1.0ex]
1 & A & $-0.03511524$ & 0.&00252(42) &&&  &  & 1.87 & 0.497(6) \\ 
  & B & $-0.03511524$ & 0.&0085(18) & $-$0.&00217(63) & & & 1.03 & 0.490(7) \\ 
  & C & $-0.03511524$ & 0.&01531(2) & $-$0.&00448(15) & & & 2.06 & 0.479(6) \\ 
  & D & $-0.03511524$ & 0.&0154(54) & $-$0.&0082(46) & 0.&00122(91) & 0.96 & 0.486(7) \\ 
  & E & $-0.03511524$ & 0.&01531(2) & $-$0.&00815(92) & 0.&00120(30) & 0.88 & 0.486(6) \\ 
[1.0ex]\hline\\
2 & A & $-0.03511524$ & 0.&00941(69) &  && & & 0.72 & 0.449(9) \\ 
  & B & $-0.03511524$ & 0.&0069(26) & 0.&00094(93) &&  & 0.70 & 0.450(9) \\ 
  & C & $-0.03511524$ & 0.&00539(1) & 0.&00146(25) &&  & 0.67 & 0.453(8) \\ 
  & D & $-0.03511524$ & 0.&0049(77) & 0.&0028(67) & $-$0.&0004(14) & 0.76 & 0.451(10) \\ 
  & E & $-0.03511524$ & 0.&00539(1) & 0.&0023(14) & $-$0.&00029(46) & 0.69 & 0.451(9) \\ 
[1.0ex]\hline\\
3 & A & $-0.03511524$ & 0.&00934(53) &  &&&  & 0.75 & 0.401(6) \\ 
  & B & $-0.03511524$ & 0.&0113(22) & $-$0.&00071(78)& &  & 0.75 & 0.399(6) \\ 
  & C & $-0.03511524$ & 0.&01612(2) & $-$0.&00235(19) &&  & 1.05 & 0.392(5) \\ 
  & D & $-0.03511524$ & 0.&0140(68) & $-$0.&0030(57) & 0.&0005(11) & 0.80 & 0.398(7) \\ 
  & E & $-0.03511524$ & 0.&01612(2) & $-$0.&0048(11) & 0.&00080(37) & 0.74 & 0.397(6) \\ 
[1.0ex]\hline\\
4 & A & $-0.03511524$ & 0.&00407(52) &&&  &  & 0.81 & 0.541(8) \\ 
  & B & $-0.03511524$ & 0.&0042(20) & $-$0.&00006(73) &&  & 0.87 & 0.541(9) \\ 
  & C & $-0.03511524$ & 0.&00276(2) & 0.&00045(18) &  && 0.85 & 0.544(8) \\ 
  & D & $-0.03511524$ & 0.&0046(61) & $-$0.&0004(52) & 0.&0001(11) & 0.95 & 0.541(9) \\ 
  & E & $-0.03511524$ & 0.&00276(2) & 0.&0012(11) & $-$0.&00023(35) & 0.88 & 0.542(8) \\ 
[1.0ex]\hline\\
5 & A & $-0.03511524$ & 0.&00466(56) && & &  & 0.69 & 0.521(9) \\ 
  & B & $-0.03511524$ & 0.&0041(23) & 0.&00022(80) &&  & 0.74 & 0.522(9) \\ 
  & C & $-0.03511524$ & 0.&00451(1) & 0.&00007(20) & & & 0.69 & 0.521(8) \\ 
  & D & $-0.03511524$ & 0.&0044(67) & $-$0.&0001(57) & 0.&0001(11) & 0.81 & 0.522(10) \\ 
  & E & $-0.03511524$ & 0.&00451(1) & $-$0.&0002(12) & 0.&00008(39) & 0.74 & 0.522(8) \\ 
[1.0ex]\Hline\\
\end{tabular}
\caption{
Fits to the continuum step scaling functions $\sigVApAV{;s}^-$ and
results for the ratio $\zrgiVApAV{;s}^-(\mumin)$.
}
\label{tab:CLfit2}
\end{table}\addtocounter{table}{-1}

\clearpage

\begin{table}
\centering
\begin{tabular}{cclr@{\hspace{0pt}}lr@{\hspace{0pt}}lr@{\hspace{0pt}}lll}
\Hline\\[-1.0ex]
$s$ & Fit & ~~~~~~$s_1$ & \multicolumn{2}{c}{~~$s_2$} & \multicolumn{2}{c}{~~$s_3$} &  \multicolumn{2}{c}{~~$s_4$} &
$\frac{\chi^2}{{\rm d.o.f.}}$ & $\zrgi_{\rm s}^-(\mumin)$\\[1.0ex]
\hline\\[-1.0ex]
6 & A & $-0.03511524$ & 0.&00297(52) & & &&  & 0.81 & 0.551(9) \\ 
  & B & $-0.03511524$ & 0.&0044(20) & $-$0.&00052(72) &&  & 0.83 & 0.550(9) \\ 
  & C & $-0.03511524$ & 0.&00416(1) & $-$0.&00044(18) &&  & 0.77 & 0.550(8) \\ 
  & D & $-0.03511524$ & 0.&0058(60) & $-$0.&0018(52) & 0.&0003(10) & 0.90 & 0.549(9) \\ 
  & E & $-0.03511524$ & 0.&00416(1) & $-$0.&0004(11) & $-$0.&00001(35) & 0.83 & 0.550(8) \\ 
[1.0ex]\hline\\
7 & A & $-0.03511524$ & 0.&00491(44) & & & & & 1.34 & 0.462(6) \\ 
  & B & $-0.03511524$ & 0.&0094(19) & $-$0.&00162(66) & & & 0.95 & 0.457(6) \\ 
  & C & $-0.03511524$ & 0.&01531(2) & $-$0.&00362(15) &  && 1.62 & 0.447(5) \\ 
  & D & $-0.03511524$ & 0.&0154(58) & $-$0.&0068(48) & 0.&00103(95) & 0.92 & 0.453(7) \\ 
  & E & $-0.03511524$ & 0.&01531(2) & $-$0.&00673(95) & 0.&00102(31) & 0.85 & 0.453(6) \\ 
[1.0ex]\hline\\
8 & A & $-0.03511524$ & $-$0.&00017(45) & & &&  & 1.00 & 0.622(9) \\ 
  & B & $-0.03511524$ & 0.&0021(18) & $-$0.&00082(64) &&  & 0.95 & 0.620(9) \\ 
  & C & $-0.03511524$ & 0.&00194(1) & $-$0.&00078(16) & & & 0.88 & 0.621(8) \\ 
  & D & $-0.03511524$ & 0.&0048(54) & $-$0.&0033(46) & 0.&00051(94) & 1.01 & 0.618(10) \\ 
  & E & $-0.03511524$ & 0.&00194(1) & $-$0.&00089(95) & 0.&00004(31) & 0.95 & 0.621(8) \\ 
[1.0ex]\hline\\
9 & A & $-0.03511524$ & 0.&00038(50) &  & && & 0.89 & 0.600(9) \\ 
  & B & $-0.03511524$ & 0.&0022(20) & $-$0.&00065(72) &&  & 0.89 & 0.599(10) \\ 
  & C & $-0.03511524$ & 0.&00370(2) & $-$0.&00118(18) & & & 0.87 & 0.595(9) \\ 
  & D & $-0.03511524$ & 0.&0057(61) & $-$0.&0038(52) & 0.&0006(10) & 0.94 & 0.596(10) \\ 
  & E & $-0.03511524$ & 0.&00370(2) & $-$0.&0022(11) & 0.&00032(35) & 0.87 & 0.598(9) \\ 
[1.0ex]\Hline\\
\end{tabular}
\caption{
(continued)
}
\end{table}

\clearpage

\begin{figure}[p]
\centering
\vspace{178mm}
\includegraphics{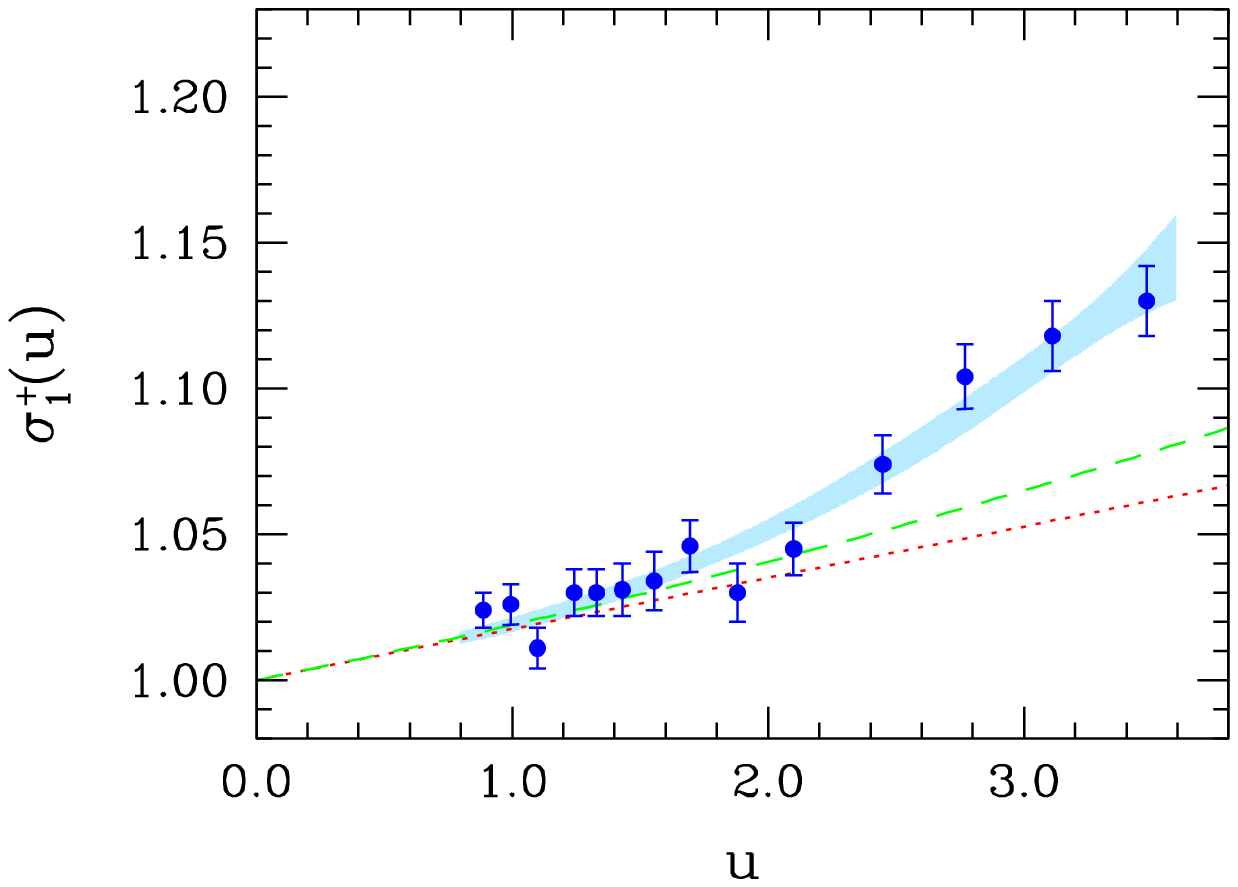}
\includegraphics{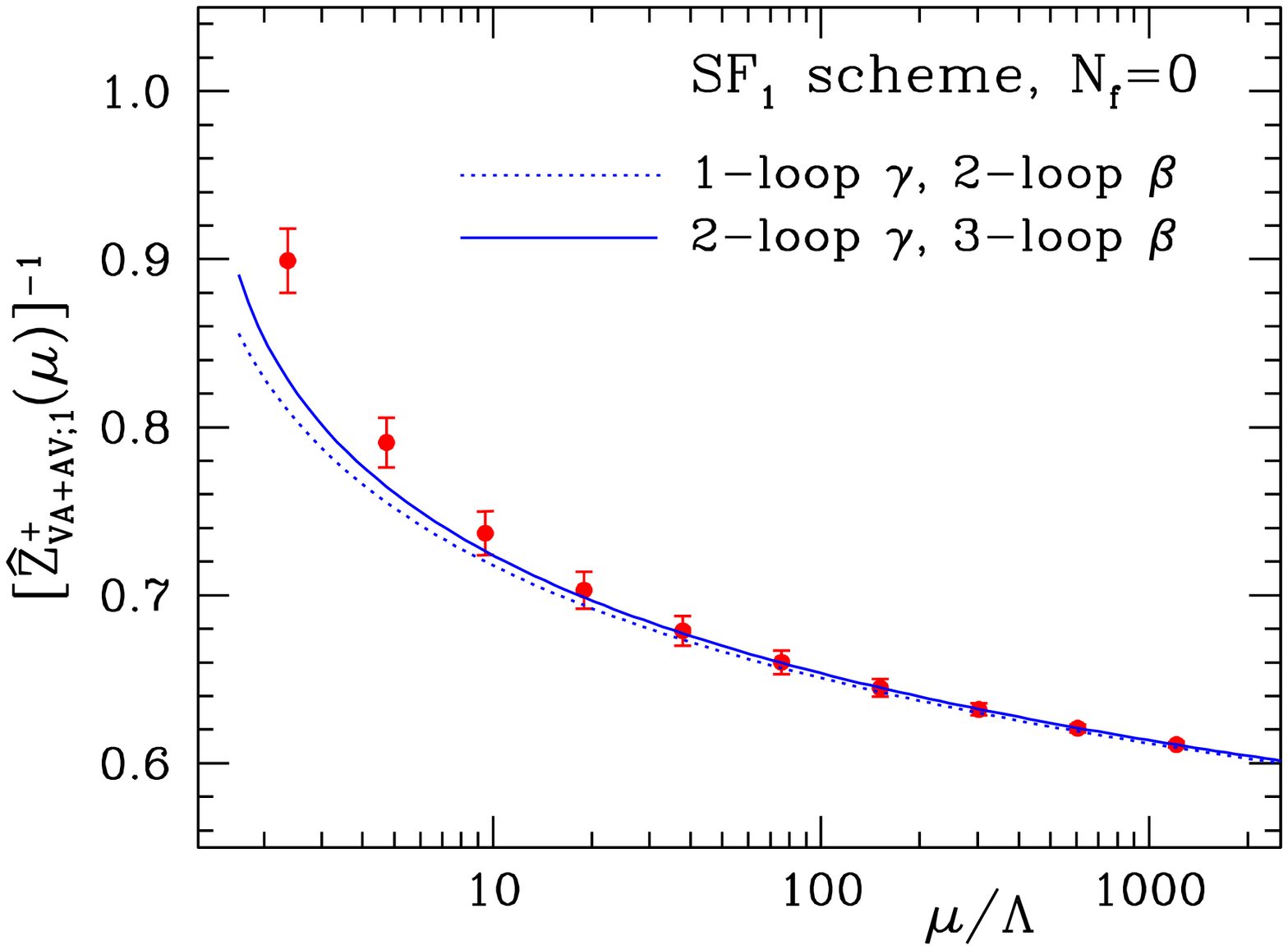}
\includegraphics{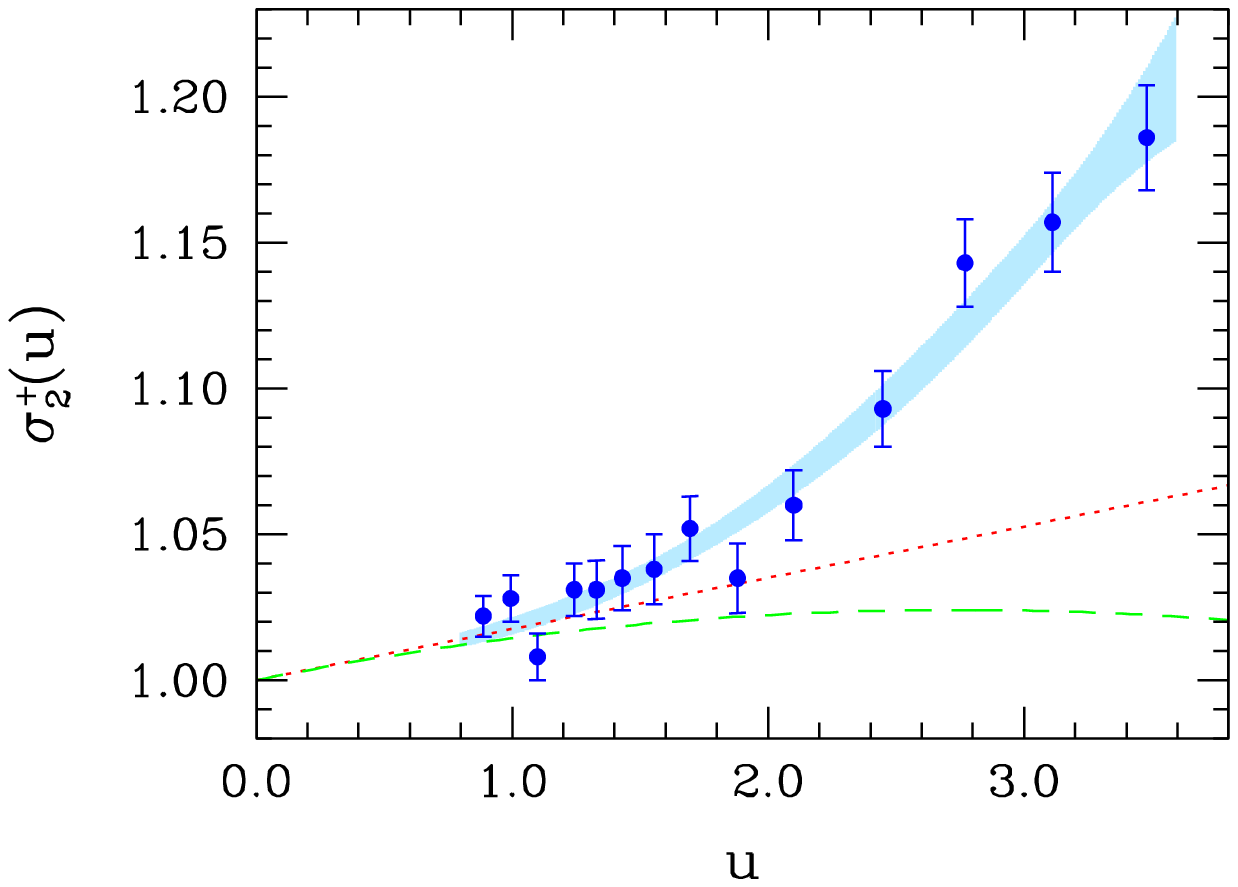}
\includegraphics{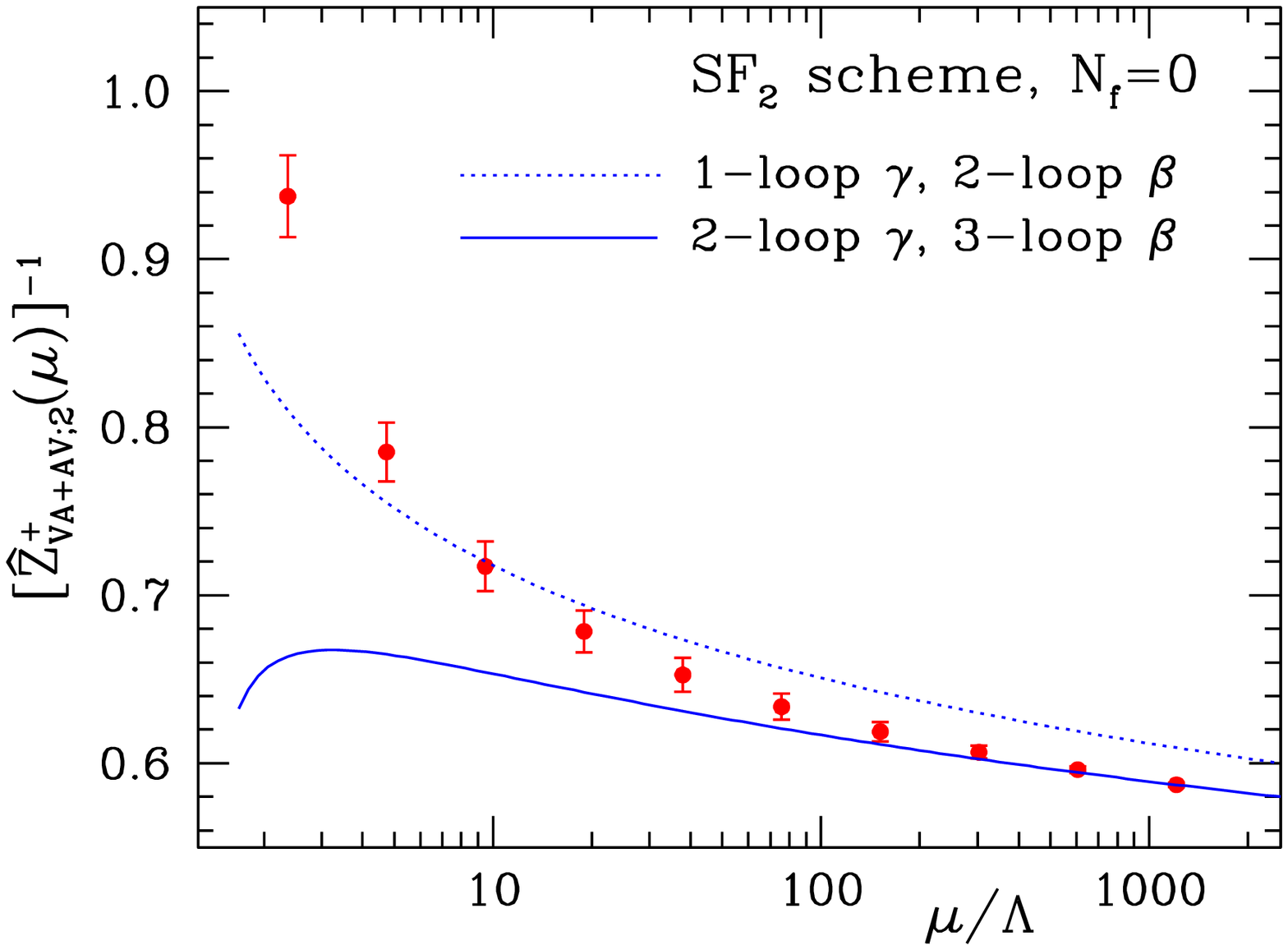}
\includegraphics{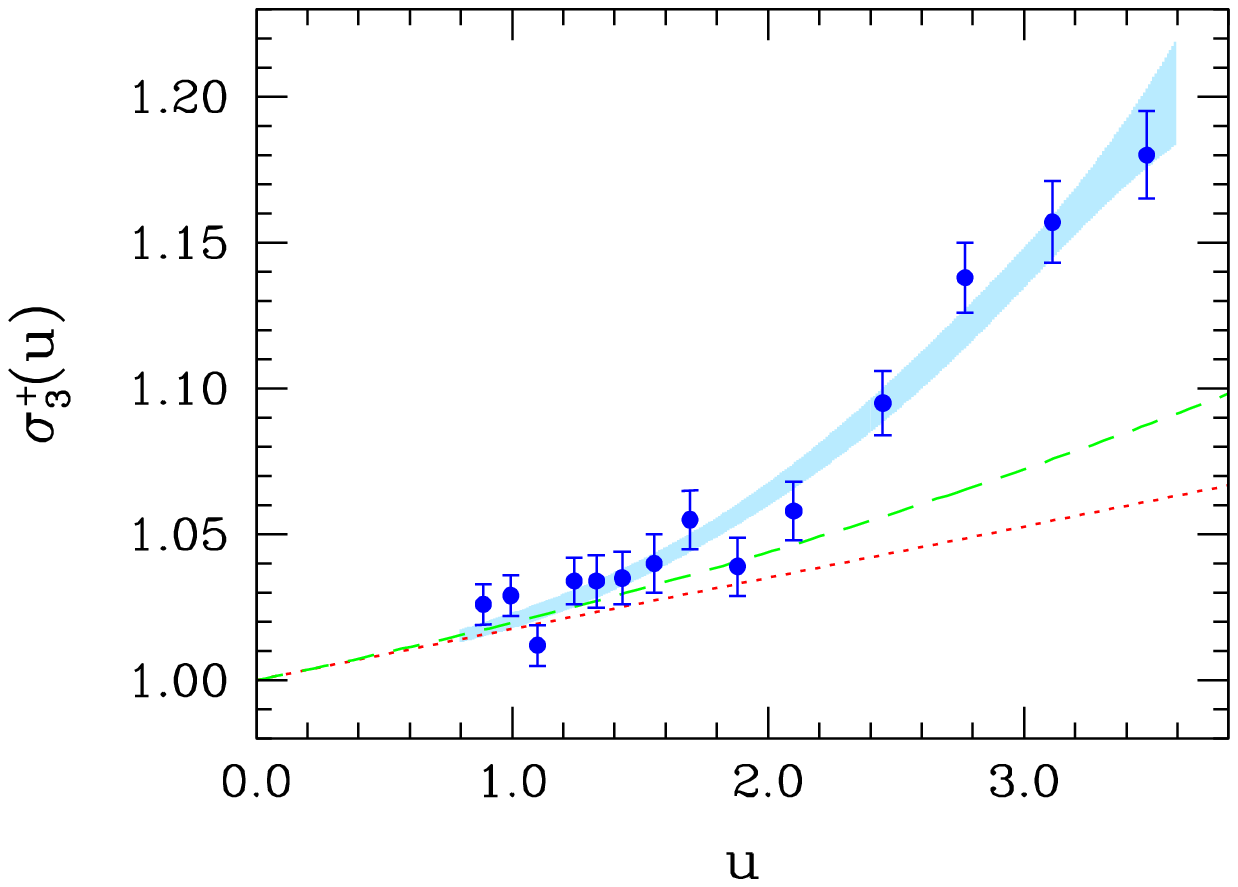}
\includegraphics{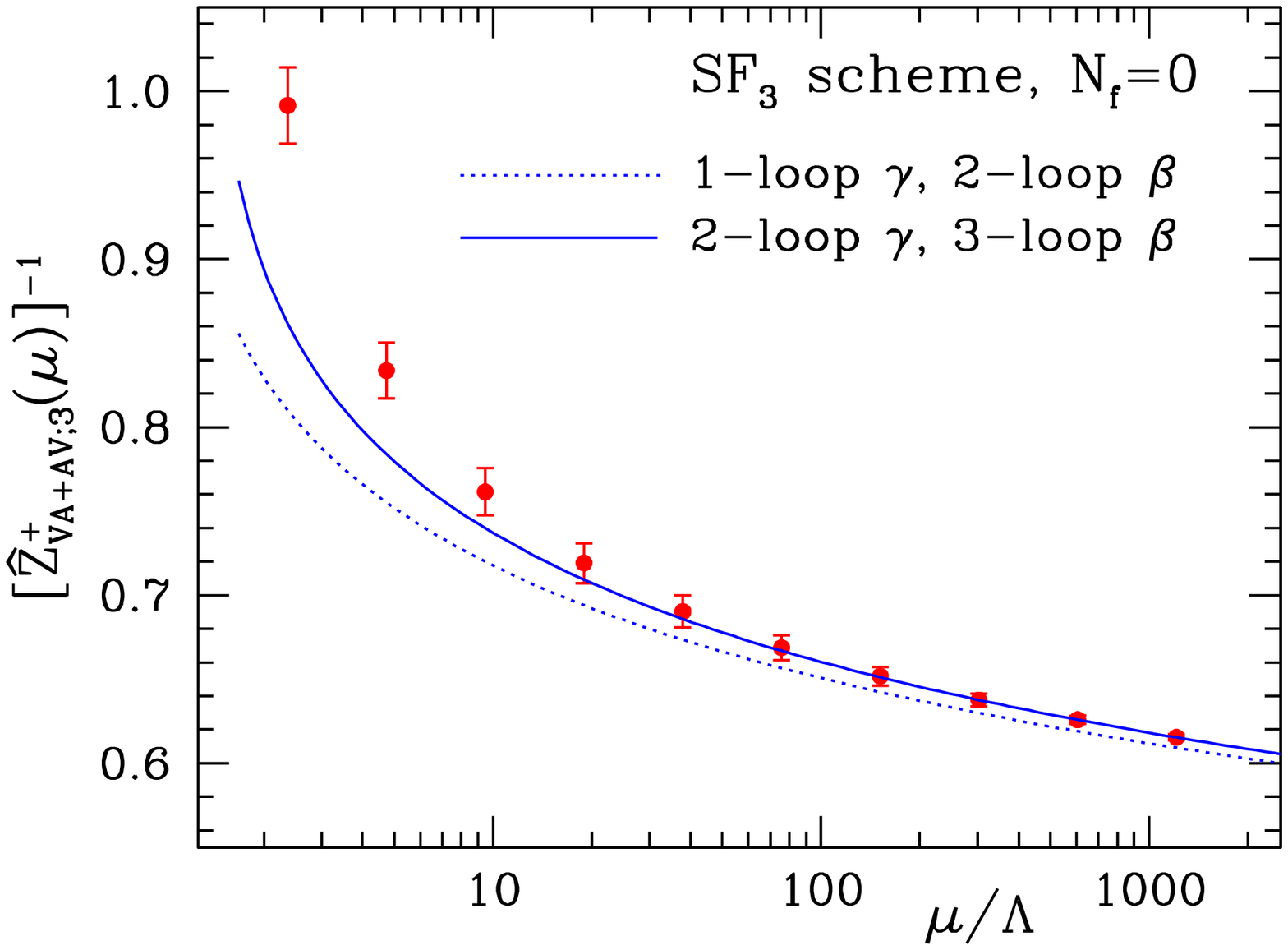}
\vspace{-18mm}
\caption{Left column: The step scaling function $\sigVApAV{;s}^+(u)$
(discrete points) as obtained non-perturbatively from combined fits
to Clover and Wilson data. The shaded area is the result of fit D
to the points (see text). The dotted (dashed) line is the LO (NLO)
perturbative result. Right column: RG running of $\oVApAV{O}^+$
obtained non-perturbatively (discrete points) at specific
values of the renormalization scale $\mu$, in units of $\Lambda$
(taken from ref.~\cite{SFmassRGI}).
The lines are perturbative results at the indicated order for the 
Callan-Symanzik $\beta$-function and the operator anomalous dimension
$\gamma$.
}
\label{fig:sigmaVApAVp}
\end{figure}\addtocounter{figure}{-1}

\clearpage

\begin{figure}[p]
\centering
\vspace{178mm}
\includegraphics{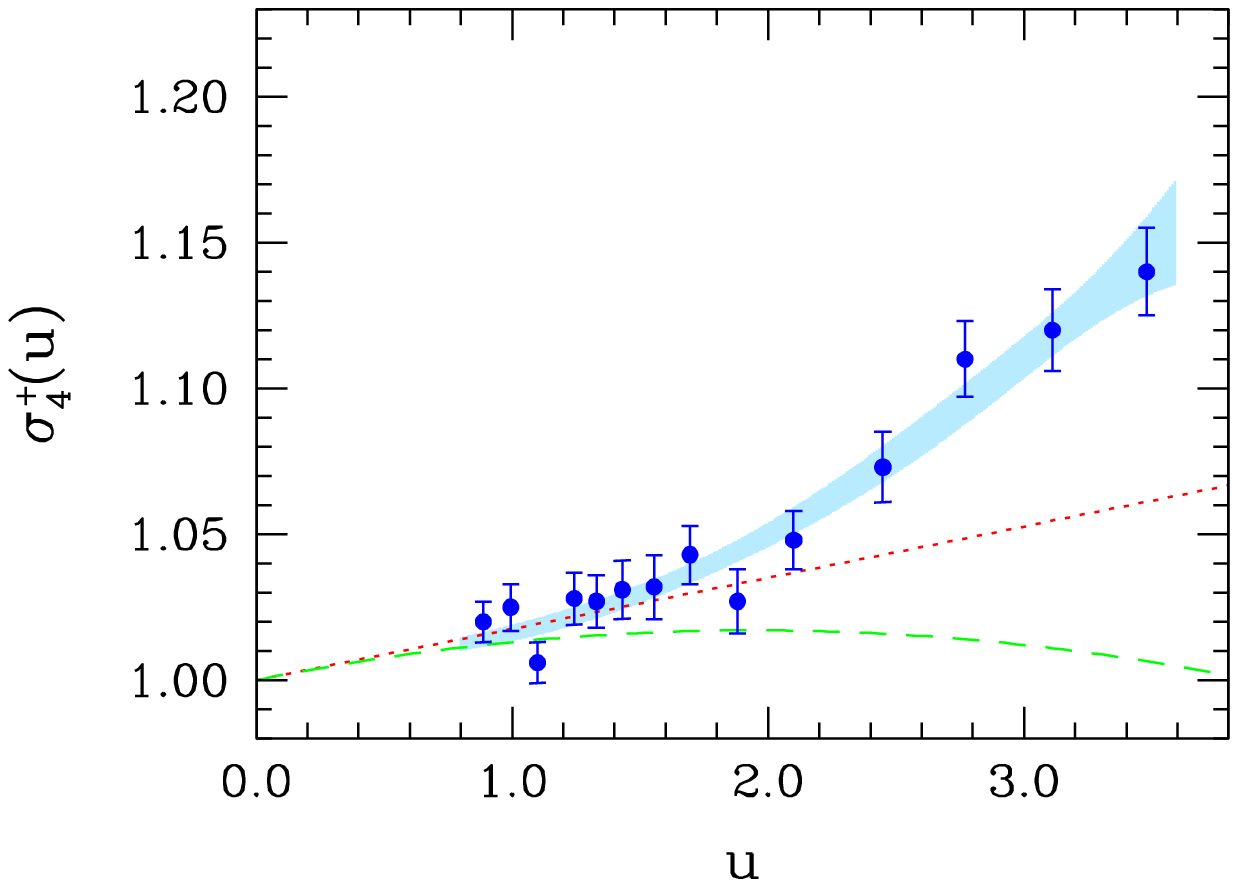}
\includegraphics{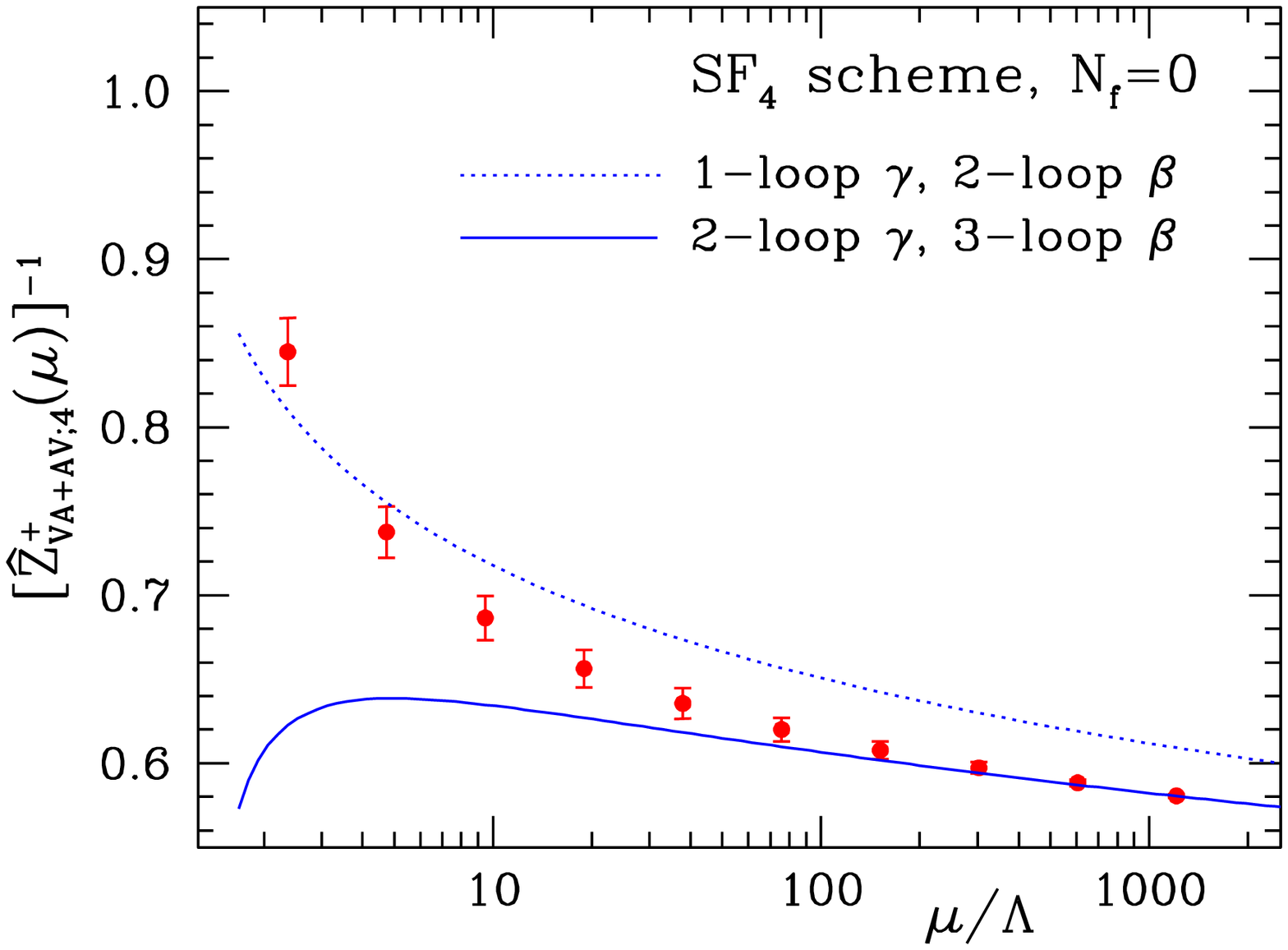}
\includegraphics{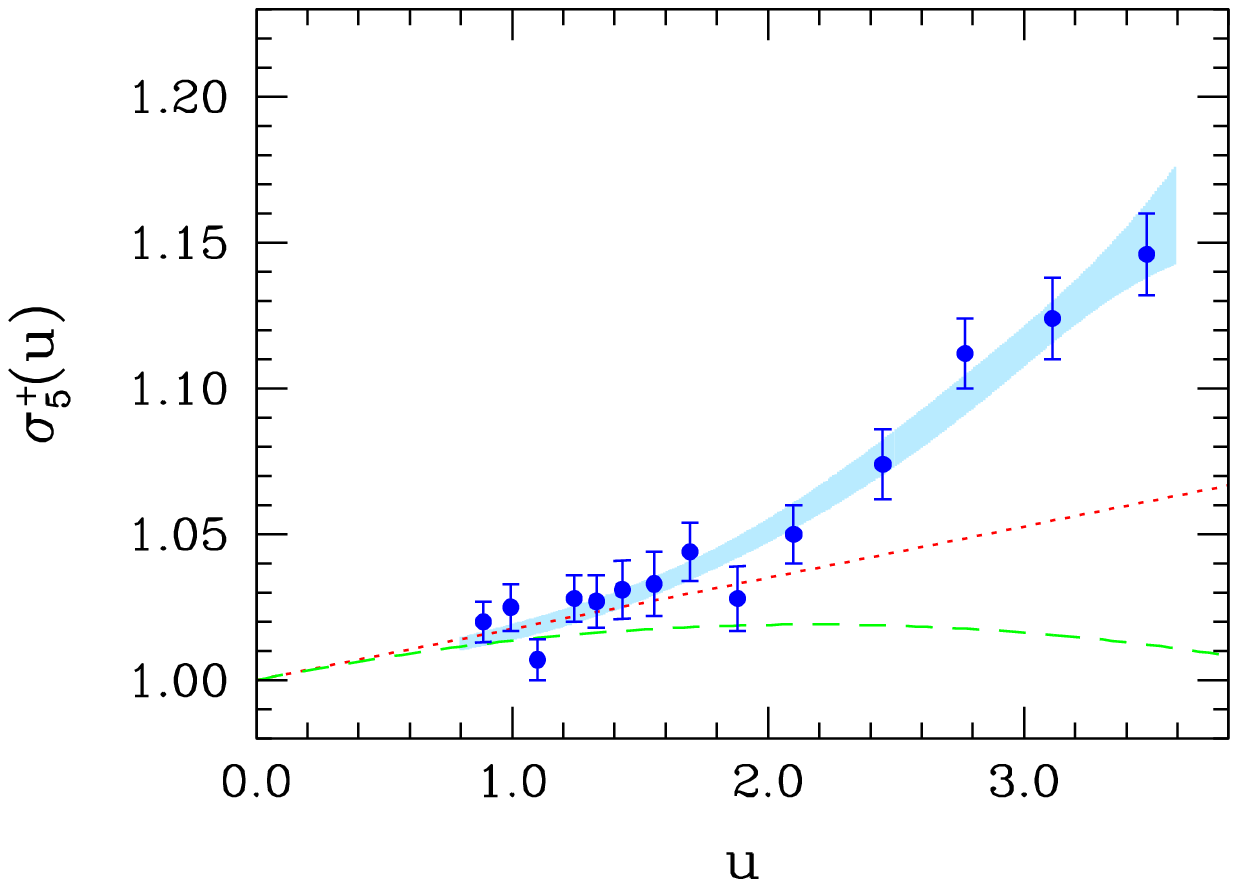}
\includegraphics{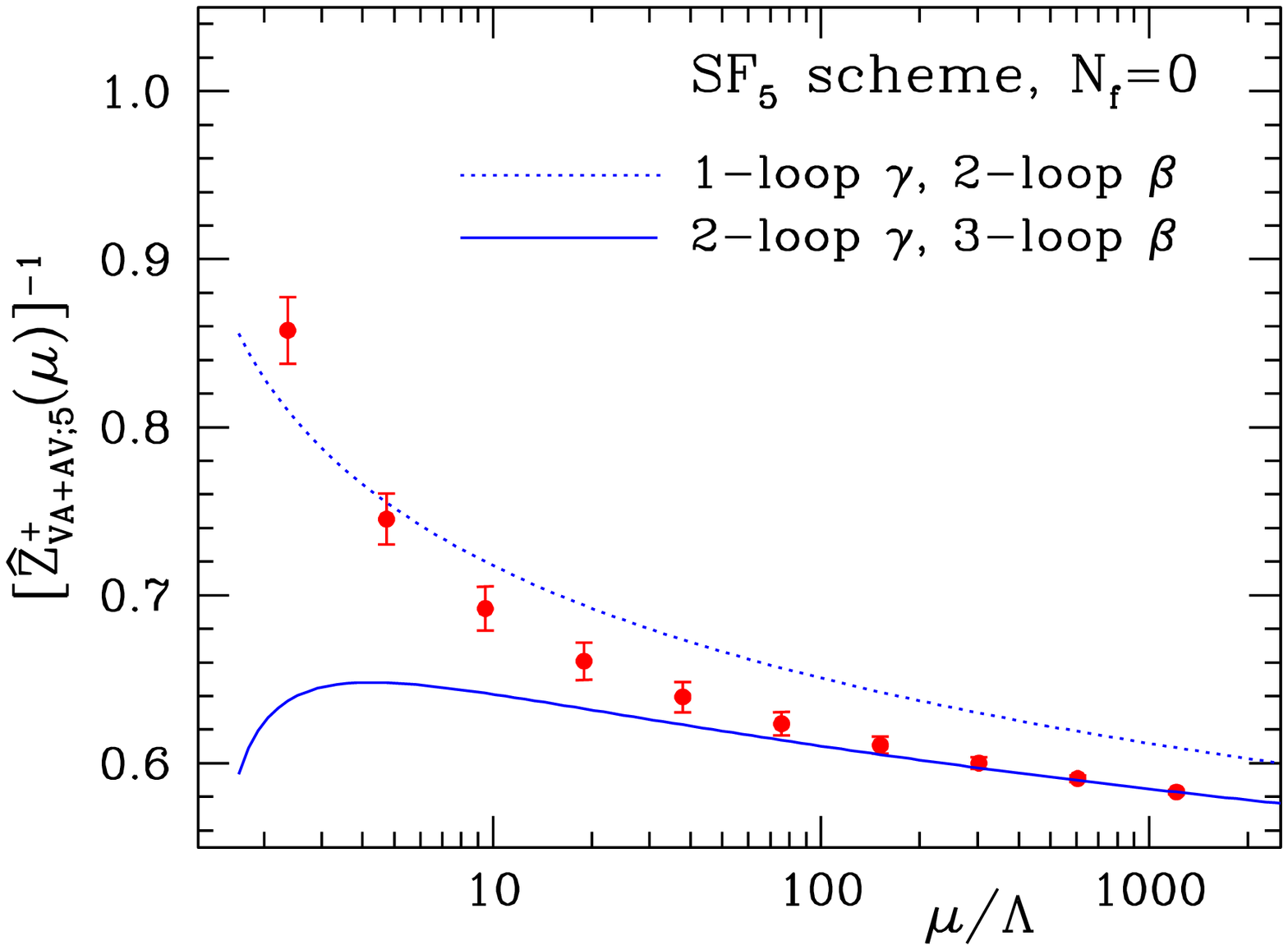}
\includegraphics{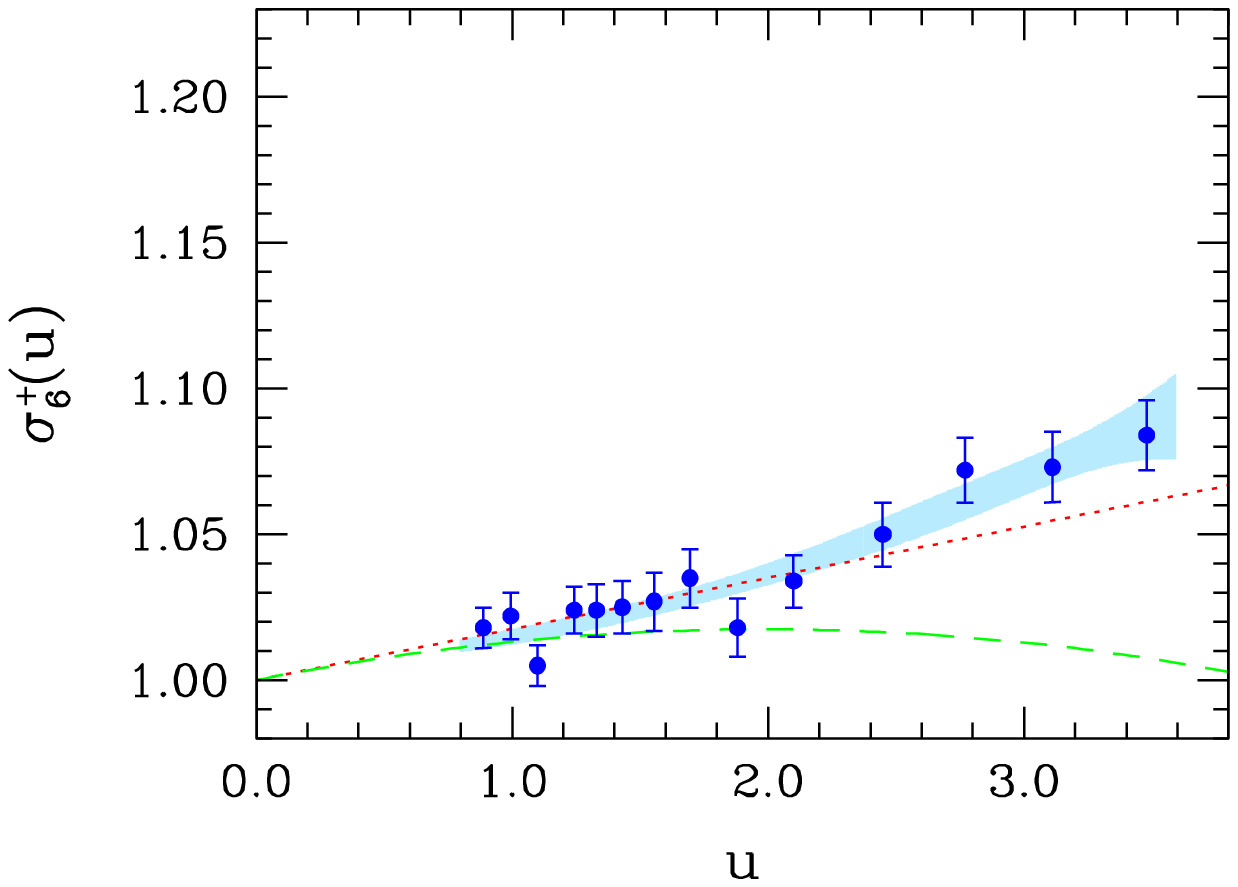}
\includegraphics{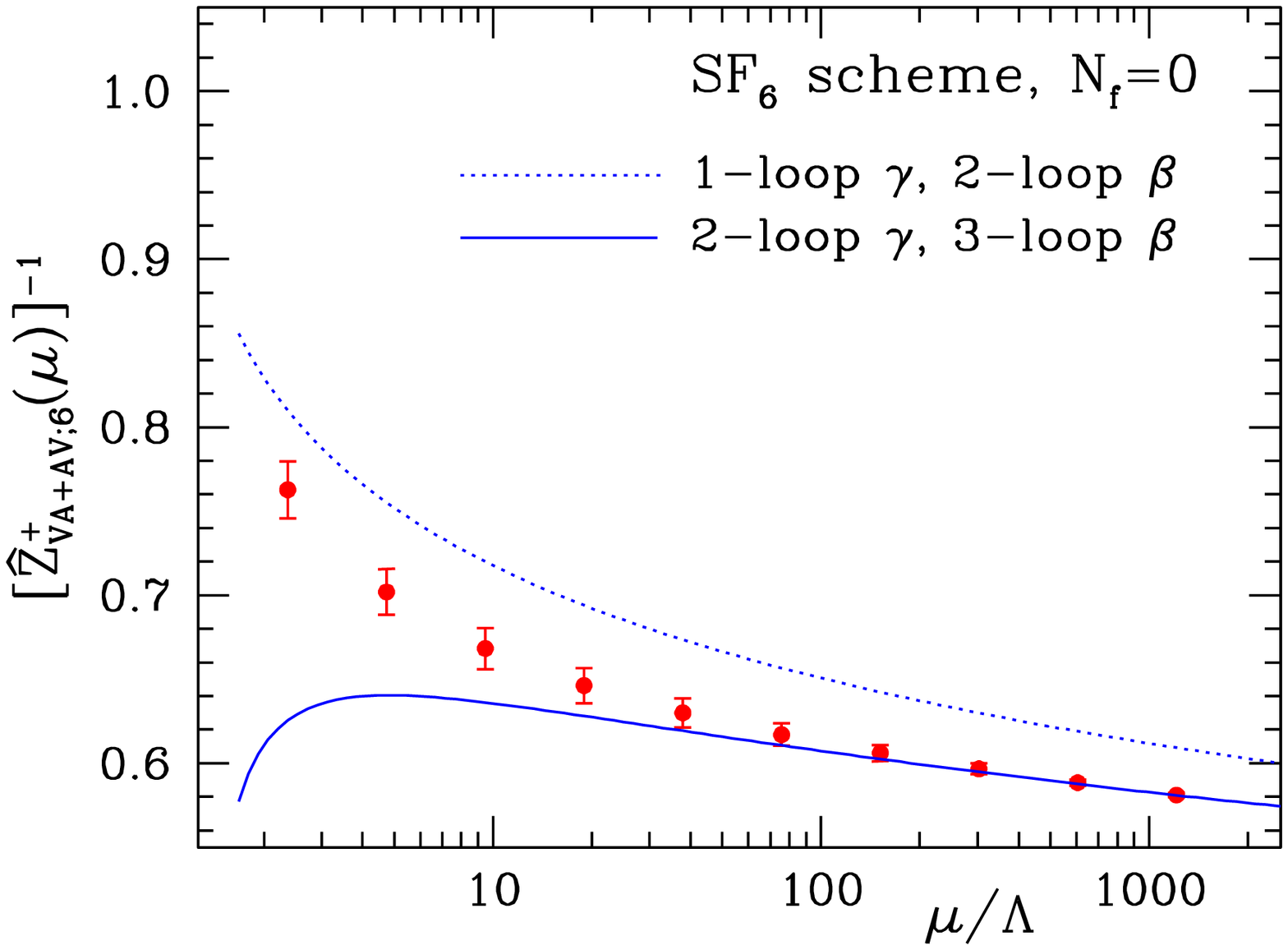}
\vspace{-18mm}
\caption{ (continued)
}
\end{figure}\addtocounter{figure}{-1}

\clearpage

\begin{figure}[p]
\centering
\vspace{178mm}
\includegraphics{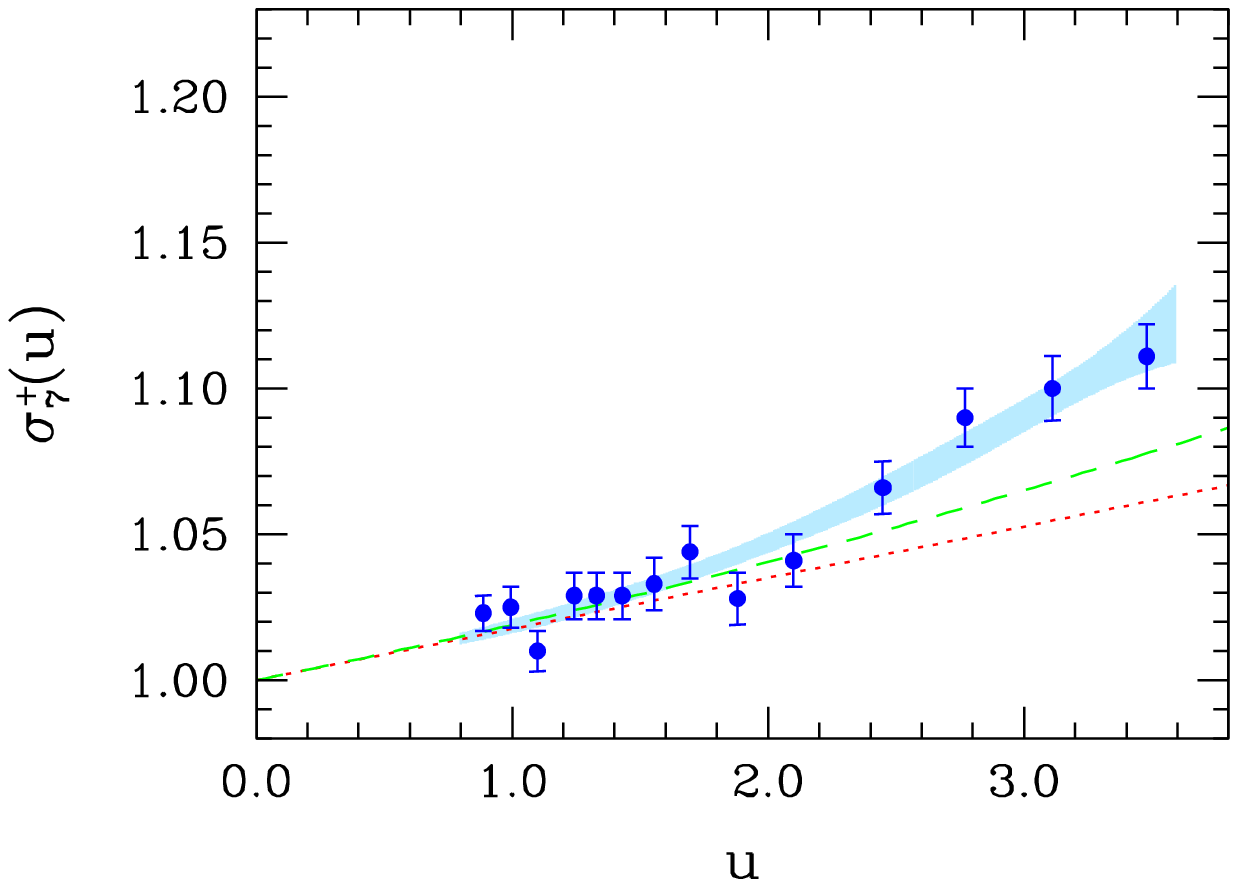}
\includegraphics{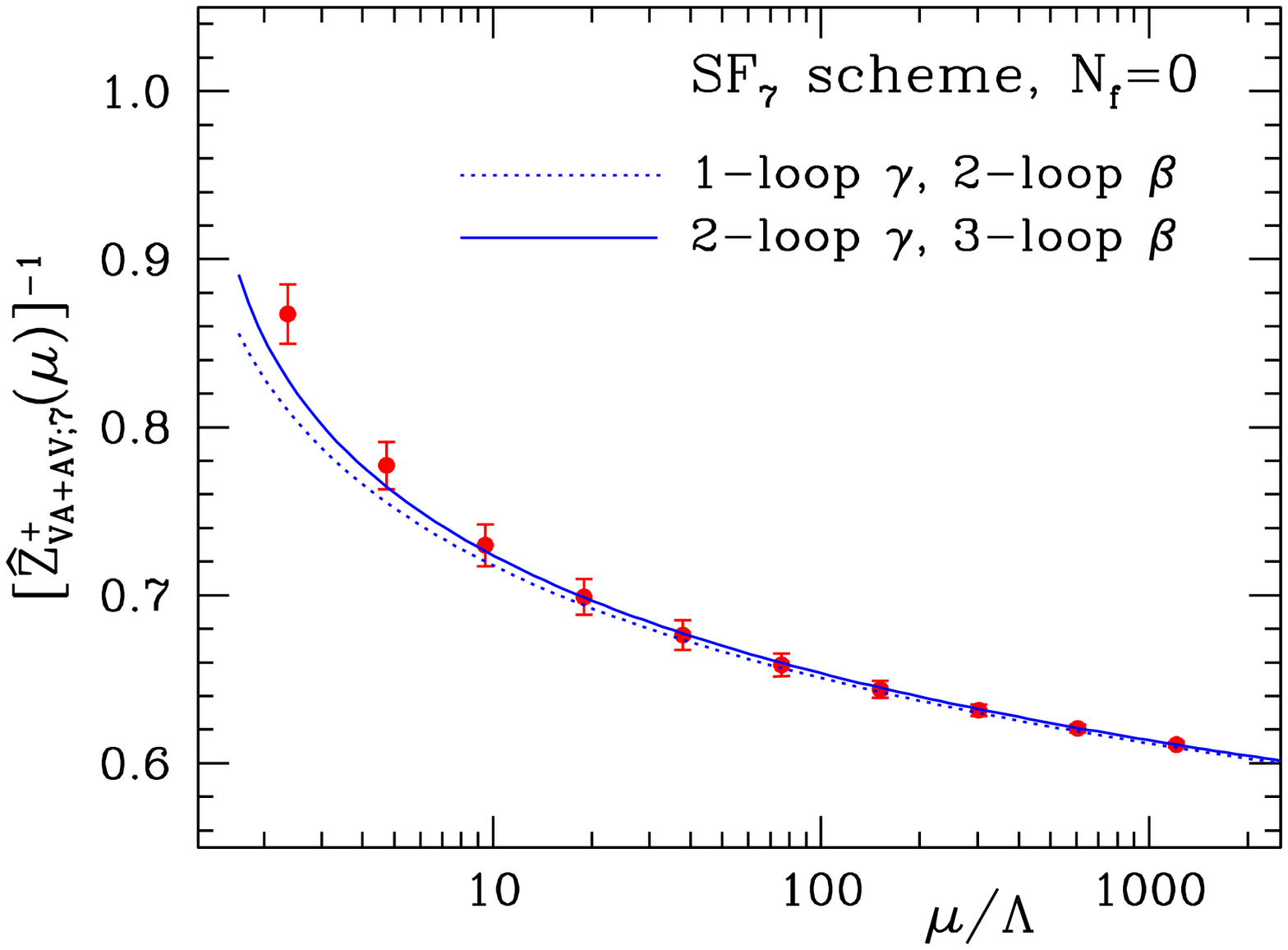}
\includegraphics{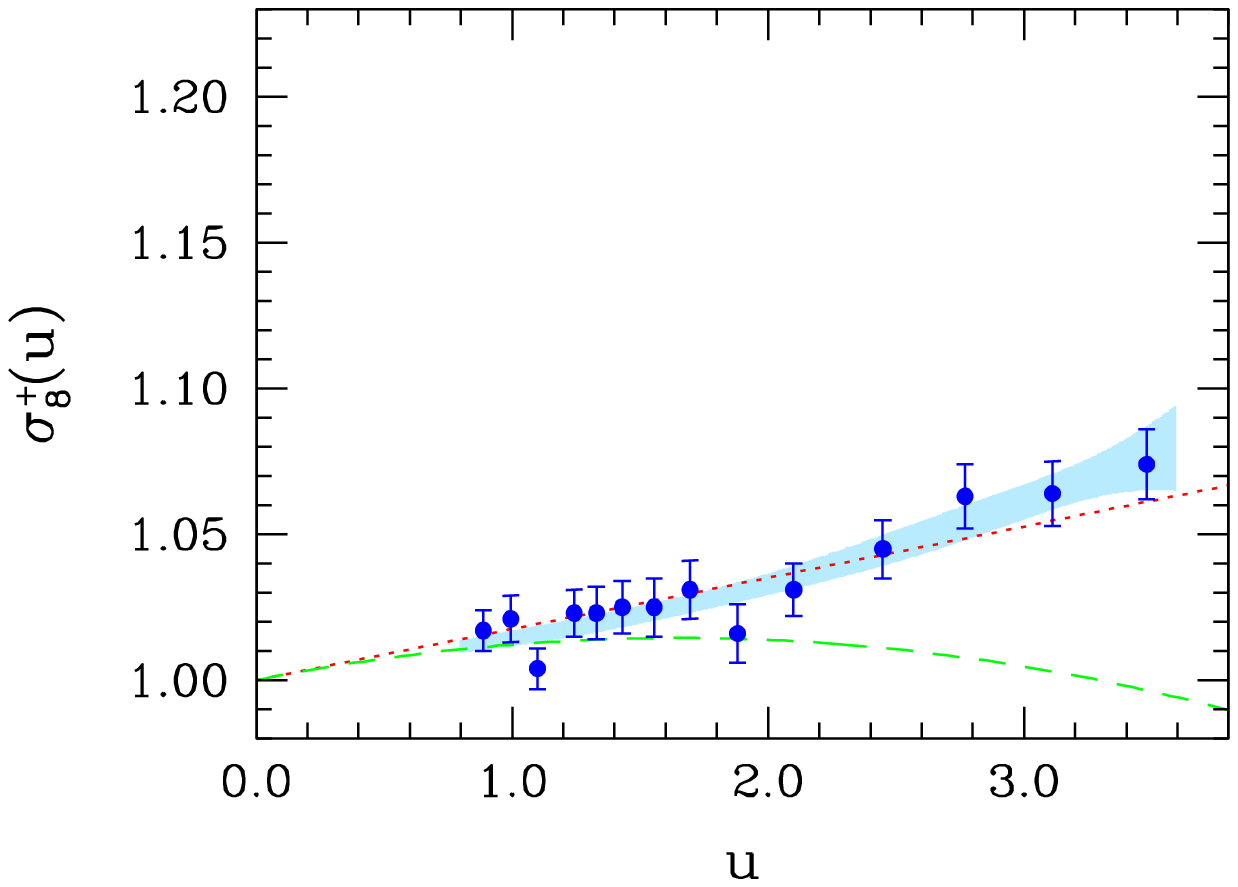}
\includegraphics{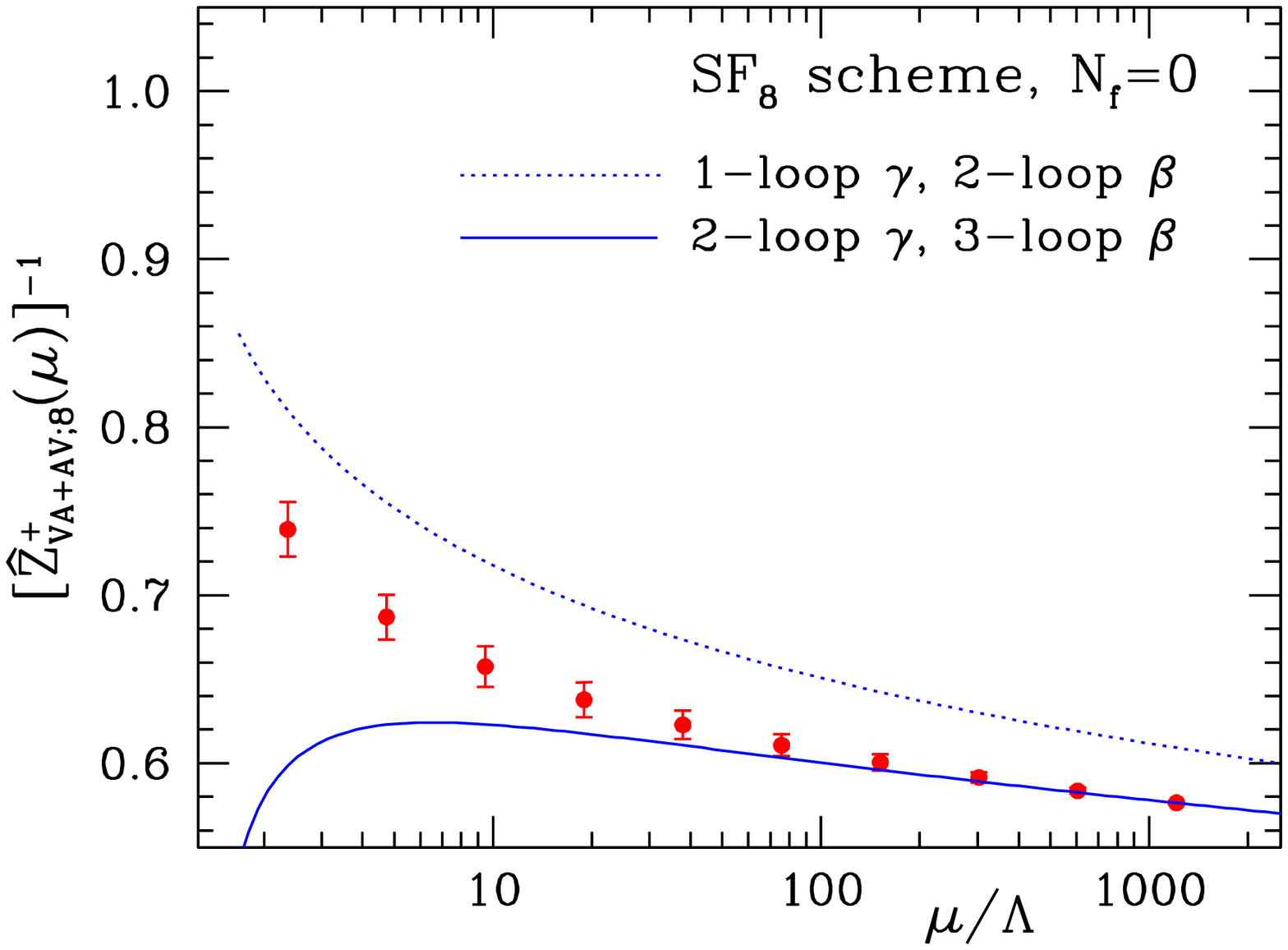}
\includegraphics{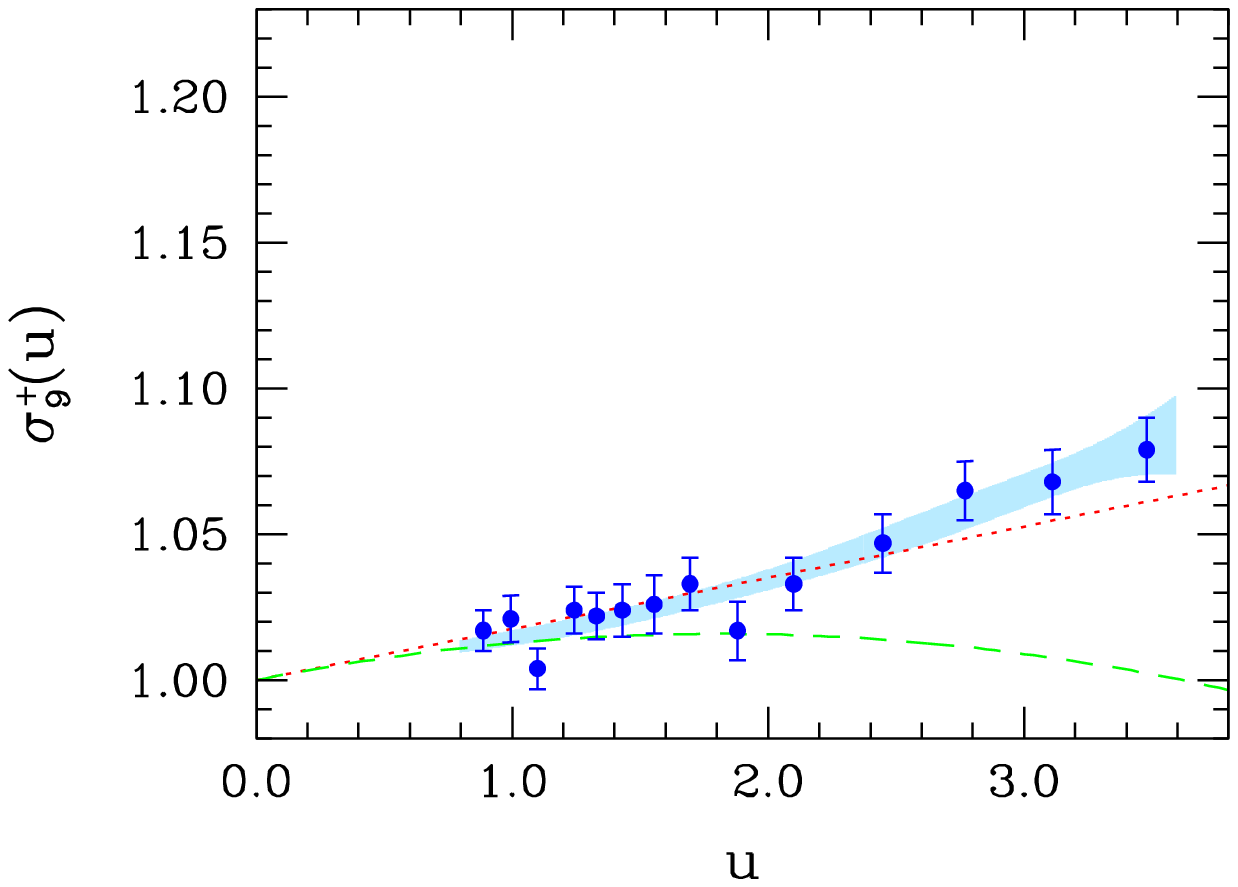}
\includegraphics{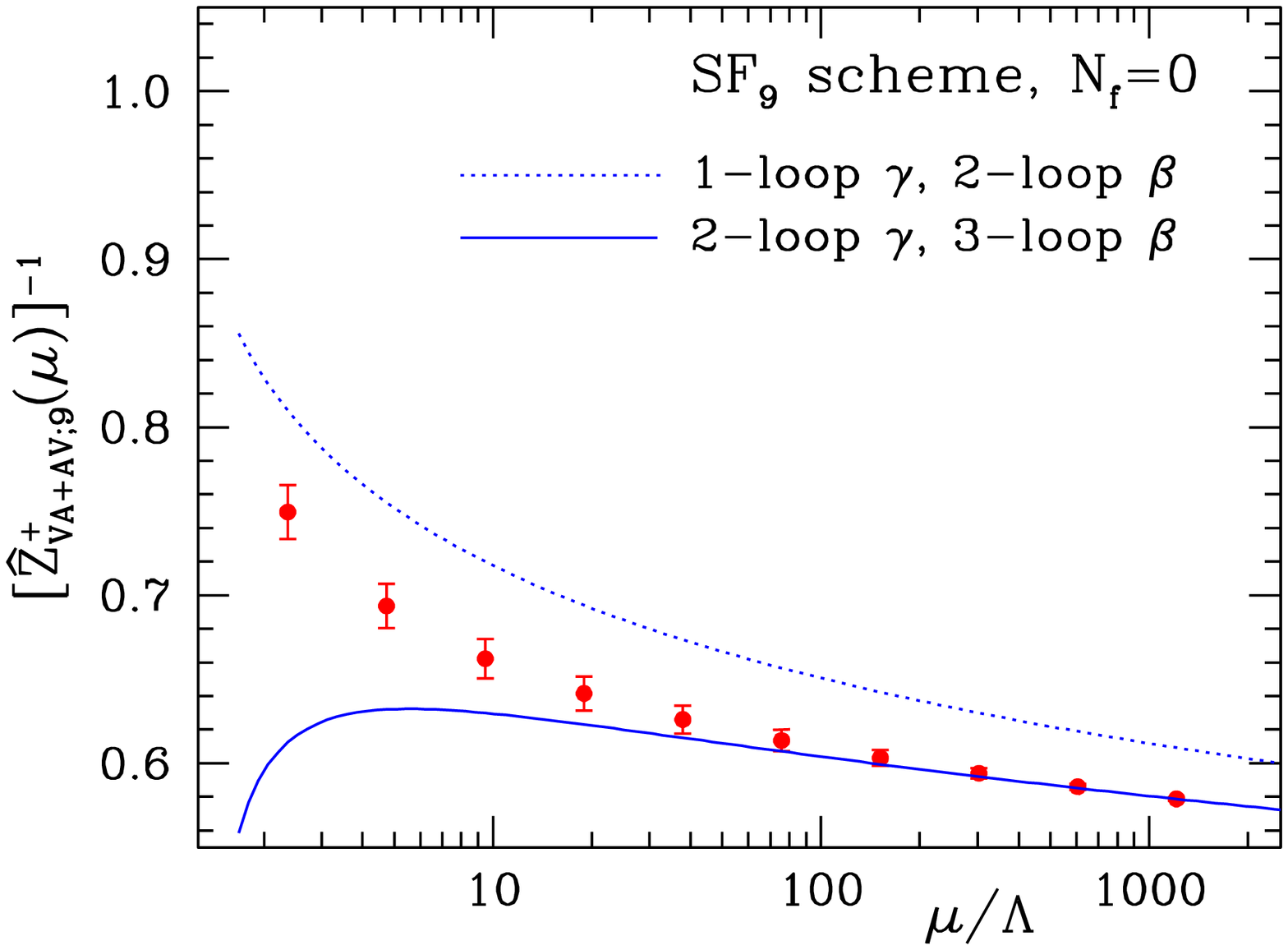}
\vspace{-18mm}
\caption{ (continued)
}
\end{figure}

\clearpage

\begin{figure}[p]
\centering
\vspace{178mm}
\includegraphics{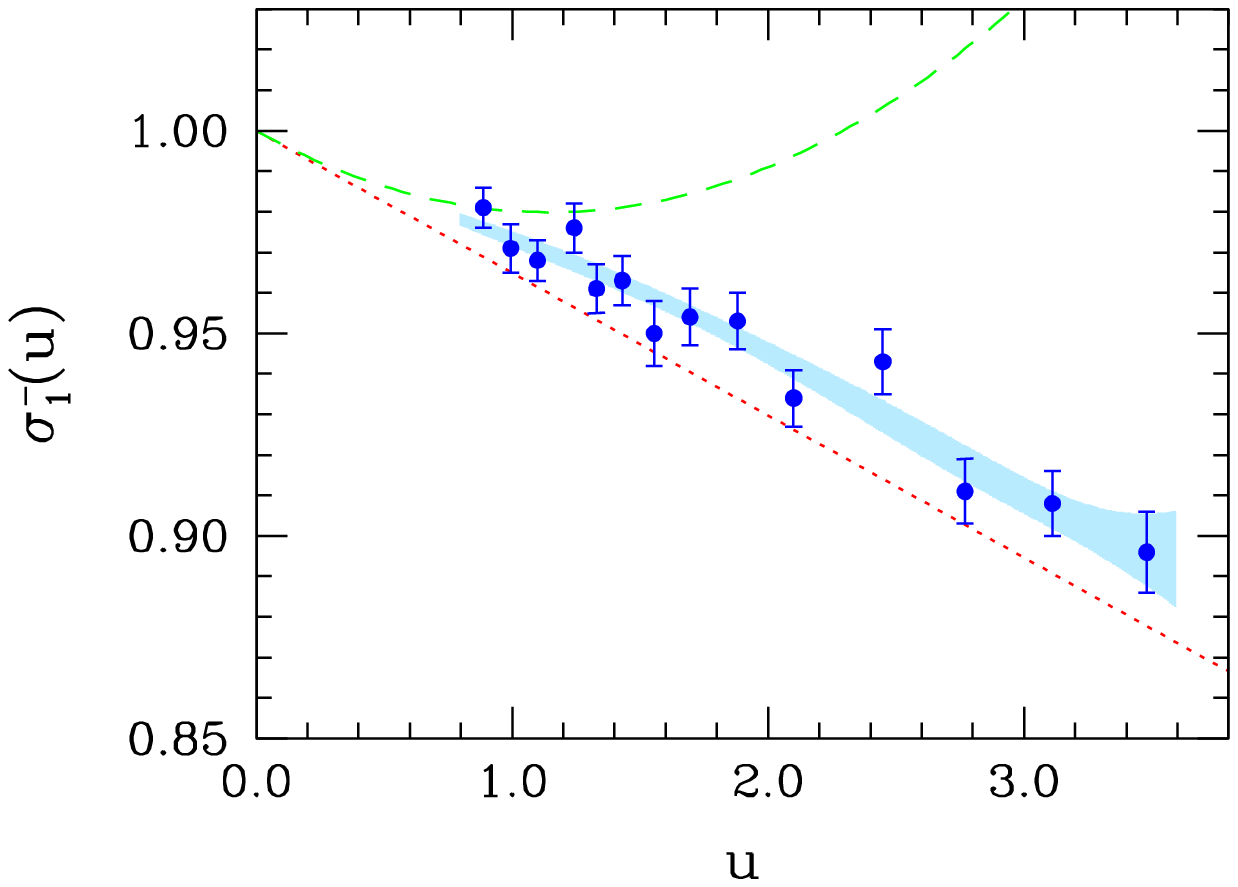}
\includegraphics{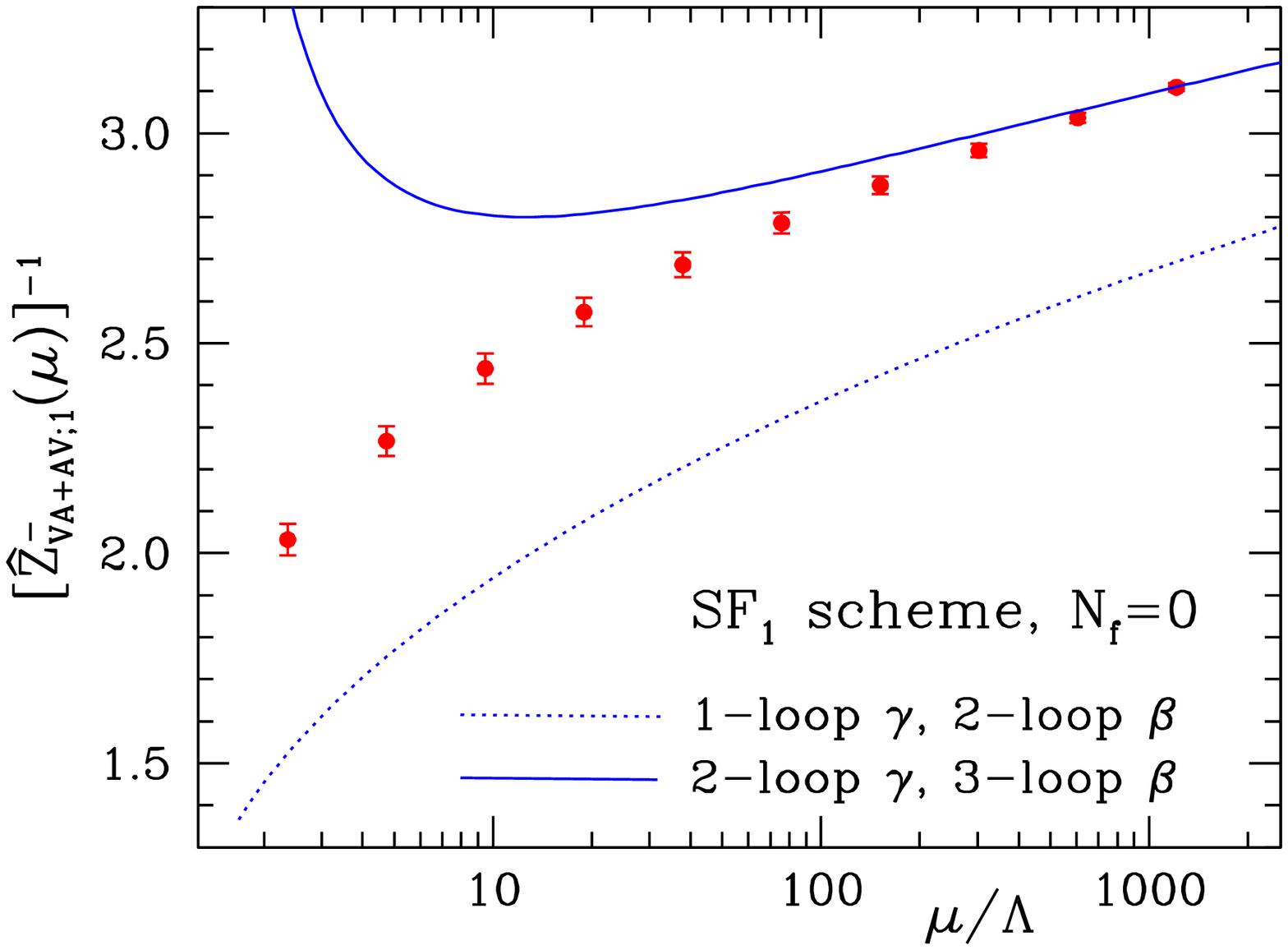}
\includegraphics{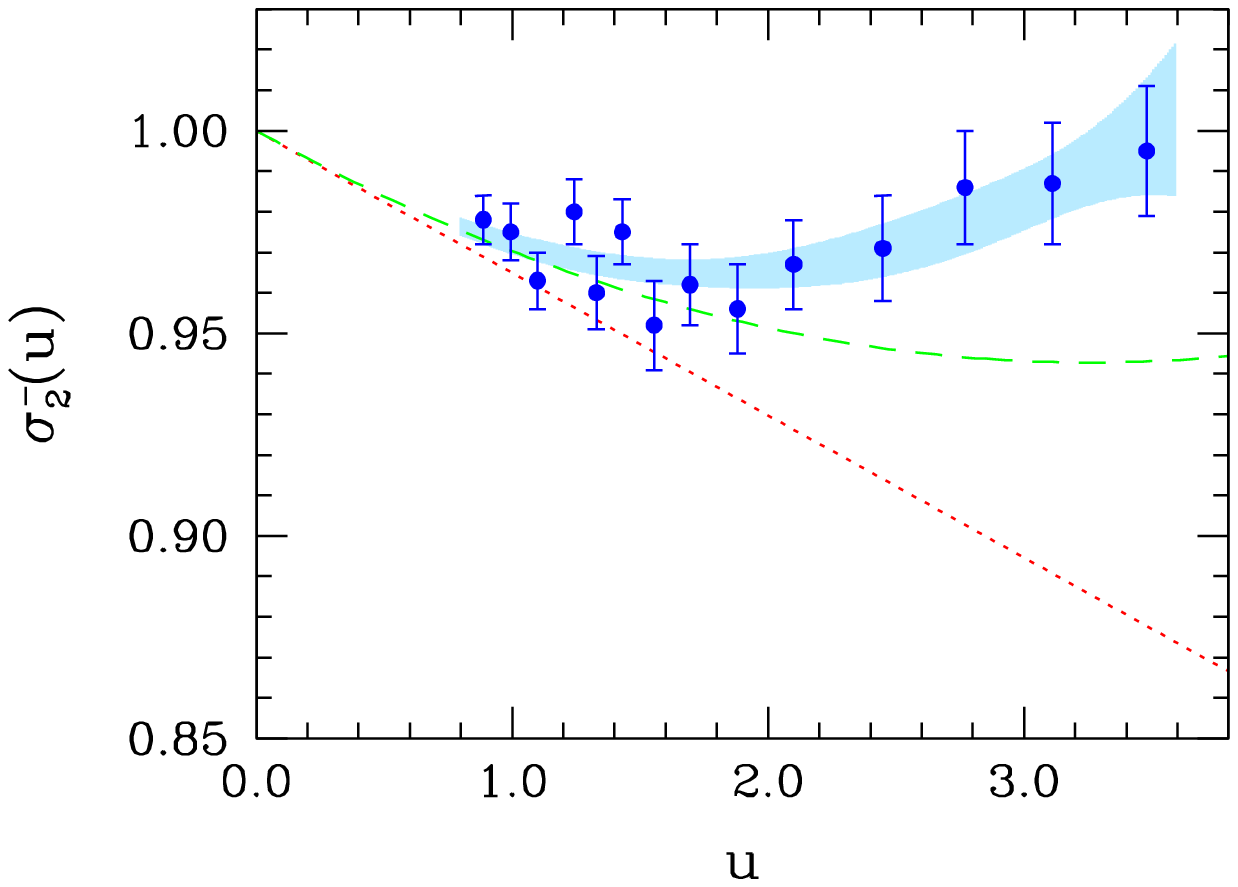}
\includegraphics{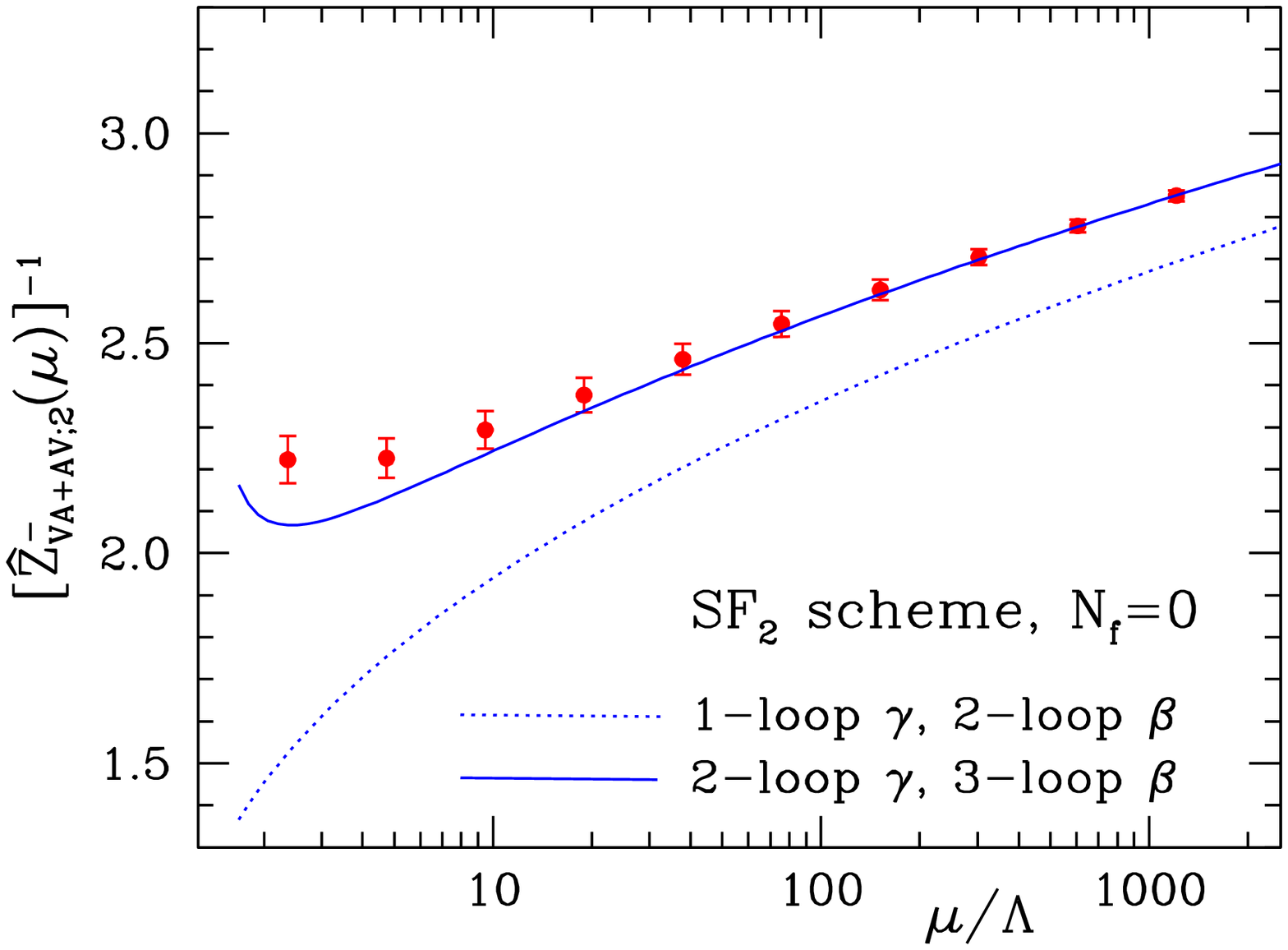}
\includegraphics{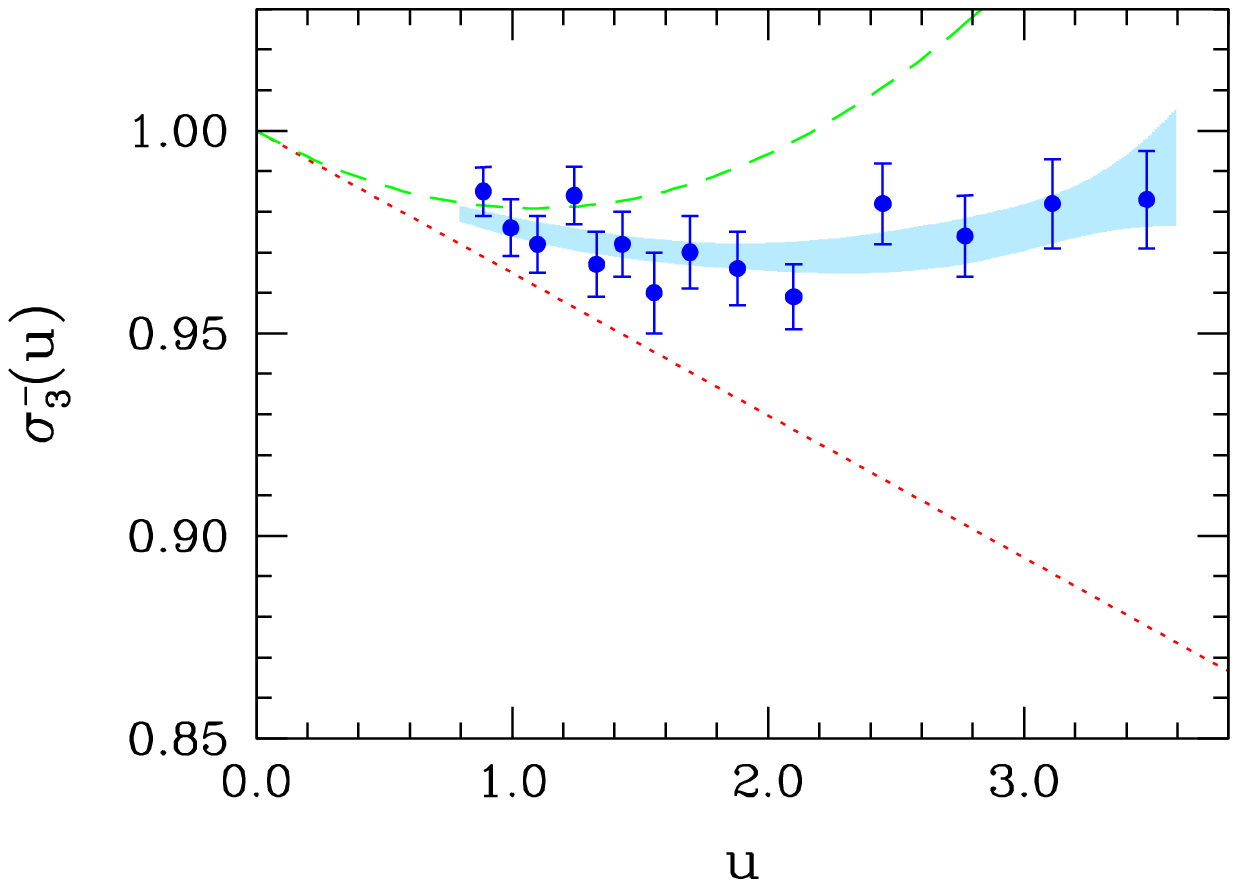}
\includegraphics{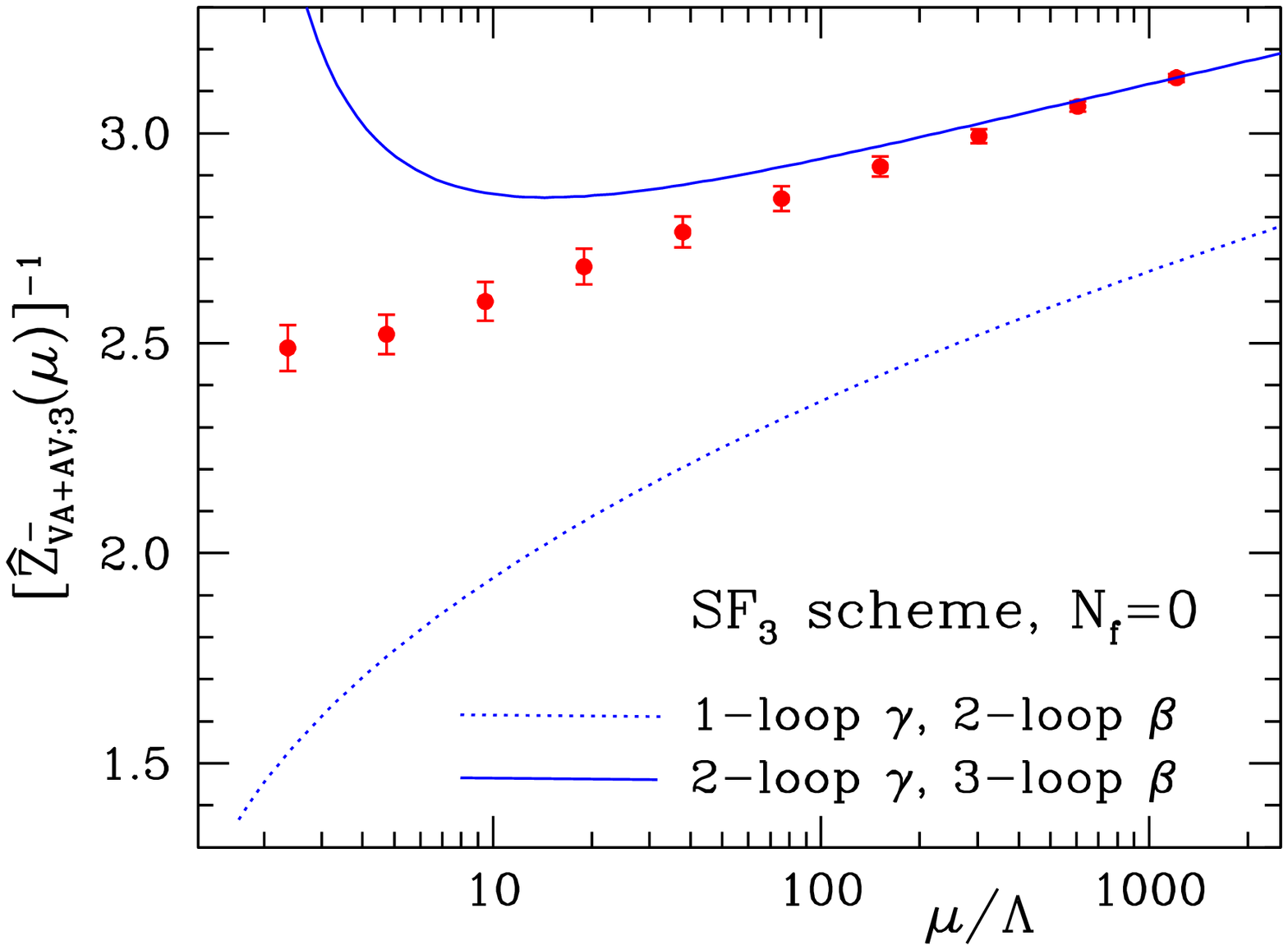}
\vspace{-18mm}
\caption{Left column: The step scaling function $\sigVApAV{;s}^-(u)$
(discrete points) as obtained non-perturbatively from combined fits
to Clover and Wilson data. The shaded area is the result of fit D
to the points (see text). The dotted (dashed) line is the LO (NLO)
perturbative result. Right column: RG running of $\oVApAV{O}^-$
obtained non-perturbatively (discrete points) at specific
values of the renormalization scale $\mu$, in units of $\Lambda$
(taken from ref.~\cite{SFmassRGI}).
The lines are perturbative results at the indicated order for the 
Callan-Symanzik $\beta$-function and the operator anomalous dimension
$\gamma$.
}
\label{fig:sigmaVApAVm}
\end{figure}\addtocounter{figure}{-1}

\clearpage

\begin{figure}[p]
\centering
\vspace{178mm}
\includegraphics{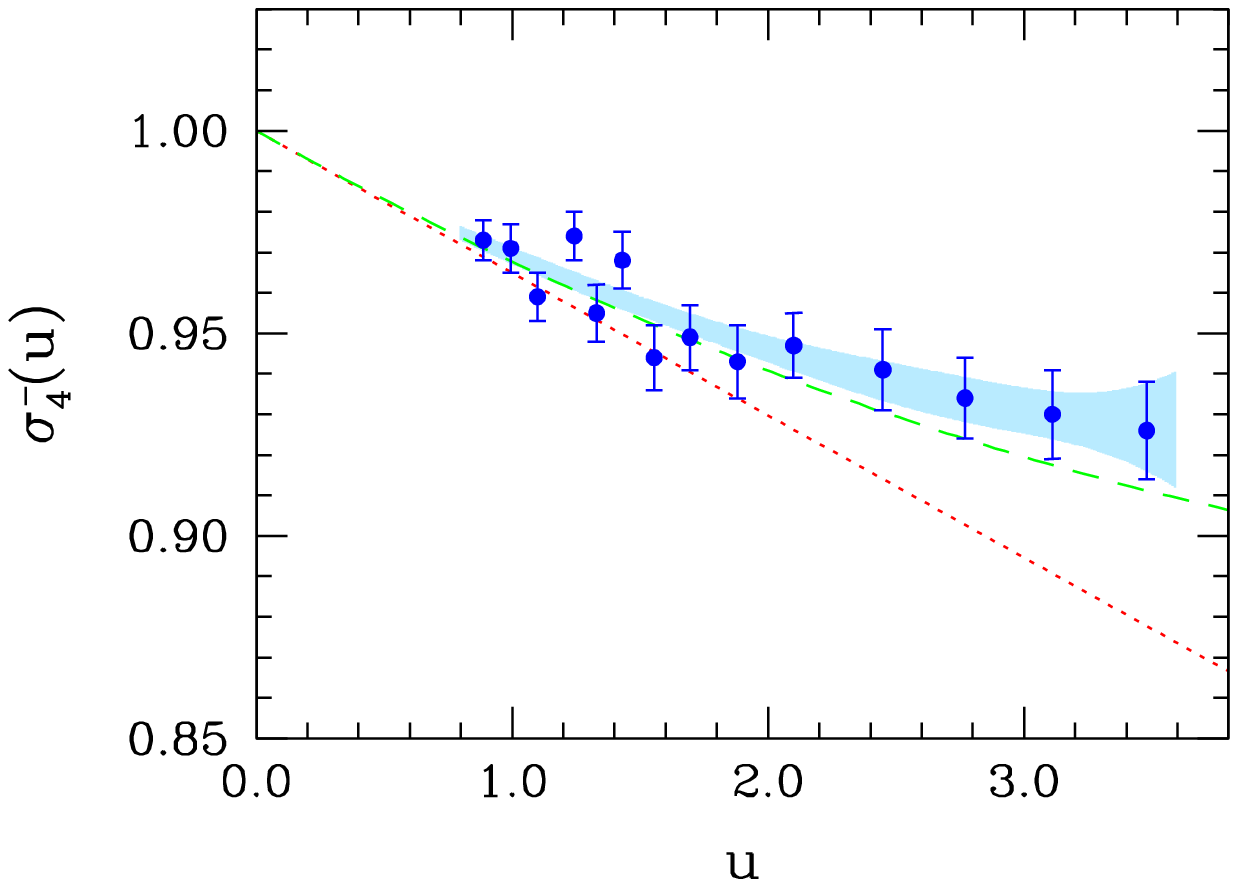}
\includegraphics{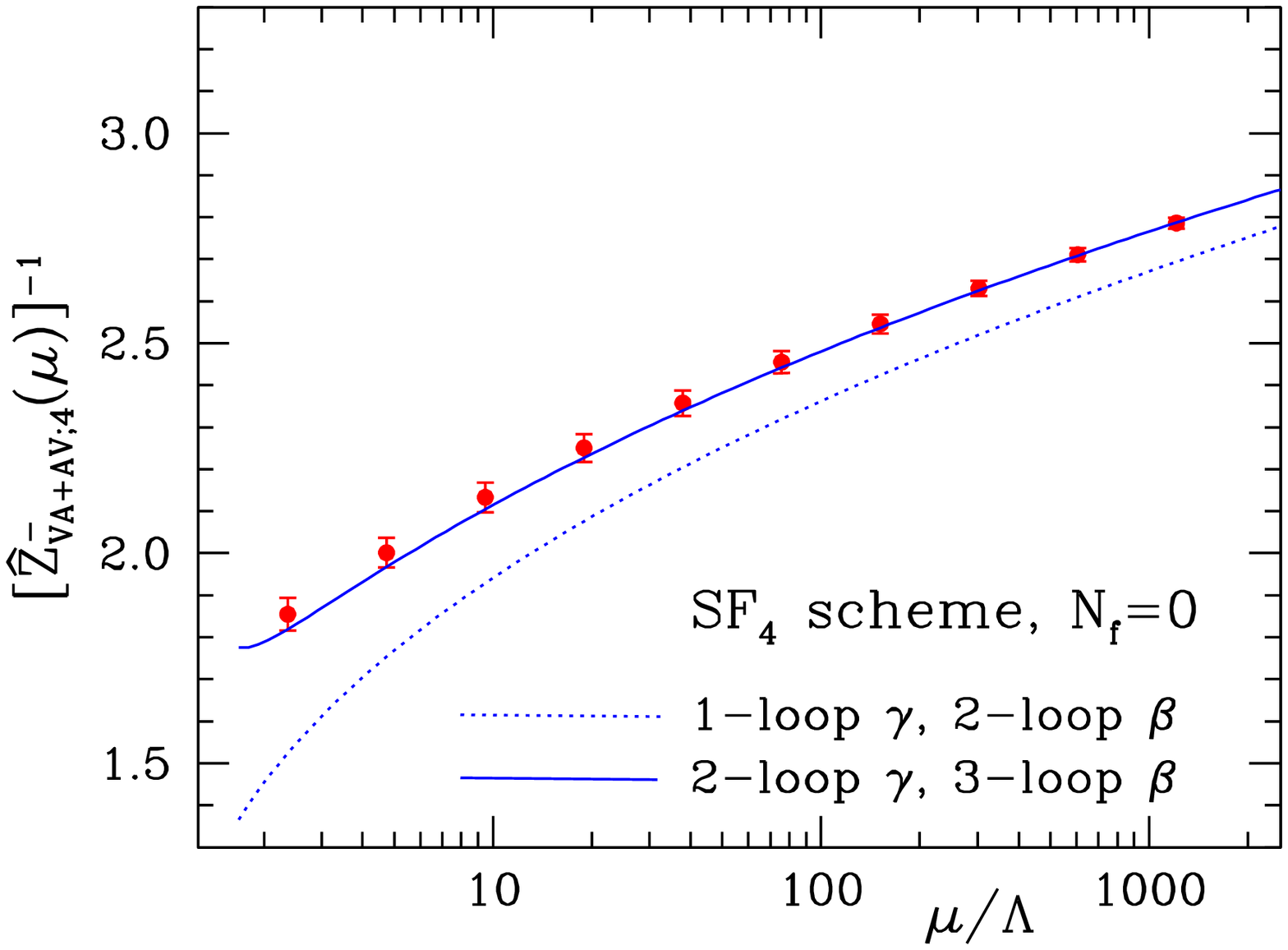}
\includegraphics{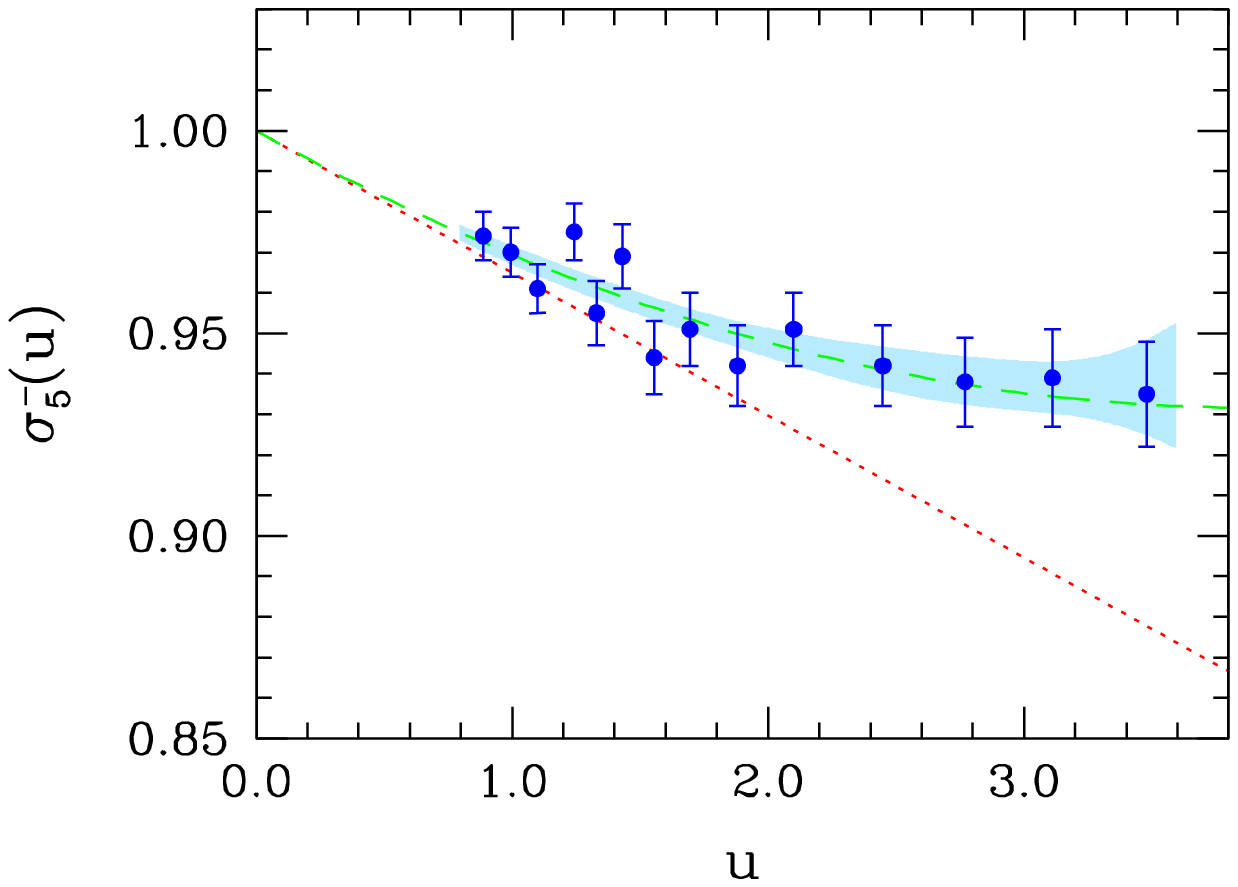}
\includegraphics{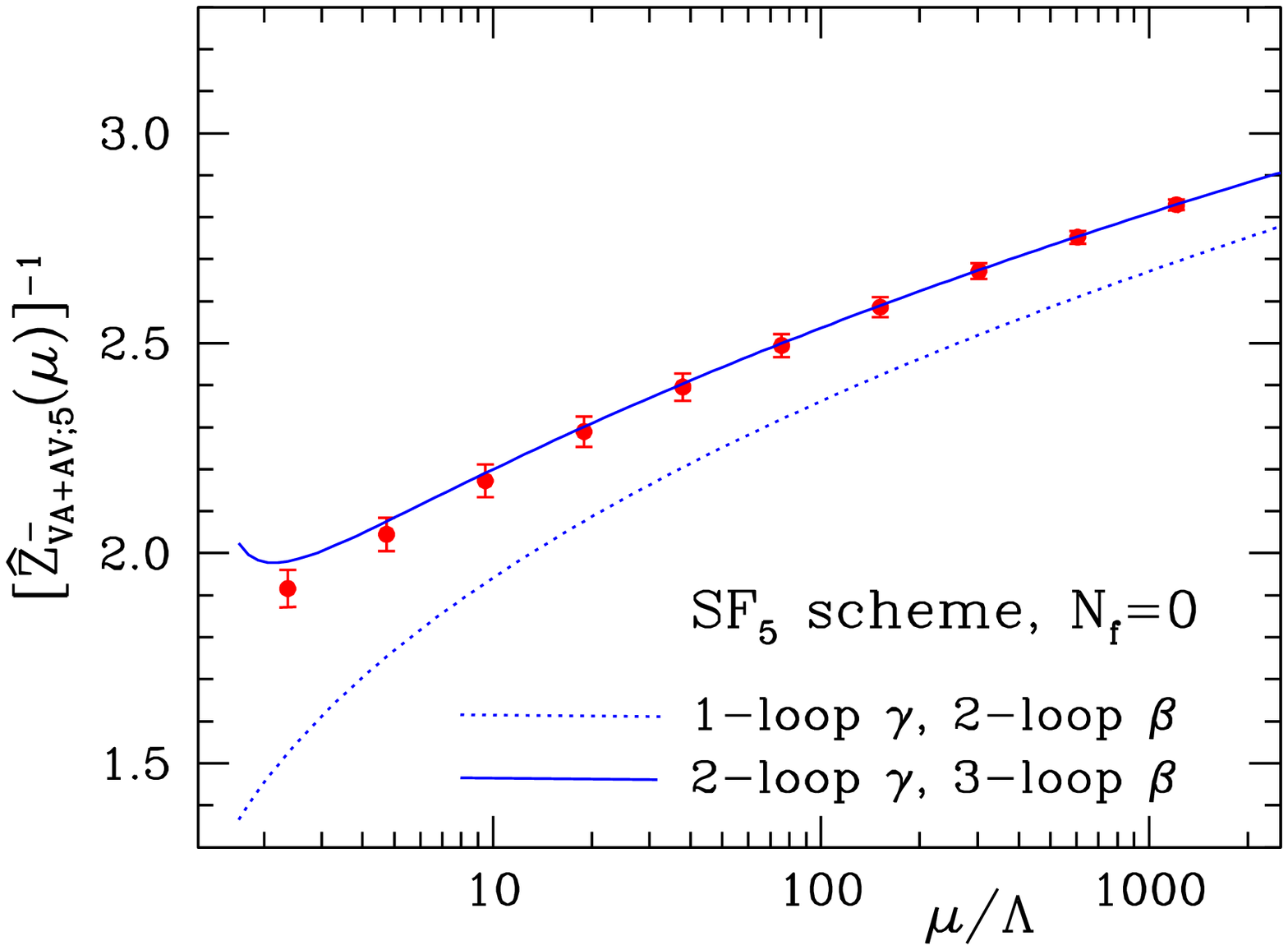}
\includegraphics{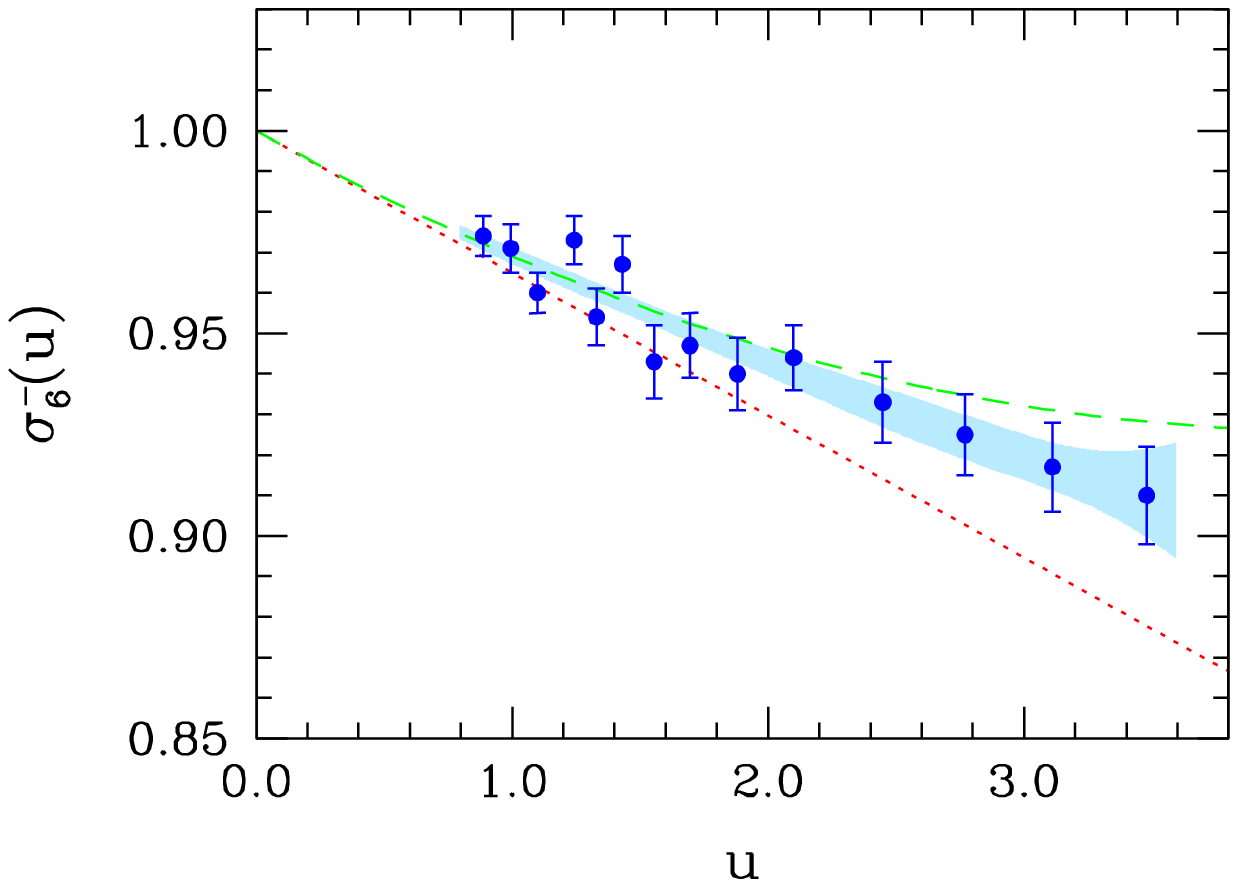}
\includegraphics{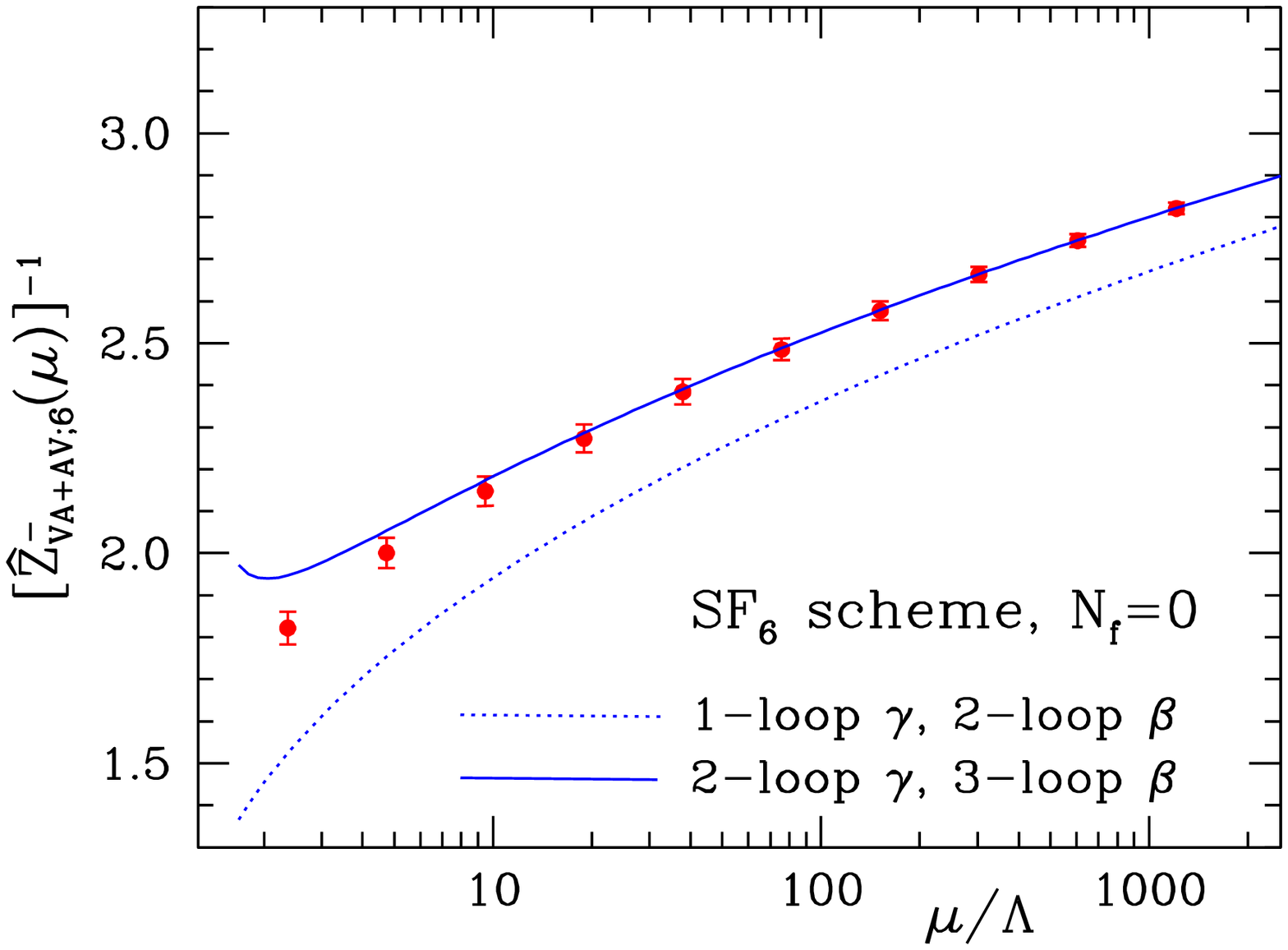}
\vspace{-18mm}
\caption{ (continued)
}
\end{figure}\addtocounter{figure}{-1}

\clearpage

\begin{figure}[p]
\centering
\vspace{178mm}
\includegraphics{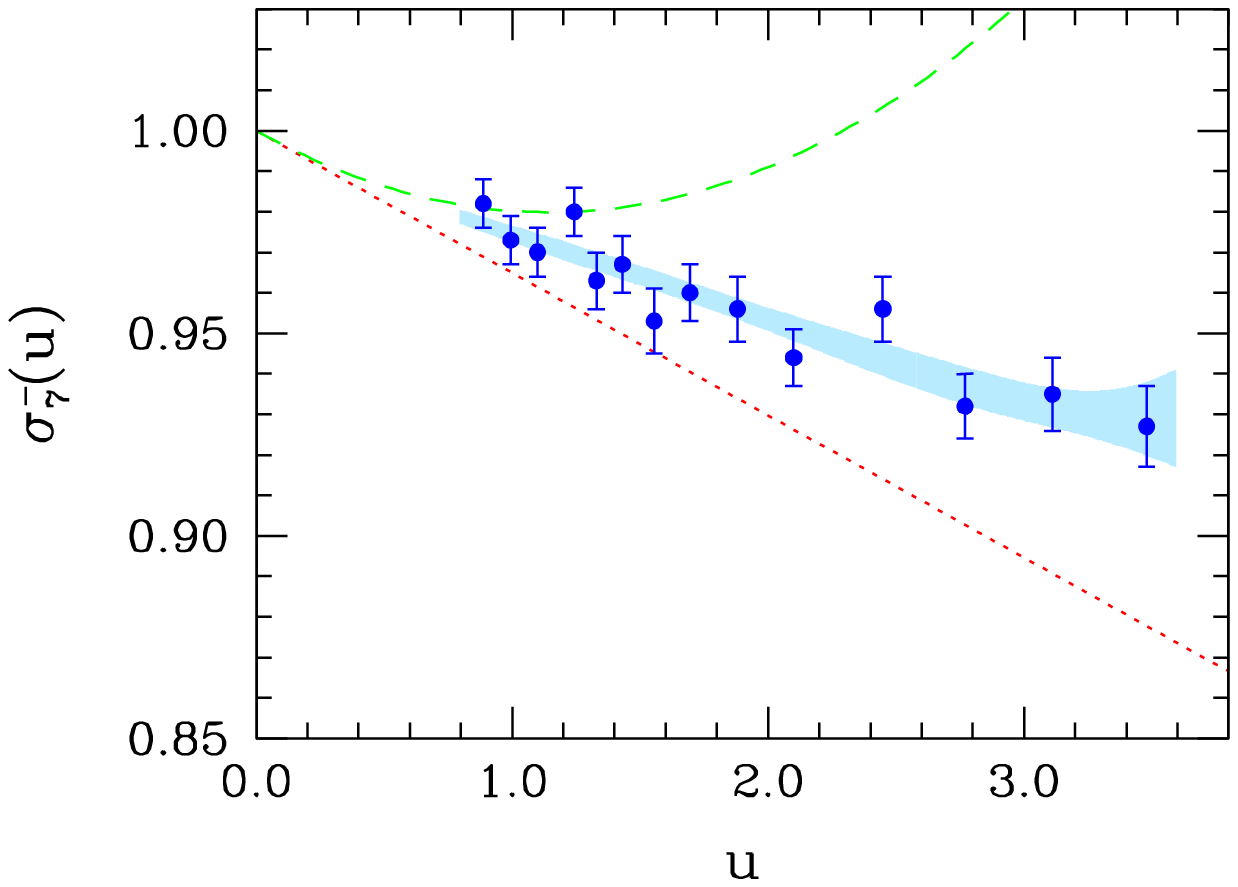}
\includegraphics{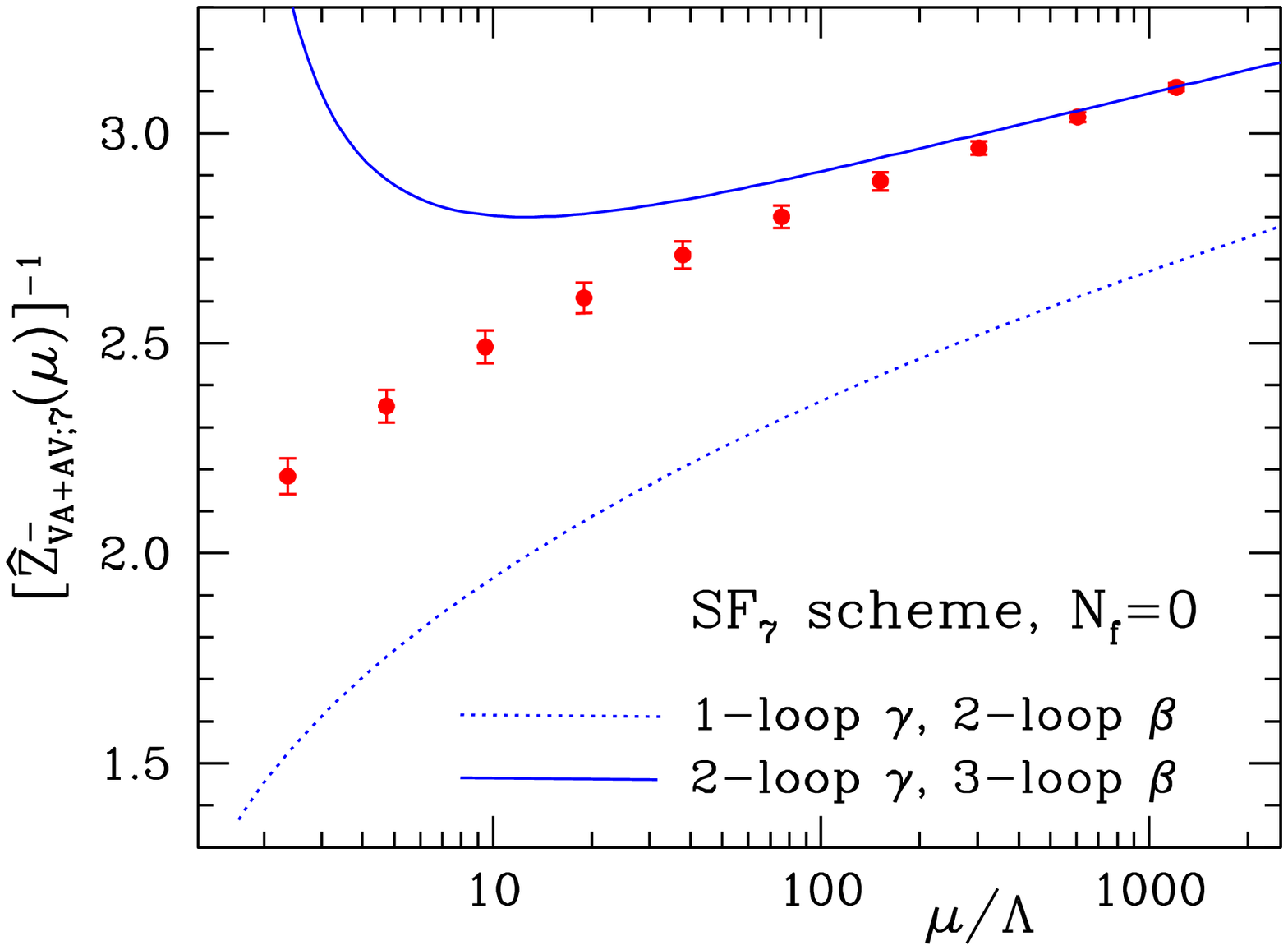}
\includegraphics{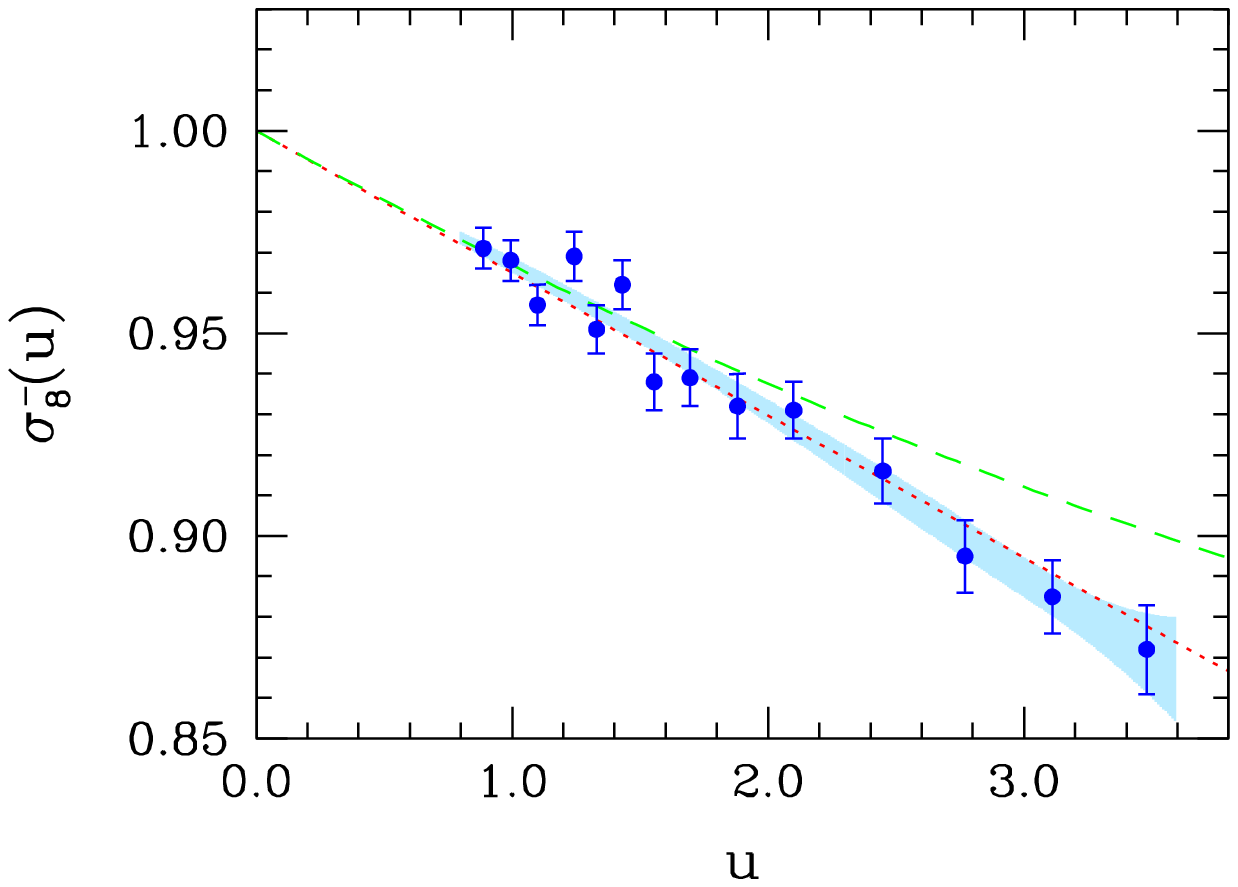}
\includegraphics{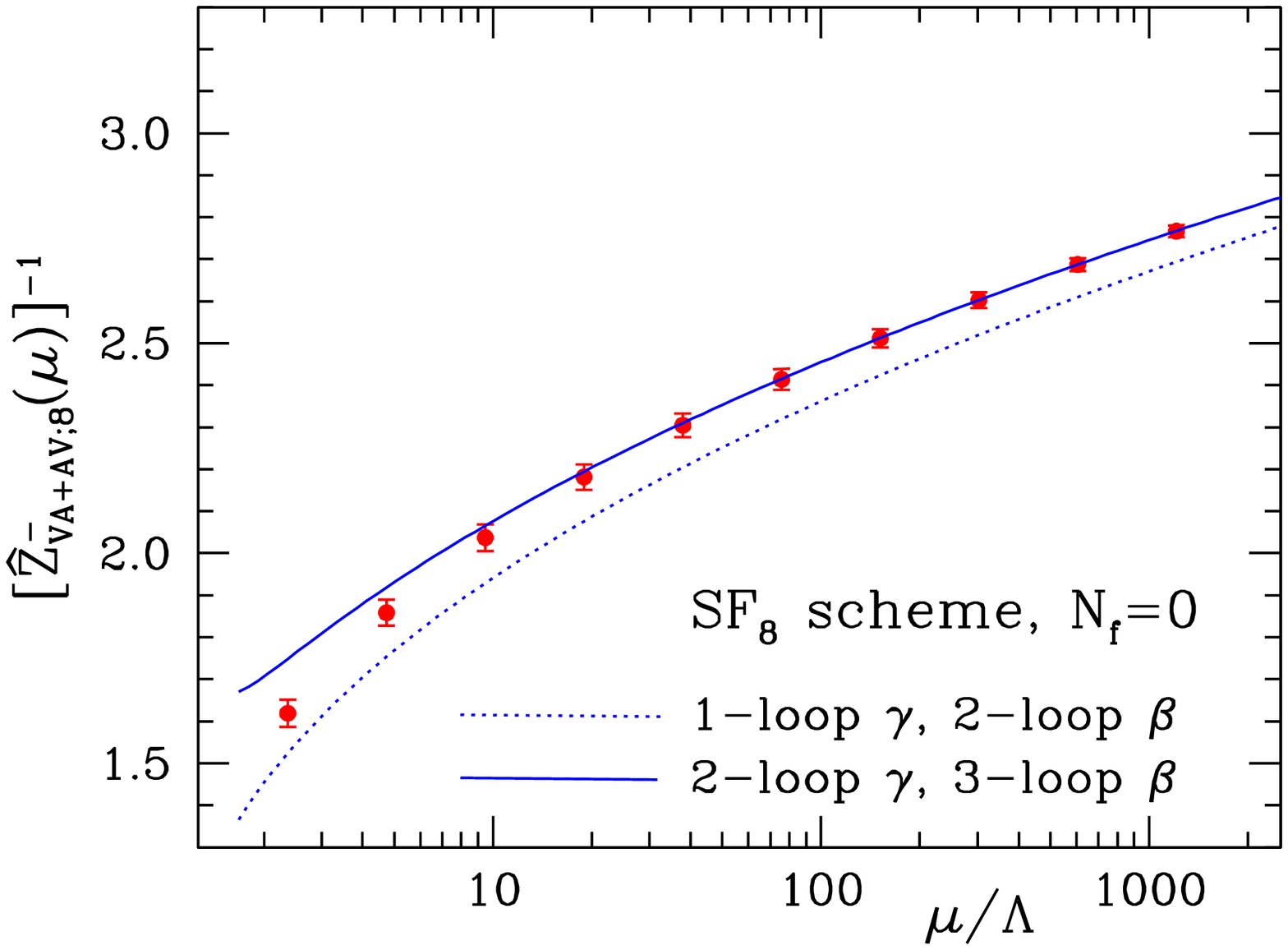}
\includegraphics{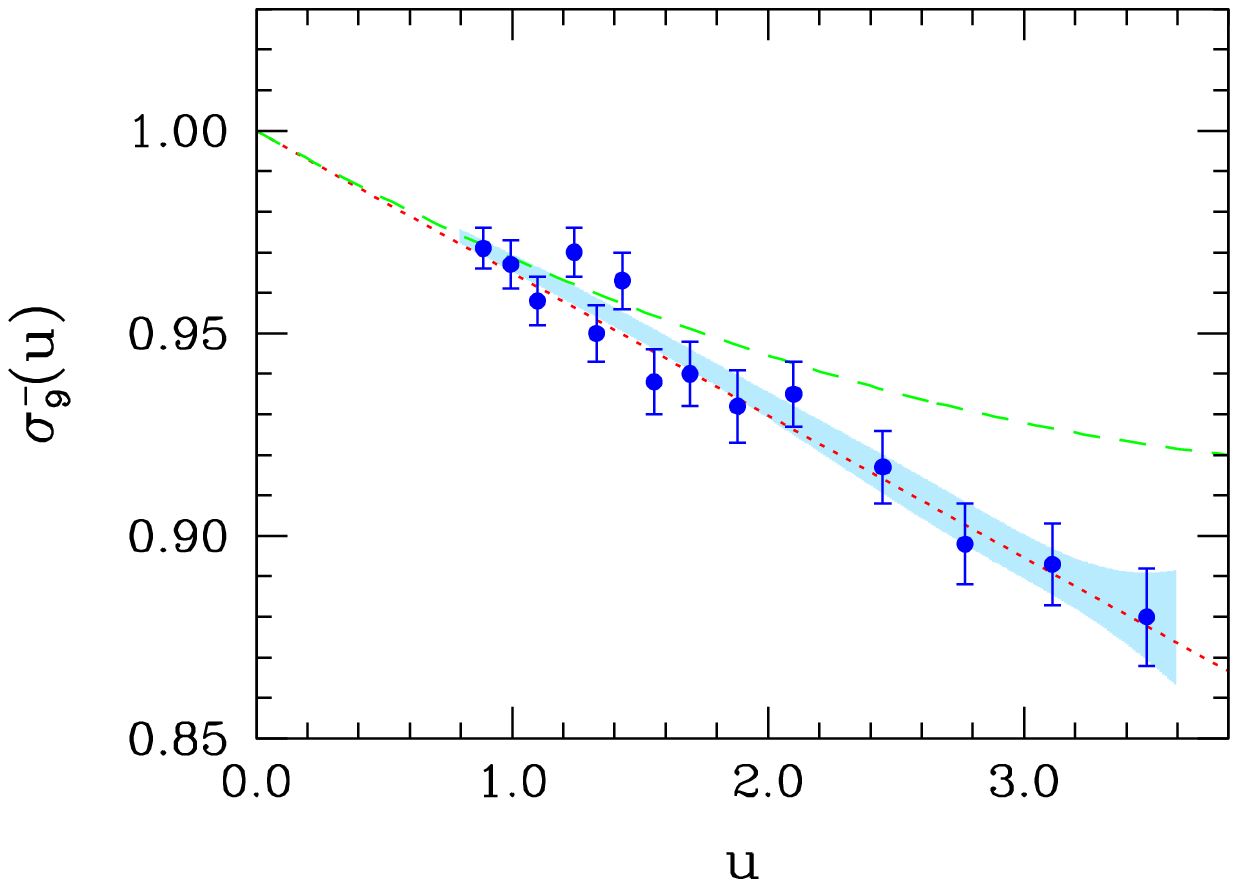}
\includegraphics{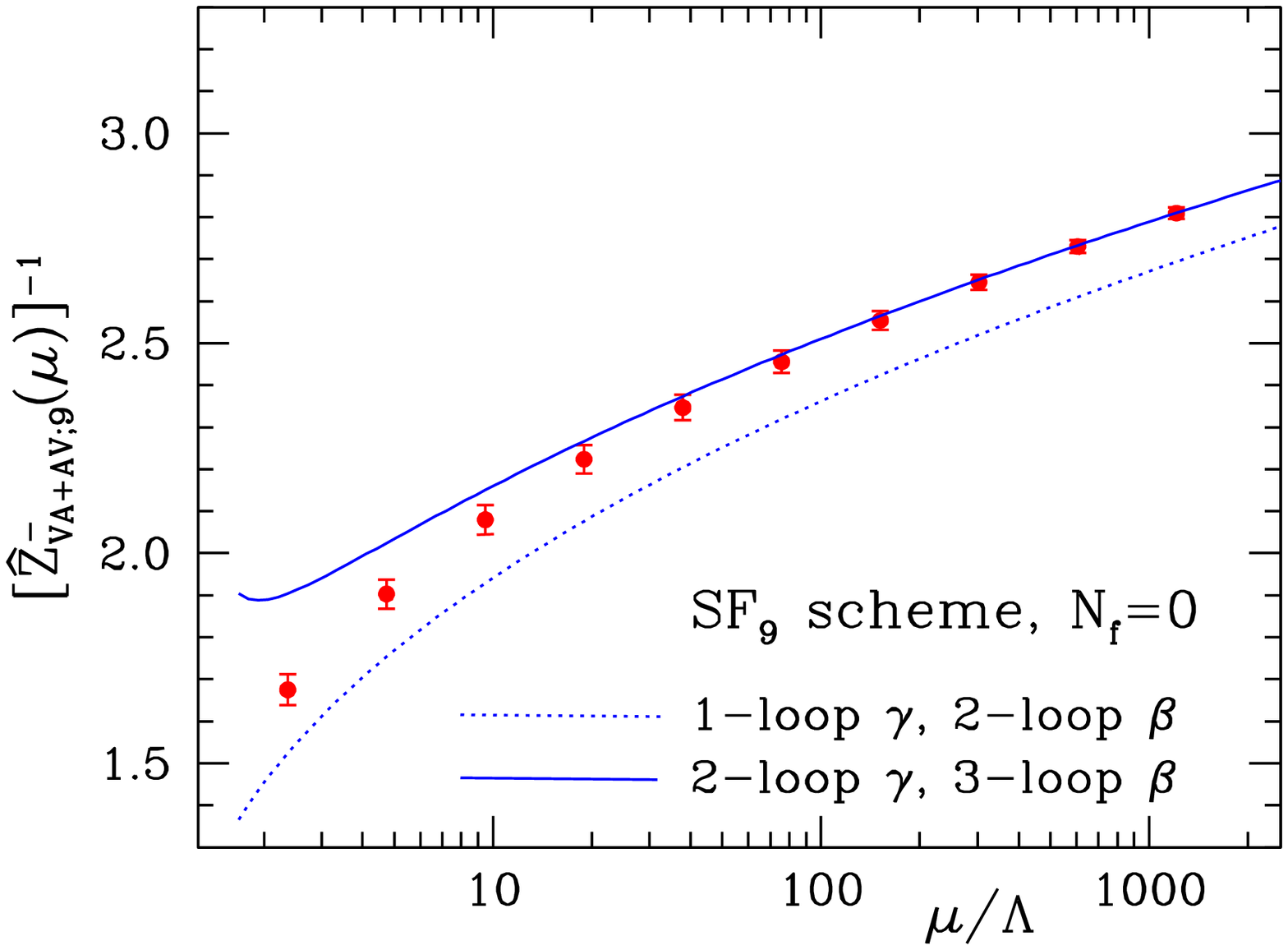}
\vspace{-18mm}
\caption{ (continued)
}
\end{figure}

\clearpage

\begin{table}
\centering
\begin{tabular}{rr@{\hspace{10mm}}ll@{\hspace{10mm}}ll}
\Hline \\[-1.0ex]
\multicolumn{2}{c}{} &
\multicolumn{2}{c}{Clover action~~~~~~~} &
\multicolumn{2}{c}{Wilson action~~~} \\[1.0ex]
$\beta~~~$ & $\frac{L}{a}$ &
$~~~~~~\hopc$ &  $\ZVApAV{;s}^+$  &
$~~~~~~\hopc$ &  $\ZVApAV{;s}^+$  \\[1.0ex]
\hline \\[-1.0ex]
6.0219 & 8 & 0.135043(17) & 0.7985(27) & 0.153371(10) & 0.6435(28) \\ 
6.1628 & 10 & 0.135643(11) & 0.8228(17) & 0.152012(7) & 0.6743(24) \\ 
6.2885 & 12 & 0.135739(13) & 0.8400(20) & 0.150752(10) & 0.6978(33) \\ 
6.4956 & 16 & 0.135577(7) & 0.8758(46) & 0.148876(13) & 0.7377(48) \\ 
[1.0ex]
6.0219 & 8 & 0.135043(17) & 0.8661(42) & 0.153371(10) & 0.6802(37) \\ 
6.1628 & 10 & 0.135643(11) & 0.8873(26) & 0.152012(7) & 0.7145(33) \\ 
6.2885 & 12 & 0.135739(13) & 0.9029(31) & 0.150752(10) & 0.7413(46) \\ 
6.4956 & 16 & 0.135577(7) & 0.9412(74) & 0.148876(13) & 0.7861(66) \\ 
[1.0ex]
6.0219 & 8 & 0.135043(17) & 0.8851(36) & 0.153371(10) & 0.7010(33) \\ 
6.1628 & 10 & 0.135643(11) & 0.9105(22) & 0.152012(7) & 0.7370(29) \\ 
6.2885 & 12 & 0.135739(13) & 0.9272(26) & 0.150752(10) & 0.7647(40) \\ 
6.4956 & 16 & 0.135577(7) & 0.9667(61) & 0.148876(13) & 0.8096(58) \\ 
[1.0ex]
6.0219 & 8 & 0.135043(17) & 0.7911(32) & 0.153371(10) & 0.6232(29) \\ 
6.1628 & 10 & 0.135643(11) & 0.8100(20) & 0.152012(7) & 0.6534(26) \\ 
6.2885 & 12 & 0.135739(13) & 0.8252(24) & 0.150752(10) & 0.6768(36) \\ 
6.4956 & 16 & 0.135577(7) & 0.8585(56) & 0.148876(13) & 0.7183(52) \\ 
[1.0ex]
6.0219 & 8 & 0.135043(17) & 0.7963(32) & 0.153371(10) & 0.6226(29) \\ 
6.1628 & 10 & 0.135643(11) & 0.8153(20) & 0.152012(7) & 0.6534(25) \\ 
6.2885 & 12 & 0.135739(13) & 0.8307(24) & 0.150752(10) & 0.6770(35) \\ 
6.4956 & 16 & 0.135577(7) & 0.8633(56) & 0.148876(13) & 0.7215(50) \\ 
[1.0ex]
\Hline
\end{tabular}
\caption{
Results for $\ZVApAV{;s}^+(g_0,L/a)$ at fixed scale $L=1.436 \,r_0$
(corresponding to $\mumin = (2\lmax)^{-1}$).
Each block contains the results from a different renormalization condition ($s=1,\ldots,5$).
}
\label{tab:Zm1}
\end{table}

\clearpage

\begin{table}
\centering
\begin{tabular}{rr@{\hspace{10mm}}ll@{\hspace{10mm}}ll}
\Hline \\[-1.0ex]
\multicolumn{2}{c}{} &
\multicolumn{2}{c}{Clover action~~~~~~~} &
\multicolumn{2}{c}{Wilson action~~~} \\[1.0ex]
$\beta~~~$ & $\frac{L}{a}$ &
$~~~~~~\hopc$ &  $\ZVApAV{;s}^+$  &
$~~~~~~\hopc$ &  $\ZVApAV{;s}^+$  \\[1.0ex]
\hline \\[-1.0ex]
6.0219 & 8 & 0.135043(17) & 0.7058(25) & 0.153371(10) & 0.5670(25) \\ 
6.1628 & 10 & 0.135643(11) & 0.7236(16) & 0.152012(7) & 0.5919(22) \\ 
6.2885 & 12 & 0.135739(13) & 0.7388(19) & 0.150752(10) & 0.6104(29) \\ 
6.4956 & 16 & 0.135577(7) & 0.7686(43) & 0.148876(13) & 0.6486(42) \\ 
[1.0ex]
6.0219 & 8 & 0.135043(17) & 0.7722(24) & 0.153371(10) & 0.6209(26) \\ 
6.1628 & 10 & 0.135643(11) & 0.7948(15) & 0.152012(7) & 0.6501(22) \\ 
6.2885 & 12 & 0.135739(13) & 0.8111(18) & 0.150752(10) & 0.6719(29) \\ 
6.4956 & 16 & 0.135577(7) & 0.8446(40) & 0.148876(13) & 0.7122(42) \\ 
[1.0ex]
6.0219 & 8 & 0.135043(17) & 0.6902(24) & 0.153371(10) & 0.5520(23) \\ 
6.1628 & 10 & 0.135643(11) & 0.7071(15) & 0.152012(7) & 0.5764(20) \\ 
6.2885 & 12 & 0.135739(13) & 0.7219(17) & 0.150752(10) & 0.5946(27) \\ 
6.4956 & 16 & 0.135577(7) & 0.7501(40) & 0.148876(13) & 0.6319(40) \\ 
[1.0ex]
6.0219 & 8 & 0.135043(17) & 0.6947(23) & 0.153371(10) & 0.5515(22) \\ 
6.1628 & 10 & 0.135643(11) & 0.7117(14) & 0.152012(7) & 0.5764(19) \\ 
6.2885 & 12 & 0.135739(13) & 0.7267(17) & 0.150752(10) & 0.5948(26) \\ 
6.4956 & 16 & 0.135577(7) & 0.7543(39) & 0.148876(13) & 0.6347(37) \\ 
[1.0ex]
\Hline
\end{tabular}
\caption{
Results for $\ZVApAV{;s}^+(g_0,L/a)$ at fixed scale $L=1.436 \,r_0$
(corresponding to $\mumin = (2\lmax)^{-1}$).
Each block contains the results from a different renormalization condition ($s=6,\ldots,9$).
}
\label{tab:Zm2}
\end{table}

\clearpage

\begin{table}
\centering
\begin{tabular}{rr@{\hspace{10mm}}ll@{\hspace{10mm}}ll}
\Hline \\[-1.0ex]
\multicolumn{2}{c}{} &
\multicolumn{2}{c}{Clover action~~~~~~~} &
\multicolumn{2}{c}{Wilson action~~~} \\[1.0ex]
$\beta~~~$ & $\frac{L}{a}$ &
$~~~~~~\hopc$ & $\ZVApAV{;s}^-$ &
$~~~~~~\hopc$ & $\ZVApAV{;s}^-$ \\[1.0ex]
\hline \\[-1.0ex]
6.0219 & 8 & 0.135043(17) & 0.5499(18) & 0.153371(10) & 0.6368(27) \\ 
6.1628 & 10 & 0.135643(11) & 0.5561(11) & 0.152012(7) & 0.6212(21) \\ 
6.2885 & 12 & 0.135739(13) & 0.5573(13) & 0.150752(10) & 0.6115(29) \\ 
6.4956 & 16 & 0.135577(7) & 0.5592(30) & 0.148876(13) & 0.6006(40) \\ 
[1.0ex]
6.0219 & 8 & 0.135043(17) & 0.6474(36) & 0.153371(10) & 0.7150(40) \\ 
6.1628 & 10 & 0.135643(11) & 0.6472(21) & 0.152012(7) & 0.7023(32) \\ 
6.2885 & 12 & 0.135739(13) & 0.6452(25) & 0.150752(10) & 0.6922(44) \\ 
6.4956 & 16 & 0.135577(7) & 0.6524(57) & 0.148876(13) & 0.6895(67) \\ 
[1.0ex]
6.0219 & 8 & 0.135043(17) & 0.6743(25) & 0.153371(10) & 0.7719(35) \\ 
6.1628 & 10 & 0.135643(11) & 0.6814(16) & 0.152012(7) & 0.7570(28) \\ 
6.2885 & 12 & 0.135739(13) & 0.6809(18) & 0.150752(10) & 0.7497(38) \\ 
6.4956 & 16 & 0.135577(7) & 0.6866(41) & 0.148876(13) & 0.7311(51) \\ 
[1.0ex]
6.0219 & 8 & 0.135043(17) & 0.5549(24) & 0.153371(10) & 0.6056(26) \\ 
6.1628 & 10 & 0.135643(11) & 0.5538(14) & 0.152012(7) & 0.5944(21) \\ 
6.2885 & 12 & 0.135739(13) & 0.5529(17) & 0.150752(10) & 0.5854(29) \\ 
6.4956 & 16 & 0.135577(7) & 0.5558(37) & 0.148876(13) & 0.5869(46) \\ 
[1.0ex]
6.0219 & 8 & 0.135043(17) & 0.5516(26) & 0.153371(10) & 0.6057(29) \\ 
6.1628 & 10 & 0.135643(11) & 0.5514(15) & 0.152012(7) & 0.5941(23) \\ 
6.2885 & 12 & 0.135739(13) & 0.5520(18) & 0.150752(10) & 0.5858(32) \\ 
6.4956 & 16 & 0.135577(7) & 0.5562(40) & 0.148876(13) & 0.5868(50) \\ 
[1.0ex]
\Hline
\end{tabular}
\caption{
Results for $\ZVApAV{;s}^-(g_0,L/a)$ at fixed scale $L=1.436 \,r_0$
(corresponding to $\mumin = (2\lmax)^{-1}$).
Each block contains the results from a different renormalization condition ($s=1,\ldots,5$).
}
\label{tab:Zm3}
\end{table}

\clearpage

\begin{table}
\centering
\begin{tabular}{rr@{\hspace{10mm}}ll@{\hspace{10mm}}ll}
\Hline \\[-1.0ex]
\multicolumn{2}{c}{} &
\multicolumn{2}{c}{Clover action~~~~~~~} &
\multicolumn{2}{c}{Wilson action~~~} \\[1.0ex]
$\beta~~~$ & $\frac{L}{a}$ &
$~~~~~~\hopc$ & $\ZVApAV{;s}^-$ &
$~~~~~~\hopc$ & $\ZVApAV{;s}^-$ \\[1.0ex]
\hline \\[-1.0ex]
6.0219 & 8 & 0.135043(17) & 0.5276(23) & 0.153371(10) & 0.5960(27) \\ 
6.1628 & 10 & 0.135643(11) & 0.5278(14) & 0.152012(7) & 0.5819(22) \\ 
6.2885 & 12 & 0.135739(13) & 0.5279(16) & 0.150752(10) & 0.5700(29) \\ 
6.4956 & 16 & 0.135577(7) & 0.5328(36) & 0.148876(13) & 0.5689(44) \\ 
[1.0ex]
6.0219 & 8 & 0.135043(17) & 0.5883(19) & 0.153371(10) & 0.6837(30) \\ 
6.1628 & 10 & 0.135643(11) & 0.5948(12) & 0.152012(7) & 0.6677(23) \\ 
6.2885 & 12 & 0.135739(13) & 0.5956(14) & 0.150752(10) & 0.6586(32) \\ 
6.4956 & 16 & 0.135577(7) & 0.5999(32) & 0.148876(13) & 0.6432(41) \\ 
[1.0ex]
6.0219 & 8 & 0.135043(17) & 0.4841(20) & 0.153371(10) & 0.5364(22) \\ 
6.1628 & 10 & 0.135643(11) & 0.4835(12) & 0.152012(7) & 0.5243(17) \\ 
6.2885 & 12 & 0.135739(13) & 0.4837(14) & 0.150752(10) & 0.5143(24) \\ 
6.4956 & 16 & 0.135577(7) & 0.4856(30) & 0.148876(13) & 0.5163(38) \\ 
[1.0ex]
6.0219 & 8 & 0.135043(17) & 0.4812(21) & 0.153371(10) & 0.5365(23) \\ 
6.1628 & 10 & 0.135643(11) & 0.4813(12) & 0.152012(7) & 0.5241(18) \\ 
6.2885 & 12 & 0.135739(13) & 0.4829(14) & 0.150752(10) & 0.5147(26) \\ 
6.4956 & 16 & 0.135577(7) & 0.4860(31) & 0.148876(13) & 0.5162(40) \\ 
[1.0ex]
\Hline
\end{tabular}
\caption{
Results for $\ZVApAV{;s}^-(g_0,L/a)$ at fixed scale $L=1.436 \,r_0$
(corresponding to $\mumin = (2\lmax)^{-1}$).
Each block contains the results from a different renormalization condition ($s=6,\ldots,9$).
}
\label{tab:Zm4}
\end{table}

\appendixend

\end{document}